\documentclass{aa}

\pdfoutput=1
\usepackage{graphicx,url,twoopt,natbib}
\usepackage{hyperref}                

\hypersetup{
  colorlinks=true,   
  urlcolor=blue,     
  linkcolor=blue     
}

\bibpunct{(}{)}{;}{a}{}{,}    

\makeatletter
\newcommand{\bibnote}[2]{\global\@namedef{#1note}{#2}}
\newcommand{\biblink}[2]{\global\@namedef{#1link}{#2}}
\makeatother

\makeatletter
  \protected\def\stonyslink{%
     \def\hyper@linkstart##1##2{}\let\hyper@linkend\@empty}
  \newcommandtwoopt{\citeads}[3][][]{%
   \href{http://ui.adsabs.harvard.edu/abs/#3/abstract}%
        {\stonyslink \citealp[#1][#2]{#3}}
   \biblink{#3}{\href{http://ui.adsabs.harvard.edu/abs/#3/abstract}{ADS}}}
 \newcommandtwoopt{\citepads}[3][][]{%
   \href{http://ui.adsabs.harvard.edu/abs/#3/abstract}%
        {\stonyslink \citep[#1][#2]{#3}}
   \biblink{#3}{\href{http://ui.adsabs.harvard.edu/abs/#3/abstract}{ADS}}}
 \newcommandtwoopt{\citetads}[3][][]{%
   \href{http://ui.adsabs.harvard.edu/abs/#3/abstract}%
        {\stonyslink \citet[#1][#2]{#3}}
  \biblink{#3}{\href{http://ui.adsabs.harvard.edu/abs/#3/abstract}{ADS}}}
 \newcommandtwoopt{\citeyearads}[3][][]{%
   \href{http://ui.adsabs.harvard.edu/abs/#3/abstract}%
        {\stonyslink \citeyear[#1][#2]{#3}}
   \biblink{#3}{\href{http://ui.adsabs.harvard.edu/abs/#3/abstract}{ADS}}}
\makeatother

\usepackage{graphicx}

\usepackage{savesym}
\usepackage{amsmath}
\savesymbol{iint}
\usepackage{txfonts}
\restoresymbol{TXF}{iint}

\usepackage[utf8]{inputenc} 
\usepackage[T1]{fontenc}    
\usepackage{hyperref}       
\usepackage{url}            
\usepackage{booktabs}       
\usepackage{amsfonts}       
\usepackage{nicefrac}       
\usepackage{microtype}      
\usepackage{lipsum}
\usepackage{multicol}       
\usepackage{hyperref}       
\hypersetup{colorlinks=true,linkcolor=blue, linktocpage}
\usepackage{textcomp}
\usepackage{gensymb}        
\usepackage{graphicx}       
\usepackage{amssymb}        
\usepackage{float}          
\usepackage{subcaption}
\usepackage{stfloats}       
\usepackage[export]{adjustbox} 
\usepackage[notquote]{hanging}        
\usepackage[labelfont=bf]{caption}  
\usepackage[font=scriptsize]{caption}   

\usepackage{pdflscape}
\usepackage{afterpage}
\usepackage{capt-of}

\usepackage{times}

\begin{document} 

    \title{An underlying clock in the extreme flip-flop state transitions of the black hole transient Swift J1658.2-4242}

\author{D. Bogensberger \inst{1}, G. Ponti \inst{2,1}, C. Jin \inst{3}, T. M. Belloni \inst{2}, H. Pan \inst{3}, K. Nandra \inst{1}, T. D. Russell \inst{4}, J. C. A. Miller-Jones \inst{5}, T.~Mu\~noz-Darias \inst{6,7}, P. Vynatheya \inst{8}, and F. Vincentelli \inst{9}}
\institute{Max Planck Institut für Extraterrestrische Physik, Gie{\ss}enbachstra{\ss}e 1, 85748 Garching bei München, Germany \\
    \email{dbogen@mpe.mpg.de}
    \and
        INAF - Osservatorio Astronomico di Brera, via E. Bianchi 46, 23807 Merate, Italy
    \and
        National Astronomical Observatories, Chinese Academy of Sciences, A20 Datun Road, Beijing 100101, China
    \and
        Anton Pannekoek Institute of Astronomy, Science Park 904, 1098 XH Amsterdam, Netherlands
    \and
        International Centre for Radio Astronomy Research – Curtin University, GPO Box U1987, Perth, WA 6845, Australia
    \and
        Instituto de Astrof\'isica de Canarias, E-38205 La Laguna, Tenerife, Spain
    \and 
        Departamento de Astrof\'\i{}sica, Universidad de La Laguna, E-38206 La Laguna, Tenerife, Spain
    \and 
        IISER Kolkata, Mohanpur, Nadia district, West Bengal - 741246, India
    \and
        Department of Physics and Astronomy, University of Southampton, Highfield, Southampton SO17 1BJ, UK
        }


\abstract{
        \textit{\textbf{Aims}}:  Flip-flops are top-hat-like X-ray flux variations which have been observed in some transient accreting black hole binary systems, and feature simultaneous changes in the spectral hardness and the Power Density Spectrum (PDS). They occur at a crucial time in the evolution of these systems, when the accretion disk emission starts to dominate over coronal emission. Flip-flops remain a poorly understood phenomenon, so we aim to thoroughly investigate them in a system featuring several such transitions.
        
		\textit{\textbf{Methods}}: Within the multitude of observations of \object{Swift J1658.2-4242} during its outburst in early 2018, we detected 15 flip-flops, enabling a detailed analysis of their individual properties, and the differences between them. We present observations by \textit{XMM-Newton}, \textit{NuSTAR}, \textit{Astrosat}, \textit{Swift}, \textit{Insight}-HXMT, \textit{INTEGRAL}, and \textit{ATCA}. We analyse their light curves, search for periodicities, compute their PDS, and fit their X-ray spectra, to investigate the source behaviour during flip-flop transitions, and how the interval featuring flip-flops differs from the rest of the outburst. 

		\textit{\textbf{Results}}: The flip-flops of Swift J1658.2-4242 are of an extreme variety, exhibiting flux differences of up to 77\% within $\sim100~\mathrm{s}$, much larger than has been seen previously. We observed radical changes in the PDS simultaneous with the sharp flux variations, featuring transitions between the Quasi-Periodic Oscillation types C and A, which have never been observed before. Changes to the PDS are delayed, but more rapid than changes in the light curve. Flip-flops occur in two intervals within the outburst, separated by about two weeks in which these phenomena were not seen. Transitions between the two flip-flop states occurred at random integer multiples of a fundamental period, of $2.761~\mathrm{ks}$ in the first interval, and $2.61~\mathrm{ks}$ in the second. Spectral analysis reveals the high and low flux flip-flop states to be very similar, but distinct from intervals lacking flip-flops. A change in the inner temperature of the accretion disk is responsible for most of the flux difference in the flip-flops. We also highlight the importance of correcting for the influence of the dust scattering halo on the X-ray spectra. }


\titlerunning{Extreme flip-flops of Swift J1658.2-4242}
\authorrunning{David Bogensberger}

\maketitle 

\section{Introduction}

\subsection{Spectral-timing states of a black hole transient}

Low-mass transient black hole binaries usually spend most of their time in a quiescent state, with a low accretion rate, and hence a low luminosity. These periods of quiescence are interspersed by outbursts in which the accretion rate and source flux increase by many orders of magnitude (see e.g. \citeads{1997ApJ...491..312C}, \citeads{2006ARA&A..44...49R}). Outbursts typically last for a few months, with the source passing through several well-defined states (see e.g. \citeads{2011BASI...39..409B}). These are distinguished, and classified via their intensity, spectral, and timing properties (\citeads{2005Ap&SS.300..107H}). These different states can be visualized on a Hardness Intensity Diagram (HID), which shows the X-ray intensity against the flux ratio between a hard and a soft X-ray energy band. Black hole transients (BHTs) evolve through a hysteresis `q'-shaped path, traversed in an anti-clockwise fashion (see e.g. \citeads{2004MNRAS.355.1105F}).

At the start and end of the outburst, the system is located in the lower right region of the HID. It has a low luminosity and a hard X-ray spectrum, which is dominated by a power law component produced via Compton up-scattering of soft thermal X-ray photons from the accretion disk by high energy electrons in the corona (\citeads{2010LNP...794...17G}). In the standard truncated disk model (\citeads{1997ApJ...489..865E}, and \citeads{1997MNRAS.292L..21P}), the geometrically thin, optically thick accretion disk is terminated at a large radius, so that the disk black body component only makes a small contribution to the X-ray spectrum. The system also features a steady, compact jet (see e.g. \citeads{2001ApJ...554...43C}). This is known as the Low Hard State (LHS). 
As the accretion rate and luminosity increase, the truncation radius of the accretion disk starts to decrease. This causes the emission of the disk to become stronger, and the spectrum overall to be characterized by a combination of a multi-temperature black body, as well as a power law component. This part of the outburst is classified into two types: the Hard Intermediate and Soft Intermediate States (HIMS and SIMS, respectively), which can for example be distinguished by their variability. The transition from the HIMS to the SIMS is marked by the steady compact jet switching off, being quenched by more than 3.5 orders of magnitude (\citeads{2019ApJ...883..198R}) after a short-lived, bright and rapidly flaring transient jet is launched  (see e.g. \citeads{2004MNRAS.355.1105F}). Sometimes a BHT temporarily reaches an anomalous state (AS), which branches off of the HID at the SIMS, towards even higher luminosities and greater hardness. 

The truncation radius continues to decrease as the luminosity starts to drop slightly. When the accretion disk extends all the way to the innermost stable orbit, the spectrum is dominated by black body emission from the accretion disk, whereas the Comptonized power law component becomes relatively insignificant. This part of the outburst is characterized as the High Soft State (HSS), during which no jet is launched (see e.g. \citeads{1999ApJ...519L.165F}). Eventually, after the source has faded enough, the hardness starts to increase again, and the BHT returns to the LHS, via the SIMS and HIMS, before finally returning to quiescence. This hardening in the intermediate states occurs at a lower luminosity than the initial softening, and features the re-establishment of the compact jet over a few weeks (see e.g. \citeads{2014MNRAS.439.1390R}). Multiple transitions between the HIMS and the SIMS usually occur in both of these horizontal regions of the HID. 

The above describes the standard picture of the evolution of a BHT in an outburst, but it is by no means universal. For example, some outbursts fail to reach some of the softer states, before returning to the LHS, and then to quiescence (see e.g. \citeads{2016ASSL..440...61B}).  

\subsection{Quasi-periodic oscillations}
\label{sec:QPOIntro}

Temporal analysis of the X-ray light curves of many BHTs has revealed the existence of strong stochastic noise, along with Quasi-Periodic Oscillations (QPOs). QPOs appear as Lorentzian peaks in the Power Density Spectrum (PDS), which is the distribution of squared Fourier frequencies of the light curve. Greater variability and QPO amplitude at higher X-ray energies suggests that the QPOs originate very close to the black hole; in the corona or the inner accretion flow (see e.g. \citeads{1997ApJ...488L.109B}). Accordingly, QPOs provide an additional useful window for studying this region, and are essential in distinguishing different parts of an outburst. They come in two classes: Low (LFQPOs), and High Frequency QPOs (HFQPOs) (\citeads{2016AN....337..398M}). LFQPOs have frequencies in the range $0.01-30~\mathrm{Hz}$, but are typically found below $10~\mathrm{Hz}$, and have been detected and studied much more than HFQPOs, which commonly have frequencies of several hundreds of $\mathrm{Hz}$. The LFQPOs have themselves been grouped into three different types: A, B, and C (\citeads{1999ApJ...526L..33W}, and for a detailed analysis, see e.g. \citeads{2005ApJ...629..403C}, or \citeads{2016AN....337..398M}). The three types are distinguished via their PDS properties. PDSs are usually fitted with a sum of several Lorentzian functions (\citeads{2002ApJ...572..392B}), which describe both the QPO and the stochastic noise continuum:
\begin{equation*}
L(\nu) = \frac{r^2\nu_0}{2\pi Q}\left(\left(\nu-\nu_0\right)^2+\left(\frac{\nu_0}{2Q}\right)^2\right)^{-1}
\end{equation*}
Here, $\nu_0$ is the centroid frequency, at which the Lorentzian reaches its maximum. $Q=\nu_0/\mathrm{FWHM}$ is the quality factor, which is used to characterize the width of the Lorentzian. The temporal variability in the light curve is parameterized by the fractional root mean square variability (rms). $r$ describes the variability of a single Lorentzian. If the PDS is normalized according to \citeads{1990A&A...227L..33B}, the rms is equal to the square root of the PDS integral: $r=\sqrt{\int_{-\infty}^{\infty}L(\nu)\,d\nu}$. The total rms of a PDS is determined by taking the square root of its properly normalized integral between two frequencies: $\nu_1$, and $\nu_2$, and is typically expressed as a percentage.  $\mathrm{rms}=\sqrt{\int_{\nu_1}^{\nu_2}\left(\sum_{i}^nL_i(\nu)\right)\,d\nu}$, where $n$ is the number of Lorentzian components used in the fit.

Type C QPOs are observed most frequently, and are usually detected in the LHS and HIMS, although they can appear in all spectral-timing states (\citeads{2016AN....337..398M}). They have frequencies spanning the entire LFQPO frequency range: $0.01~\mathrm{Hz}\lesssim\nu_0\leq30~\mathrm{Hz}$, are very narrow: $Q \gtrsim 10$, and have a large rms variability: $5\%\lesssim \mathrm{rms}\lesssim 20\%$. The strong broad-band continuum associated with type C QPOs contributes significantly towards their large rms. The continuum appears as a flat top noise extending up to a break frequency, above which it drops steeply. The break frequency has a similar value to, and correlates with the QPO frequency (\citeads{1999ApJ...514..939W}). The type C frequency also roughly correlates with the X-ray flux, and anti-correlates with the broad-band rms (\citeads{2011MNRAS.418.2292M}).

Type B QPOs are the defining characteristic of the SIMS, distinguishing it from the HIMS. (\citeads{2016ASSL..440...61B}). They are typically located at: $\nu_0\sim6~\mathrm{Hz}$ in the early bright phase of the outburst, but have also been found at lower frequencies, especially in the low flux horizontal branch traversed in the return to the LHS (\citeads{2011MNRAS.418.2292M}). They are also quite narrow: $Q\gtrsim 6$, and have low variability amplitude: $\mathrm{rms}\lesssim 4\%$. The rms is predominantly due to the QPO peak, with a comparatively small contribution from the power law continuum noise (see e.g. \citeads{2016AN....337..398M}).

Type A QPOs are observed even less frequently than type Bs and are the hardest to detect. This is due to their broad and very shallow peak, with: $Q\lesssim 3$, at frequencies of $6 ~\mathrm{Hz}\lesssim \nu_0 \lesssim 8\, \mathrm{Hz}$. Type A QPOs have the lowest variability amplitude, of $\mathrm{rms}\lesssim 3\%$, a result of the weak QPO and a small contribution from the continuum noise. Type A QPOs do not feature harmonics, unlike types B, and C. They are commonly found in the HSS (\citeads{2005Ap&SS.300..107H}).

Despite considerable analysis of LFQPOs, there is no consensus as to their physical origin. QPOs are thought to originate close to the black hole, but LFQPO frequencies are substantially smaller than the Keplerian orbital frequencies at these radii, which are several hundred Hz. \citeads{2011MNRAS.418.2292M}, and \citeads{2012MNRAS.427..595M} argue that QPO types B and C have different physical origins, whereas types A and C could be due to the same process. They point to the different correlations between QPO frequency and power law flux of the three QPO types, along with the observation of simultaneous type B and C QPOs in the PDS of GRO J1655-40. 

The relativistic precession model (\citeads{1998ApJ...492L..59S}) identifies the centroid frequency of LFQPOs with the frequency of  Lense-Thirring precession at a single radius in the accretion disk. Within this model, pairs of HFQPOs are identified as corresponding to periastron precession and Keplerian frequency at the same radius (\citeads{1999ApJ...524L..63S}).

\citeads{2009MNRAS.397L.101I}, \citeads{2013MNRAS.434.1476I}, and \citeads{2016MNRAS.461.1967I} describe the precessing inner flow model. In this model, the entire inner flow undergoes Lense-Thirring precession as a rigid body. A variation in inclination of the inner flow leads to a modulation of the Doppler boost, which subsequently generates the QPO.

The accretion ejection instability model (\citeads{1999A&A...349.1003T}, \citeads{2002A&A...387..497V}, \citeads{2002A&A...394..329V}, and \citeads{2016A&A...591A..36V}) hypothesizes an instability with which energy and angular momentum are transported from a magnetized accretion disk to the corona, via a spiral density wave and a Rossby vortex. In this model, QPO types A, B, and C are distinguished by being produced in a relativistic, semi-relativistic, and non-relativistic regime of the instability, respectively. 

Another model for the QPOs is the transition layer model (\citeads{2004ApJ...612..988T}). In this model, a transition layer separates the outer parts of the accretion disk, which orbit at a Keplerian frequency, from the inner regions, which have a sub-Keplerian orbital frequency. The orbital frequency at this transition layer would equal the QPO frequency. 

A fourth hypothesis by \citeads{2010MNRAS.404..738C} considers a magneto-acoustic wave in the corona, which causes a variation in the efficiency of comptonization, thereby generating a QPO. 

In summary, QPO properties provide an additional mode for ascertaining the physical processes occurring in the vicinity of the black hole, and are an effective way of classifying the state of the source, providing information which the flux and energy spectra alone cannot supply. However, the physical origin of the QPOs is still debated, with several hypotheses positing vastly different mechanisms. While observations suggest a clear distinction between QPO types A, B, and C, it remains unknown whether these are the result of different physical processes, or merely different regimes of the same fundamental mechanism.

\subsection{Flip-flops}

Flip-flops, first noted by \hypertarget{M91}{\citeads{1991ApJ...383..784M}} (hereafter M91), are rare phenomena which appear as top-hat like flux variations in the light curve of some black hole transients in outburst. Flip-flops are distinguished from absorption dips by their longer duration, their top-hat like shape, and the positive correlation between flux and hardness during transitions (in contrast, dips have softer spectra at higher fluxes). Flip-flops are also seen to accompany changes in the PDS and are considered to be associated with some state transitions. Most state transitions however do not involve a flip-flop. 

Due to the marked and abrupt change in flux which defines the flip flop, we refer to the higher and lower flux levels as the ``bright state'' and ``dim state'' respectively, and analyse them separately.

Out of a population of $\sim 60$ galactic BHTs (\citeads{2016A&A...587A..61C}), only seven systems exhibited properties fitting the description of a flip-flop: GX 339-4 (\hyperlink{M91}{M91}, \hypertarget{N03}{\citeads{2003A&A...412..235N}}, hereafter N03), GS 1124-683 (\hypertarget{T97}{\citeads{1997ApJ...489..272T}}, hereafter T97), XTE J1550-564 (\hypertarget{H01}{\citeads{2001ApJS..132..377H}}, hereafter H01, \citeads{2016ApJ...823...67S}), XTE J1859+226 (\hypertarget{C04}{\citeads{2004A&A...426..587C}}, hereafter C04, \citeads{2013ApJ...775...28S}), H1743-322 (\hypertarget{H05}{\citeads{2005ApJ...623..383H}}, hereafter H05), XTE J1817-330 (\hypertarget{S12}{\citeads{2012A&A...541A...6S}}), and MAXI J1659-152 (\hypertarget{K11}{\citeads{2011ApJ...731L...2K}}, hereafter K11, \citeads{2013A&A...552A..32K}). There are other systems in which similar properties have been observed, such as in GRS 1915+105 (\citeads{2008MNRAS.383.1089S}), or MAXI J1535-571 (\citeads{2018ApJ...866..122H}), but for which an identification as a flip-flop is not certain. 

Flip-flops exhibit a large variety of different properties, not only between different systems, or different outbursts of the same system, but also within a single outburst of a BHT. Some flip-flops are seen at, or very close to the peak of the outburst (\hyperlink{M91}{M91},  \hyperlink{T97}{T97}, \hyperlink{H01}{H01}, \hyperlink{N03}{N03}, and \hyperlink{C04}{C04}), while others are observed somewhat later, when the flux was dropping back down again (\hyperlink{C04}{C04}, \hyperlink{H05}{H05}, \hyperlink{K11}{K11}, and \hyperlink{S12}{S12}). 

The observed flux change during a flip-flop also differs from system to system. Bright states can have an X-ray flux of between $3\%$ and $33\%$ greater than the flux of neighbouring dim states. Transition times have also been observed to span a large range, from fractions of a second, up to more than $1~\mathrm{ks}$. A large spread of values is also seen in the duration of each state. The time between adjacent flip-flops can be anywhere between a few tens of seconds, and several $\mathrm{ks}$. There seems to be a relation between those two parameters, in that short-lived flip-flop states have fast transitions (\hyperlink{M91}{M91}, \hyperlink{T97}{T97}, and \hyperlink{S12}{S12}), and long-lived states have slow transitions (\hyperlink{N03}{N03}, \hyperlink{H01}{H01}, \hyperlink{C04}{C04}, and \hyperlink{H05}{H05}).

All flip-flops observed so far have involved a type B QPO in either the bright or the dim state. Therefore, flip-flops seem to require transitions between the SIMS and either the HIMS or the AS. 

\hyperlink{C04}{C04} noticed a hierarchy of QPO states in the flip-flops of XTE J1859+226, with type As occurring at the highest luminosities, type Cs at the lowest luminosities, and type Bs in between. The limiting fluxes separating the different QPO types were found to decrease exponentially with time. All observed flip-flops between QPO types B and C agree with this hierarchy (\hyperlink{T97}{T97}, \hyperlink{H01}{H01}, \hyperlink{C04}{C04}, and \hyperlink{K11}{K11}), but flip-flops between QPO types A and B disagree with it (\hyperlink{M91}{M91}, \hyperlink{N03}{N03}, and \hyperlink{S12}{S12}) more often than they agree with it (\hyperlink{C04}{C04}, and \hyperlink{H05}{H05}). Despite the fascinating phenomenology of flip-flops, a clear physical interpretation is still lacking. 

\subsection{Swift J1658.2-4242}

At 05:37:23 UTC on February 16, 2018 (MJD 58165), the Burst Alert Telescope on the \textit{Neil Gehrels Swift Observatory} (\textit{Swift}/BAT, \citeads{2004ApJ...611.1005G}) was triggered by an X-ray burst from a point source in the galactic plane, (\citeads{2018GCN.22416....1B}). This X-ray transient was subsequently called Swift J1658.2-4242. This first detection was quickly followed up with an observation by the \textit{Swift X-Ray Telescope} (\textit{Swift}/XRT), by \citeads{2018GCN.22417....1D}, and \citeads{2018GCN.22419....1L}, as well as one observation with the \textit{Mobile Astronomical System of TElescope Robots}, at the Observatorio Astronomico Felix Aguilar (\textit{MASTER-OAFA}, \citeads{2010AdAst2010E..30L}), by \citeads{2018GCN.22418....1L}. \citeads{2018ATel11306....1G} performed subsequent examinations of data gathered by \textit{Swift}/BAT (\citeads{2018GCN.22419....1L}), and by the \textit{International Gamma-Ray Astrophysics Laboratory} (\textit{INTEGRAL}, \citeads{2003A&A...411L...1W}), which revealed that the source had started flaring on February 13, 2018 (MJD 58162). \citeads{2018GCN.22419....1L} determined a more accurate location of it, at $\mathrm{RA}=16\mathrm{h}\, 58\mathrm{m}\, 12.64\mathrm{s}$, $\mathrm{Dec}=-42\degree\, 41'\, 54.4 ''$, and found its spectrum to be well described by a highly absorbed power law. Radio observations with the \textit{Australia Telescope Compact Array} (\textit{ATCA}) strongly supported the notion that Swift J1658.2-4242 is a black hole X-ray binary, rather than a neutron star X-ray binary (\citeads{2018ATel11322....1R}).

\citeads{2018ApJ...865...18X} described the first observation of the source by the \textit{Nuclear Spectroscopic Telescope Array} (\textit{NuSTAR}, \citeads{2013ApJ...770..103H}). They fitted the spectrum using a highly absorbed power law with a relativistic reflection component, and a Gaussian component to fit an absorption line at $7.03~\mathrm{keV}$. They found the properties of the source to be consistent with known properties of black hole binaries, and therefore described it as a black hole candidate. They also noted a type C QPO, whose centroid frequency increased continuously along with the flux, from $0.14~\mathrm{Hz}$ to $0.21~\mathrm{Hz}$ throughout this observation.  

\citeads{2019ApJ...875..157J} analysed the spectra detected by the \textit{Chandra X-ray Observatory} (\textit{Chandra}, \citeads{2000SPIE.4012....2W}), and by the \textit{X-ray Multi-Mirror Mission} (\textit{XMM-Newton}, \citeads{2001A&A...365L...1J}). They detected a strong X-ray Dust Scattering Halo (DSH), which significantly affected the observed source spectrum, and could be fitted with three dust layers at different distances from the BHT. Multiple methods were used to determine the distance to Swift J1658.2-4242, which was estimated to be $\sim 10~\mathrm{kpc}$.

\citeads{2019ApJ...879...93X} examined the first flip-flop transition at the brightest part of the outburst, using data from both \textit{NuSTAR} and \textit{XMM-Newton}. They described a $45\%$ flux decrease in $\sim 40~\mathrm{s}$. The lower flux level contained a type C QPO, but no QPO was detected at the high flux level. Spectral fitting revealed only small differences in the high and low flux spectra. The strong relativistic reflection effect detected in the LHS by \citeads{2018ApJ...865...18X} was no longer present in these observations. 

Detailed timing analysis of the HIMS of Swift J1658.2-4242, using observations from the \textit{Hard X-ray Modulation Telescope} (\textit{Insight}-HXMT, \citeads{2018SPIE10699E..1UZ}), the \textit{Neutron Star Interior Composition Explorer} (\textit{NICER}, \citeads{2012SPIE.8443E..13G}), and \textit{Astrosat} (\citeads{2014SPIE.9144E..1SS}) was performed by \citeads{2019JHEAp..24...30X}. They traced the increase in the QPO frequency alongside a decrease in the rms, determined the independence of QPO frequency and photon energy, and found the phase lag to be close to 0. Their results are consistent with a high inclination for Swift J1658.2-4242. 

\citeads{2019ApJ...887..101J} made a timing and spectral analysis of the three \textit{Astrosat} observations of Swift J1658.2-4242, examining the properties of the detected QPO and its harmonics, their energy dependence, and time lag. The first observation featured a type C QPO which increased in frequency from $1.56~\mathrm{Hz}$ to $1.74~\mathrm{Hz}$. The second observation featured additional flux variations reminiscent of the one observed by \citeads{2019ApJ...879...93X}. A $6.6~\mathrm{Hz}$ type C QPO was only observed at low fluxes. By the final observation, the QPO frequency had dropped to $4.0~\mathrm{Hz}$. The QPO fractional rms was found to increase with photon energy. Positive time lags suggested that the QPO originated in the corona, and propagated outwards.  

\citeads{2018ATel12072....1B} reported that Swift J1658.2-4242 had returned to quiescence by October 2018. 

\section{Observations}

Following the initial detection, Swift J1658.2-4242 was observed by \textit{NuSTAR}, \textit{XMM-Newton}, \textit{Astrosat}, \textit{Chandra}, \textit{Swift}, Insight-\textit{HXMT}, \textit{INTEGRAL}, and \textit{NICER} in the X-ray band. \textit{ATCA} also provided coverage in the radio band. We utilized measurements from all these instruments to obtain the best possible understanding of the phenomena occurring during this outburst. 

\textit{XMM-Newton}, \textit{NuSTAR}, \textit{Astrosat}, and \textit{Chandra} provided detailed high quality information of the properties of the source at a few distinct parts of the outburst, enabling comprehensive analysis of the light curve, PDS, and energy spectrum. \textit{Swift}/XRT and BAT, \textit{Insight}-HXMT LE, ME, and HE, \textit{NICER}, and \textit{INTEGRAL} provided short snapshots at many different times, allowing an investigation of day-to-day variations, and an overall impression of the entire outburst, thereby complementing the sparse, yet long duration observations by other instruments. 

Out of the eight \textit{NuSTAR} observations, six were performed simultaneously with \textit{XMM-Newton}, and one simultaneously with \textit{Chandra}. There was also an overlap between the second combined \textit{NuSTAR} and \textit{XMM-Newton} observation analysed here, with the second observation by \textit{Astrosat}. These long duration observations by \textit{NuSTAR}, \textit{XMM-Newton}, \textit{Astrosat}, and \textit{Chandra} are summarized in Table \ref{listobsX}.

\textit{Swift}/XRT performed 78 observations of Swift J1658.2-4242, spanning more than seven months. \textit{Swift}/BAT produces daily average count rates. 42 observations by \textit{NICER} were used to examine the timing evolution of the BHT. \textit{INTEGRAL} obtained 151 observations of the BHT during its outburst, providing good hard X-ray coverage on shorter timescales than \textit{Swift}/BAT.

Swift J1658.2-4242 was also observed with \textit{Insight}-HXMT in 21 individual pointing observations, from MJD 58169 to 58196, with a total net exposure time of $\sim700~ks$, at low, medium, and high X-ray energies.

\textit{ATCA} examined the radio emission from Swift J1658.2-4242 at 10 different times within the outburst. These observations are summarized in Table \ref{listobsR} 

\begin{table*}[h]
\centering
\setlength{\tabcolsep}{4pt}
\def\arraystretch{1.1}
\begin{tabular}{l l l l l l l}
    \textbf{Instrument} & \textbf{ObsID} & \textbf{Start time (MJD)} & \textbf{Exposure (ks)} & \textbf{State} & \textbf{QPO} \\ \hline \hline
    
    \textit{NuSTAR} & 90401307002 & 58165.97658 & 33.337 & LHS & C \\ \hline
    
    \textit{Astrosat} & T02\_004T01\_9000001910 & 58169.76510 & 9.961 & HIMS & C \\ \hline
    
    \begin{tabular}{@{}l}
        \textit{NuSTAR}\\
        \textit{XMM-Newton}\\
    \end{tabular} & 
    \begin{tabular}{@{}l}
        80301301002\\
        0802300201\\
    \end{tabular} & 
    \begin{tabular}{@{}l}
        58174.14666\\
        58174.25471\\
    \end{tabular} & 
    \begin{tabular}{@{}l}
        31.537\\
        70.526\\
    \end{tabular} & 
    \begin{tabular}{@{}l}
    Flip-flops\\
    HIMS$\leftrightarrow$AS
    \end{tabular} &
    C and A \\ \hline

    \begin{tabular}{@{}l}
        \textit{NuSTAR}\\
        \textit{XMM-Newton}\\
    \end{tabular} & 
    \begin{tabular}{@{}l}
        80302302002\\
        0811213401\\
    \end{tabular} & 
    \begin{tabular}{@{}l}
        58176.75478\\
        58176.79290\\
    \end{tabular} & 
    \begin{tabular}{@{}l}
        24.025\\
        60.000\\
    \end{tabular} & 
    \begin{tabular}{@{}l}
    Flip-flops\\
    HIMS$\leftrightarrow$AS
    \end{tabular} &
    C and A \\ \hline

    \begin{tabular}{@{}l}
        \textit{Astrosat}\\
        \textit{NuSTAR}\\
        \textit{XMM-Newton}\\
    \end{tabular} & 
    \begin{tabular}{@{}l}
        T02\_011T01\_9000001940\\
        80302302004\\
        0805200201\\
    \end{tabular} & 
    \begin{tabular}{@{}l}
        58180.78589\\
        58181.31859\\
        58181.33765\\
    \end{tabular} & 
    \begin{tabular}{@{}l}
        29.578\\
        25.076\\
        59.800\\
    \end{tabular} & 
    \begin{tabular}{@{}l}
    Flip-flops\\
    HIMS$\leftrightarrow$AS
    \end{tabular} &
    C and A \\ \hline

    \begin{tabular}{@{}l}
        \textit{NuSTAR}\\
        \textit{XMM-Newton}\\
    \end{tabular} & 
    \begin{tabular}{@{}l}
        80302302006\\
        0805200301\\
    \end{tabular} & 
    \begin{tabular}{@{}l}
        58188.16432\\
        58188.21508\\
    \end{tabular} & 
    \begin{tabular}{@{}l}
        28.245\\
        51.200\\
    \end{tabular} & 
    HIMS & C \\ \hline

    \begin{tabular}{@{}l}
        \textit{NuSTAR}\\
        \textit{XMM-Newton}\\
    \end{tabular} & 
    \begin{tabular}{@{}l}
        80302302008\\
        0805200401\\
    \end{tabular} & 
    \begin{tabular}{@{}l}
        58192.46058\\
        58192.54771\\
    \end{tabular} & 
    \begin{tabular}{@{}l}
        26.714\\
        50.140\\
    \end{tabular} & 
    HIMS & C \\ \hline
    
    \begin{tabular}{@{}l}
        \textit{NuSTAR}\\
        \textit{XMM-Newton}\\
    \end{tabular} & 
    \begin{tabular}{@{}l}
        80302302010\\
        0805201301\\
    \end{tabular} & 
    \begin{tabular}{@{}l}
        58205.01353\\
        58205.07581\\
    \end{tabular} & 
    \begin{tabular}{@{}l}
        29.102\\
        50.900\\
    \end{tabular} & 
    \begin{tabular}{@{}l}
    Late flip-flops\\
    HIMS$\leftrightarrow$SIMS
    \end{tabular} &
    No QPO\\ \hline    
    
    \begin{tabular}{@{}l}
        \textit{NuSTAR}\\
        \textit{Chandra}\\
    \end{tabular} & 
    \begin{tabular}{@{}l}
        90401317002\\
        21083\\
    \end{tabular} & 
    \begin{tabular}{@{}l}
        58235.95971\\
        58236.05065\\
    \end{tabular} & 
    \begin{tabular}{@{}l}
        28.493\\
        30.080\\
    \end{tabular} & 
    SIMS & No QPO\\ \hline 
    
\end{tabular}
\vspace{3 mm}
\caption{List of X-ray observations of Swift J1658.2-4242 by \textit{Astrosat}, \textit{Chandra}, \textit{NuSTAR}, and \textit{XMM-Newton} (in Timing mode). Observations with overlapping exposures have been grouped together. \label{listobsX}}
\end{table*}

\section{Data analysis}

\subsection{X-ray data}
We produced \textit{Swift}/XRT light curves and spectra using the University of Leicester's online data analysis software\footnote{\href{http://www.swift.ac.uk/user_objects/}{http://www.swift.ac.uk/user\_objects/}}, using standard input parameters to create one spectrum for each observation. Dead pixel, pileup, and vignetting corrections were applied. The BAT light curve was obtained from the \textit{Swift}/BAT transient monitor website\footnote{\href{https://swift.gsfc.nasa.gov/results/transients/}{https://swift.gsfc.nasa.gov/results/transients/}} (\citeads{2013ApJS..209...14K}), which bins observations made throughout each day into one datapoint.

\textit{NuSTAR} data were reduced using the standard procedure with \texttt{nupipeline} and \texttt{nuproducts}, using NuSTARDAS version 1.8.0, CALDB version 4.7.8 and the general functionality of HEASoft version 6.25. In all observations, the lower left corner of Focal Plane Module B (FPMB) of \textit{NuSTAR} was contaminated by stray light from a nearby source. To prevent this from affecting the extracted source signal, we restricted the size of the source extraction region, as there was some overlap between the source point spread function, and the affected region of the CCD. We chose a circular region with a radius of 155" centred at the location of the source, for FPMB. This selection excluded virtually all the stray light seen in the CCD and included almost all the source photons. Focal Plane Module A (FPMA) did not suffer the same problem, so we chose a much larger source extraction region of a circle with a radius of 270" instead, also centred on the source. This different selection enabled us to test whether our choice of source extraction region for FPMB affected its spectrum. There are slight differences between the spectra of the two instruments at low energies, but they are not significant to this analysis. Slight normalization differences between the two instruments are as expected. Background extraction regions were selected manually for each observation and focal plane module, as the area of a circle of radius 30", placed far away from the source, and any other possible contaminants in the field of view. We restricted the \textit{NuSTAR} energy range to $3\textup{--}79~\mathrm{keV}$. \textit{NuSTAR} spectra were also rebinned with GRPPHA to contain at least 20 counts per energy bin. 

Observations by \textit{XMM-Newton} were taken by the pn, and MOS2 detectors of the European Photon Imaging Camera (EPIC, \citeads{2001A&A...365L..18S}). The EPIC-pn camera was set to Timing and Burst mode, in which the pnCCD achieves high temporal resolution, but loses spatial resolution along the $y$ direction of the CCD. Thus, instead of a circular region, we selected individual pixel columns. As Swift J1658.2-4242 was very bright throughout all \textit{XMM-Newton} observations, the entire CCD was dominated by source photons. We therefore used all observed photons in the creation of the light curves and spectra and did not apply a background correction. The contribution from the background was estimated by comparing the spectra in the strip of $4\times200$ pixels furthest away from the source to the spectra of the remaining $60\times200$ pixels of the CCD. The four pixel wide column had a comparatively larger soft spectral component, than the 60 pixel column. We interpret this as being primarily due to background contamination. As the four pixel column was still dominated by the source, it would be inadvisable to use this as a background estimate. Instead, we decided to ensure minimal background contamination of the source spectra, by ignoring all energies for which the four pixel column had a flux equal to at least $5\%$ of the flux in the remaining pixels. This limit was exceeded at all energies below $2.7\, \mathrm{keV}$, but not at any energies higher than that. We therefore restricted the \textit{XMM-Newton} spectra to the energy range $2.7\textup{--}10~\mathrm{keV}$. 

We used version 16.1.0 of \textit{XMM-Newton}'s Science Analysis System (SAS), implementing \texttt{epchain} to create event files, then using the standard procedures to produce spectra and light curves. We used \texttt{epiclccorr} to correct the light curves for telemetry dropouts. A few soft proton flares were detected when plotting the \textit{XMM-Newton} light curves at energies exceeding $12 ~\mathrm{keV}$. All time intervals for which the count rate at these energies exceeded $4~\mathrm{counts}~\mathrm{s}^{-1}$ were excluded from the light curves. Photons were binned into $6 ~\mathrm{ms}$ intervals, limited to that value by the frame time for EPIC-pn in Timing mode. We compared the count rates of the observations made in Timing mode with those made in Burst mode shortly before or afterwards, to test whether the \textit{XMM-Newton} observations were suffering from pile up. Such comparisons indicated a negligible level of pileup. 

\citeads{2018GCN.22419....1L} determined that Swift J1658.2-4242 is heavily obscured, with a hydrogen column density of $N_H\sim10^{23} ~\mathrm{cm^{-2}}$. \citeads{2019ApJ...875..157J} also detected a strong DSH around the black hole candidate. This provides a challenge for spectral fitting, as the DSH results effectively in a significant, energy dependent modification of the point spread function of the source. Spectra extracted from insufficiently large regions are therefore distorted, in particular at low energies. \citeads{2017MNRAS.468.2532J}, and \citeads{2019ApJ...875..157J} developed the spectral fitting model \texttt{dscor} to account for scattering of X-rays by a DSH along the line of sight. However, these models only work for circular source extraction regions and could therefore not be used for the \textit{XMM-Newton} observations in Timing mode. In order to fit these spectra accurately as well, we developed a new spectral fitting model to describe the effect of the DSH on rectangular extraction regions which are not centred on the source.

We used the \textit{Chandra} Grating-Data Archive and Catalogue webpage\footnote{\href{http://tgcat.mit.edu/}{http://tgcat.mit.edu/}} to calibrate the data from the \textit{Chandra} High Energy Transmission Grating (HETG) observation of Swift J1658.2-4242, and extract an energy spectrum for the positive and negative grating directions. 

The two \textit{Astrosat} observations with the Large Area X-ray Proportional Counter (LAXPC) were first reduced using the \texttt{laxpc\_make\_event} code inside the laxpcsoft package. From the resulting level-2 event files, we generated light curves using the task \texttt{laxpc\_make\_lightcurve}. The \textit{Astrosat} observations had a significant background count rate, which was determined by the chain of analysis tasks \texttt{laxpc\_make\_spectra}, \texttt{laxpc\_make\_backspectra}, and finally \texttt{laxpc\_make\_backlightcurve}.

Swift J1658.2-4242 was observed with \textit{Insight}-HXMT in pointed observation mode starting on MJD 58169. There are 21 available individual pointing observations with a total net exposure time of $\sim700~\mathrm{ks}$. All the data were reduced following standard procedures using the \textit{Insight}-HXMT data analysis software package HXMTDAS v2.01. The good time intervals (GTIs) were filtered with the screening criteria: (1) Earth elevation angle > 10 degrees; (2) the value of the cutoff rigidity > 8; and (3) the pointing offset angle < 0.04 degrees. The events taken during satellite slews and passages through the South Atlantic Anomaly were also filtered out. For the data detected by the Medium Energy X-ray telescope (ME), the task \texttt{megti} was applied for making some corrections to the GTI file. 

The \textit{Insight}-HXMT count rate observed by the ME detector was affected by the contribution of nearby sources in the field of view (\citeads{2019JHEAp..24...30X}). Within the $5.5\textup{--}10 ~\mathrm{keV}$ energy range of interest, we determined the contamination from the brightest background source in the field of view, GX 340+0, to be $2.6~\mathrm{counts}~\mathrm{s}^{-1}$. By comparing the light curves of \textit{Insight}-HXMT ME with those of \textit{NuSTAR}, and \textit{XMM-Newton} for the same energy range, we however find the best possible agreement for an additional background count rate of $3.9~\mathrm{counts}~\mathrm{s}^{-1}$. As GX 340+0 is not the only uncorrected background source in the field of view, we used this empirical best fit value for the estimation of the additional background contribution. 

To improve our understanding of the hard X-ray evolution during the outburst, we also created light curves from observations by the IBIS/ISGRI instrument on \textit{INTEGRAL}. We extracted the count rates of individual images, each corresponding to one pointing observation, produced using the ISDC Off-line Scientific Analysis (OSA) software version 11, for five energy bands in the range $20-200 ~\mathrm{keV}$. 

\subsection{X-ray timing and spectral analysis}

X-ray observations were sorted according to whether flip-flops were detected in them. When we did not detect any flip-flops, we generated one PDS and one energy spectrum for the entire observation. But when flip-flops were detected, observations were split into individual regions, which were analysed separately. We selected intervals that started or ended at least 100 s before, or after any flip-flop transition, to be used in determining the timing, and spectral properties of each of the flip-flop states. 

Medium energy X-ray light curves were extracted for the $5.5\textup{--}10 ~\mathrm{keV}$ band. The lower limit was chosen because different telescopes disagreed on the flux densities measured at lower energies, as a result of the DSH. The upper limit was chosen because of the energy range of \textit{XMM-Newton}, and the desire to maintain consistency across different instruments. 

We generated PDSs using the GHATS software v.1.1.1 developed by T.M. Belloni, and the Stingray timing software (\citeads{2016ascl.soft08001H}). We binned individual photon detections into $10.24~\mathrm{ms}$ bins, and generated individual powerspectra for intervals of $2^{10}$ bins. These were then combined to form averaged PDSs. 

We created one PDS for every orbit within each observation by \textit{NuSTAR} and \textit{Astrosat}, as long as the orbit did not feature a flip-flop transition. If it did, the relevant orbit was subdivided further to separate the bright from the dim state. \textit{XMM-Newton} data were binned into $2~\mathrm{ks}$ segments, or shorter ones if a flip-flop transition required it. We excluded all regions with less than $500 ~\mathrm{s}$ duration, for which precise spectral fits are hampered by low statistics.

\begin{figure*}[h]
\resizebox{\hsize}{!}{\includegraphics{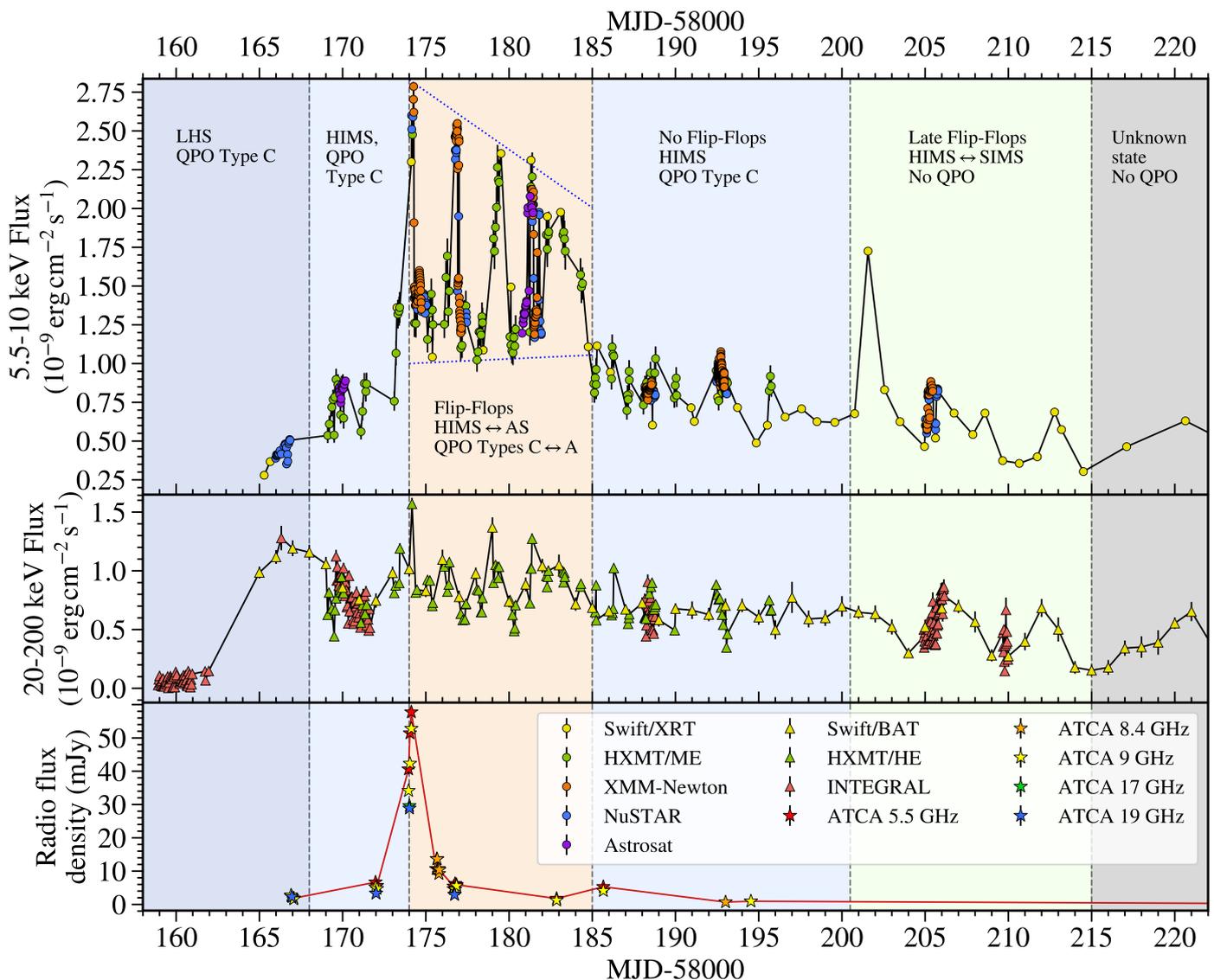}}
\caption{Light curve of the 2018 outburst of Swift J1658.2-4242 in three different energy bands. The top figure shows the flux in an intermediate energy X-ray band of $5.5\textup{--}10 ~\mathrm{keV}$, as observed by \textit{Swift}/XRT, \textit{Insight}-HXMT/ME, \textit{XMM-Newton}, \textit{NuSTAR}, and \textit{Astrosat}. In the second graph we plot the \textit{Swift}/BAT, \textit{Insight}-HXMT/HE, and \textit{INTEGRAL} light curves in the hard X-ray band of $20\textup{--}200 ~\mathrm{keV}$. The radio light curves at $5.5$, $8.4$, $9$, $17$, and $19~\mathrm{GHz}$ detected by \textit{ATCA} are plotted in the lowest graph. We split the data temporally into six regions; The initial flux increase and spectral softening in the LHS and HIMS, the flip-flops, the intermediate HIMS sandwiched between the two flip-flop intervals, the late flip-flops, and finally another interval for which we lack data to characterize accurately. The rest of the outburst was also observed, and analysed, but we primarily focused on the regions shown here. We scaled the count rates of \textit{Astrosat}, \textit{Insight}-HXMT, and \textit{Swift}/BAT to agree with observed fluxes measured by other telescopes. For both \textit{Astrosat}, and \textit{Insight}-HXMT we used the count rate in exactly the same energy band, but the \textit{Swift}/BAT light curve was obtained for the $15\textup{--}150 ~\mathrm{keV}$ range, rather than the $20\textup{--}200 ~\mathrm{keV}$ interval. We optimized the conversion factors by determining the overlap between the scaled counts and the actual fluxes measured by other instruments. Due to changes in the spectrum, a single conversion factor from count rate to flux is not accurate everywhere. This is the reason for the divergence of the \textit{Insight}-HXMT and \textit{Swift}/XRT light curves in the region between the two flip-flop intervals. \textit{NuSTAR}, \textit{XMM-Newton}, \textit{Astrosat}, \textit{Insight}-HXMT/ME, and \textit{Insight}-HXMT/HE light curves were binned into $1~\mathrm{ks}$ segments. \textit{Swift}/XRT, \textit{Swift}/BAT, INTEGRAL, and \textit{ATCA} light curves were binned to one datapoint per observation. We restricted the energy range of \textit{Swift}/XRT, \textit{NuSTAR}, and \textit{XMM-Newton} to $5.5\textup{--}10\, \mathrm{keV}$, to minimize the effects of the DSH, which has a significant impact on the spectra below $5.5~\mathrm{keV}$, and to ensure consistency between light curves of different instruments. We added 2.6\% error to all \textit{Insight}-HXMT/ME, and 1.3\% error to all \textit{Insight}-HXMT/HE measurements. \textit{Insight}-HXMT/ME was affected by a significant background. The contribution of the background to the count rate was estimated to be $3.9~\mathrm{counts}~\mathrm{s}^{-1}$ in the $5.5\textup{--}10~\mathrm{keV}$ range. The X-ray flare observed by \textit{Swift}/XRT on MJD 58201 only appeared after correcting for the removed bad pixels in the image.
 \label{LCtot}}
\end{figure*}

We subtracted the Poisson shot noise from all Leahy-normalized PDSs (\citeads{1983ApJ...266..160L}), and subsequently applied the square fractional rms normalization (\citeads{1990A&A...227L..33B}). The PDSs were rebinned logarithmically, and were then exported to XSPEC. There, we fit them using at least three different Lorentzians, at least one of which was zero-centred, following \citeads{2002ApJ...572..392B}. We fit the main QPO peak, the broad-band continuum, and all visible harmonics. We extracted the main QPO properties ($\nu_0$, $Q$, $r$, $\mathrm{rms}$) from the fits, whenever a QPO was detected. The rms was computed by integrating fitted PDSs between $0.5~\mathrm{Hz}$, and $50~\mathrm{Hz}$, and calculating the associated source and background count rates in the interval. The values of the \textit{NuSTAR} fractional rms in the QPOs were corrected for the effects of the \textit{NuSTAR} dead time (\citeads{1988tns..conf...27V}). We also applied the rms correction for suppression of high frequency power, as described by \citeads{1988tns..conf...27V}.

Spectral fits were performed using XSPEC version 12.9.1u. To model the intervening absorption we used solar abundances from \citeads{2000ApJ...542..914W}, and the photoionization cross-sections from \citeads{1996ApJ...465..487V}. The spectrum of the very first \textit{NuSTAR} observation, with ObsID 90401307002 was also analysed, but is not described here, as it was already thoroughly analysed by \citeads{2018ApJ...865...18X}, and because it differs greatly from the other \textit{NuSTAR} and \textit{XMM-Newton} spectra we investigated. It features a dominant power law component, and a negligible multicolour disk black body model (\citeads{2018ApJ...865...18X}).

\subsection{Radio data}

We obtained radio monitoring of Swift J1658.2-4242 with \textit{ATCA}, under project code C3057. Data were obtained at eight epochs from MJD 58166 -- 58235. The majority of the radio observations were taken at central frequencies of 5.5 GHz, 9 GHz, 17 GHz, and 19 GHz, where each frequency pair (5.5/9 and 17/19 GHz) was recorded simultaneously. However, our three observations after 2018-03-05 were only taken at 5.5 GHz and 9 GHz. These frequency bands were recorded with a bandwidth of 2 GHz. Additional 8.4 GHz ATCA observations were taken on 2018-02-26 and 2018-03-15 as part of observations with the Australian Long Baseline Array (project code V456). These data were recorded with a bandwidth of 64 MHz. For all radio observations the bandpass and flux calibration was performed using PKS 1934-638. Secondary phase calibration was done using the nearby (4.65 degrees away) calibrator J1714-397 for all observations except those taken on 2018-02-26 and 2018-03-15, which used J1713-4257 (2.9 degrees away).

The radio data were flagged and calibrated following standard procedures within the Common Astronomy Software Application (CASA, version 4.7.2; \citeads{2007ASPC..376..127M}). Imaging was carried out using CLEAN within CASA. Due to a significant amount of diffuse emission in the field we used a Briggs roust parameter of 0 to help mitigate effects from diffuse emission within the field. Typically, flux densities were determined by fitting for a point source in the image plane. However, due to either short on-source time, or a very compact telescope configuration, flux densities of observations on 2018-03-05, 2018-03-08, and 2018-03-17 (MJD 58182, 58185, and 58194, respectively) were determined by fitting for a point source in the uv-plane using UVMULTIFIT (\citeads{2014A&A...563A.136M}). We also used UVMULTIFIT to search for intra-observational variability within the radio observations (to identify the possibility of radio flaring associated with the X-ray flip-flops).

\section{Light curves}

\begin{figure}[h]
\resizebox{\hsize}{!}{\includegraphics{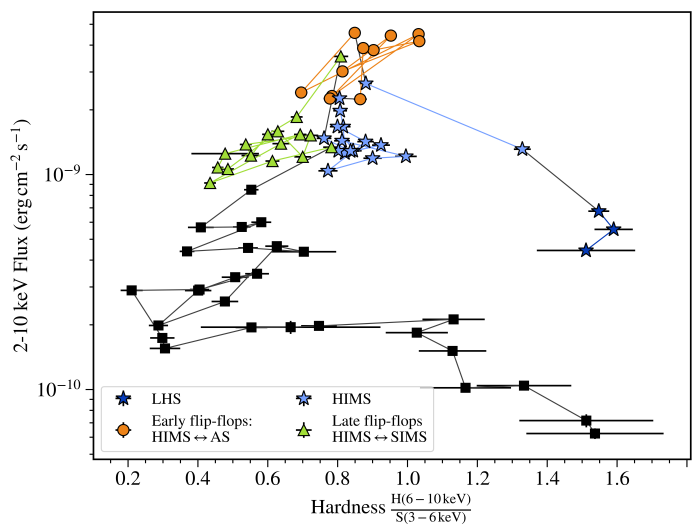}}
        \caption{HID of all \textit{Swift}/XRT observations. Colours and shapes distinguish the different states of the outburst. Hardness is defined here as the ratio of fluxes in the two energy bands $6\textup{--}10~\mathrm{keV}$, and $3\textup{--}6~\mathrm{keV}$. Each data point represents one entire observation. Black squares are observations for which a state classification was not possible. The errors in the flux are shown but are almost always smaller than the symbols used for the data points.
        \label{HIDS}}
\end{figure}

\begin{figure}[h]
\resizebox{\hsize}{!}{\includegraphics{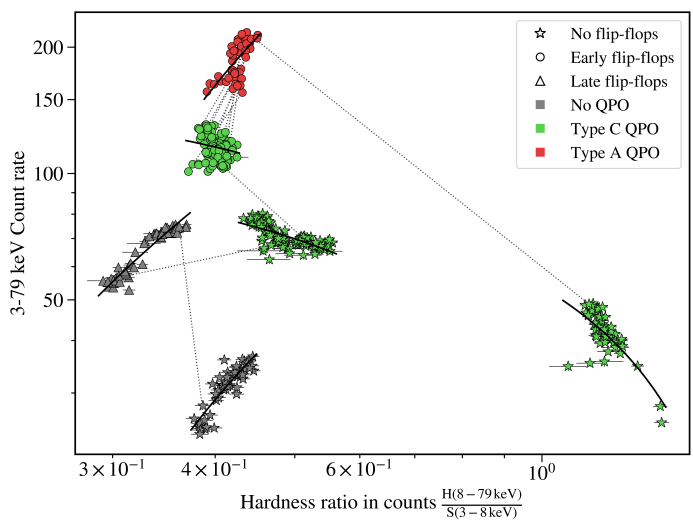}}
        \caption{\textit{NuSTAR} HID. Colours and shapes distinguish QPO types, and flip-flop detections, respectively. Hardness is defined as the ratio of count rates in the two energy bands $8\textup{--}79~\mathrm{keV}$, and $3\textup{--}8~\mathrm{keV}$. Data were binned into $200~\mathrm{s}$ intervals. Error bars have been included but are almost always smaller than the symbols used for the data points.
        \label{HIDN}}
\end{figure}

\begin{figure*}[h]
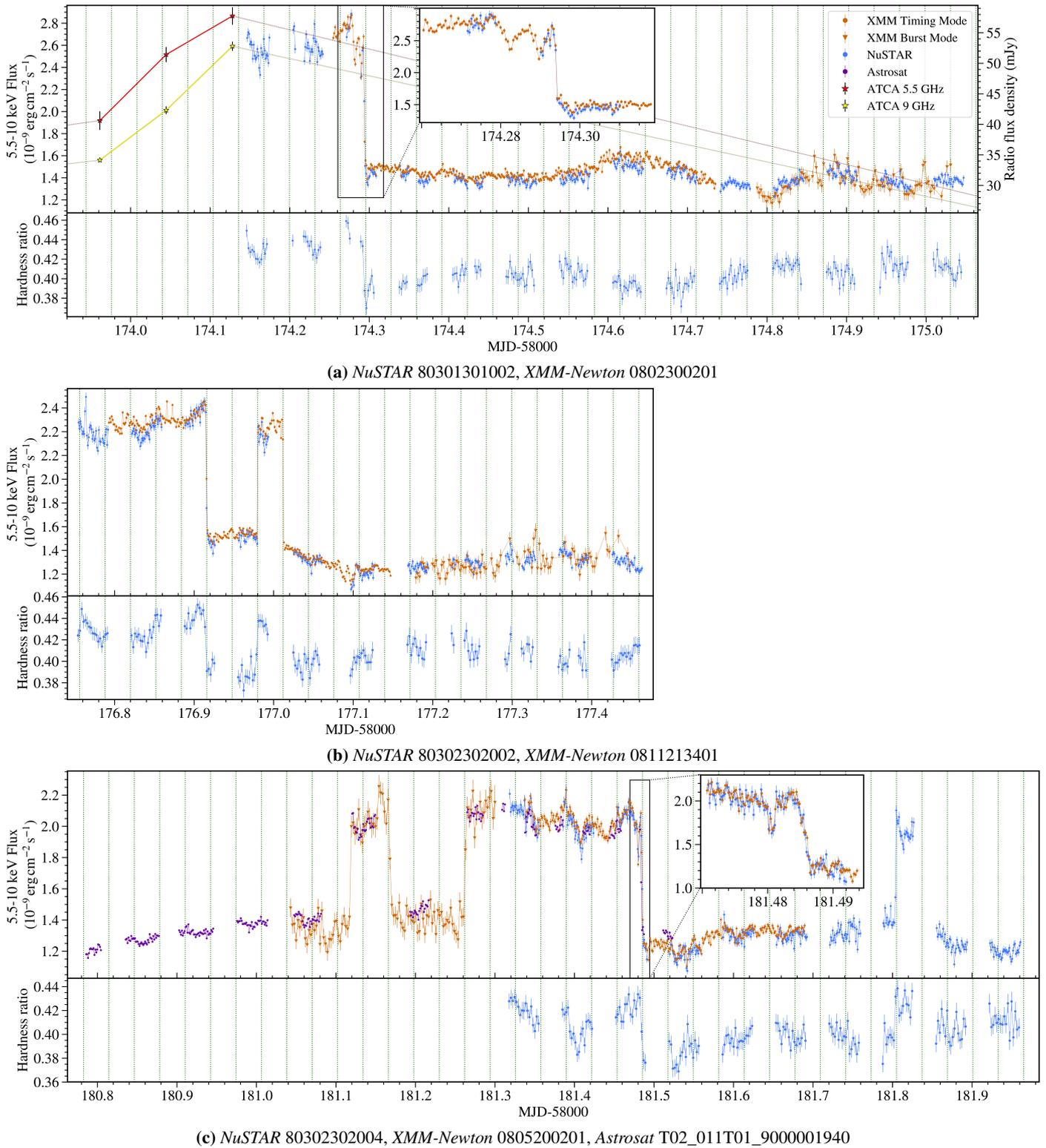

\begin{subfigure}{\textwidth}
    \resizebox{\hsize}{!}{\includegraphics{LCfh_N01_X23in1Ba.pdf}}
    \caption{\textit{NuSTAR} 80301301002, \textit{XMM-Newton} 0802300201}
    \label{LC1}
\end{subfigure}
\begin{subfigure}{\textwidth}
    \resizebox{\hsize}{!}{\includegraphics{LCfh_N02_X11Ba.pdf}}
    \caption{\textit{NuSTAR} 80302302002, \textit{XMM-Newton} 0811213401}
    \label{LC2}
\end{subfigure}
\begin{subfigure}{\textwidth}
    \resizebox{\hsize}{!}{\includegraphics{LCfh_N04_X02in1Ba.pdf}}
    \caption{\textit{NuSTAR} 80302302004, \textit{XMM-Newton} 0805200201, \textit{Astrosat} T02\_011T01\_9000001940}
    \label{LC3}
\end{subfigure}
\caption{Light curves and hardness ratios of all \textit{NuSTAR}, \textit{XMM-Newton}, and \textit{Astrosat} observations containing flip-flops. The \textit{NuSTAR} and \textit{XMM-Newton} light curves are shown for the energy range $5.5\textup{--}10~\mathrm{keV}$ in all cases. \textit{Astrosat} count rates are rescaled to the fluxes observed in the other instruments. The hardness ratios are computed as the ratio of counts in the two energy bands $8\textup{--}79~\mathrm{keV}$, and $3\textup{--}8~\mathrm{keV}$ for the \textit{NuSTAR} observations only. (a) (b) and (c) mark the three long continuous observations of the source within the early flip-flop interval. The light curves are binned into $100~\mathrm{s}$ intervals, except for the two insets, which have time bins of $40~\mathrm{s}$, and $20~\mathrm{s}$, respectively. The hardness ratios are determined for $200~\mathrm{s}$ bins. Vertical lines are plotted every 2761 s, showing that almost all flip-flop transitions occur at integer multiples of a fundamental period. This is discussed in more detail in section 5.}
\label{LCNX}
\end{figure*}

In Fig. \ref{LCtot} we show the light curve of the first 60 days of the 2018 outburst of Swift J1658.2-4242 in three different energy bands. We separated this light curve into six temporal regions based on their different features. In the first region, we detect Swift J1658.2-4242 in the LHS, with the typical flux increase from quiescence. We depict the total outburst in the HIDs of Fig. \ref{HIDS} and \ref{HIDN}. The set of data on the lower right of Fig. \ref{HIDN} is from this LHS, and shows a negative correlation between intensity and hardness, with a Pearson's correlation coefficient of $r_{xy}=-0.71$, and a two-tailed p-value of $p=7.5\times10^{-11}$.

Next we observed the HIMS. The time of transition from the LHS to the HIMS is not known, but spectral analyses of \textit{NuSTAR}, and \textit{Swift}/XRT observations indicate that it probably occurred around MJD $\sim58168$. Within this state, we detected an increase in the $5.5\textup{--}10~\mathrm{keV}$ X-ray flux, in parallel with a decrease of the $20\textup{--}200~\mathrm{keV}$ flux, showing the significant softening of the spectrum. This softening can also be seen in the upper right region of the HIDs of Fig. \ref{HIDS} and \ref{HIDN}. 

On MJD 58174, we observed the start of a major radio flare with \textit{ATCA}, which reached flux densities at least $8.7$ times brighter than those measured at any other time during the outburst. As the radio flux was still rising when the observation finished, the flare presumably reached even greater fluxes. Merely $200~\mathrm{s}$ after the end of this \textit{ATCA} observation, we measured the highest \textit{NuSTAR} and \textit{XMM-Newton} X-ray flux of the entire outburst, $5.4$ times brighter than in the previous \textit{NuSTAR} observation. This can be seen in Fig. \ref{LC1}. We also observed the brightest hard X-ray flux of the outburst at the same time, in an observation by \textit{Insight}-HXMT HE in the $20\textup{--}200 ~\mathrm{keV}$ band (see Fig. \ref{LCtot}). When fitting the \textit{NuSTAR} energy spectrum of this brightest state, we measured a total flux in the $3\textup{--}79\, \mathrm{keV}$ energy range, of $F_{3-79}=9.75\times10^{-9} ~\mathrm{erg}~\mathrm{cm}^{-2}~\mathrm{s}^{-1}$. This corresponds to an unabsorbed and unscattered flux of $F_{3-79, 0}\approx1.64\times10^{-8} ~\mathrm{erg}~\mathrm{cm}^{-2}~\mathrm{s}^{-1}$. Using the distance estimated by \citeads{2019ApJ...875..157J} of $\sim 10~\mathrm{kpc}$, we find the unabsorbed and unscattered source luminosity to be $L_{3-79} \sim 2\times 10^{38} ~\mathrm{erg}~\mathrm{s}^{-1}$.

\begin{figure}[h]
\resizebox{\hsize}{!}{\includegraphics{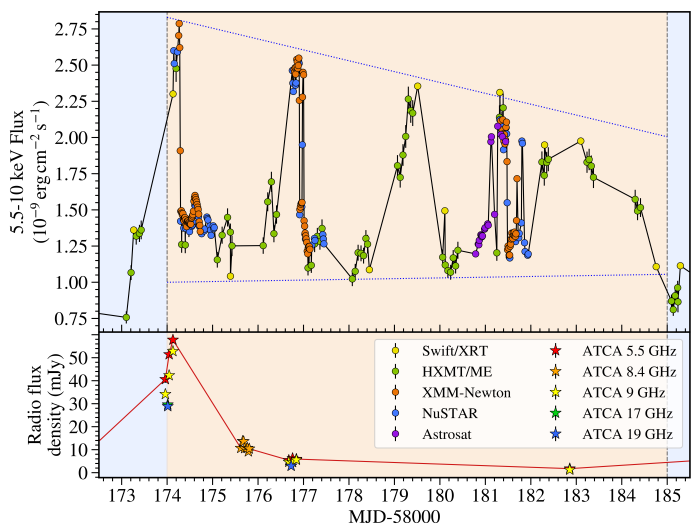}}
\caption{Light curve from Fig. \ref{LCtot}, zoomed into the flip-flop interval. We omitted the $20\textup{--}200~\mathrm{keV}$ light curve as it does not provide as much useful information as the other two light curves do, in this interval. 
\label{LCff}}
\end{figure}

Within the same simultaneous observation by \textit{NuSTAR} and \textit{XMM-Newton} in which we measured the greatest intensity of the outburst, there was a sudden flux drop of $43\%$ within $40~\mathrm{s}$. In subsequent observations by \textit{NuSTAR}, \textit{XMM-Newton}, \textit{Astrosat}, and \textit{Insight}-HXMT/LE and ME, we see similar flux increases or decreases, with the bright states having a $49\textup{--}77\%$ higher flux than the dim states (see Fig. \ref{LCNX}, and  \ref{LCff}). Transitions between the two flux levels occurred on timescales of $26\textup{--}800~\mathrm{s}$ (see Fig. \ref{LCNX}, and Table \ref{tabflux}). 

Unlike \citeads{2019ApJ...879...93X}, and \citeads{2019ApJ...887..101J}, we identify these phenomena as flip-flops, as they feature most of the characteristics of flip-flops seen before in other BHTs.

\begin{table}[h]
\centering
\setlength{\tabcolsep}{4pt}
\def\arraystretch{1.1}
\begin{tabular}{l l l l l}
       &  &  \textbf{State} & \textbf{Time of centre} & \textbf{Transition}\\ 
      \textbf{State} & \textbf{Flux} & \textbf{duration} & \textbf{of transition}& \textbf{duration} \\ \hline
       & & \textbf{(ks)}& \textbf{(MJD-5800)} & \textbf{(s)} \\ \hline \hline
    \begin{tabular}{@{}l}
        Bright \\
        Dim \\
    \end{tabular}
    &
    \begin{tabular}{@{}l}
        $9.09$ \\
        $5.15$ \\
    \end{tabular}
    &
    \begin{tabular}{@{}l}
        $\geq12.7$ \\
        $\geq65.2$ \\
    \end{tabular}
    &
    $174.2942$
    &
    $42\pm4$ \\ \hline
    
    \begin{tabular}{@{}l}
        Bright \\
        Dim \\
        Bright \\
        Dim \\
    \end{tabular}
    &
    \begin{tabular}{@{}l}
        $8.56$ \\
        $5.47$ \\
        $8.41$ \\
        $4.81$ \\
    \end{tabular}
    &
    \begin{tabular}{@{}l}
        $\geq13.9$ \\
        $5.6$ \\
        $2.7$ \\
        $\geq39.0$ \\
    \end{tabular}
    &
    \begin{tabular}{@{}l}
        $176.9158$ \\
        $176.9895$ \\
        $177.0127^{\dagger}$ \\
    \end{tabular} 
    &
    \begin{tabular}{@{}l}
        $74\pm8$ \\
        $26\pm3$ \\
        $62\pm10^{\dagger}$ \\
    \end{tabular} 
    \\ \hline
    
    \begin{tabular}{@{}l}
        Dim \\
        Bright \\
        Dim \\
        Bright \\
        Dim \\
        Bright \\
        Dim \\
    \end{tabular}
    &
    \begin{tabular}{@{}l}
         \\
         \\
         \\
        $7.25$ \\
        $4.72$ \\
        $7.04$ \\
        $4.59$ \\
    \end{tabular}
    &
    \begin{tabular}{@{}l}
        $\geq28.8$ \\
        $4.0$ \\
        $7.6$ \\
        $19.0$ \\
        $27.4$ \\
        $\geq2.0$ \\
        $\geq9.2$ \\
    \end{tabular}
    &
    \begin{tabular}{@{}l}
        $181.1200^{*}$\\
        $181.1697^{*}$\\
        $181.2635^{*}$\\
        $181.4857$ \\
        $181.8045$ \\
         \\
    \end{tabular} 
    &
    \begin{tabular}{@{}l}
        $120\pm40^{*}$\\
        $800\pm200^{*}$\\
        $320\pm80^{*}$\\
        $160\pm20$ \\
        $36\pm5$ \\
         \\
    \end{tabular} 
    \\ \hline \hline
    
    \begin{tabular}{@{}l}
        Dim \\
        Bright \\
        Dim \\
        Bright \\
        Dim \\
        Bright \\
        Dim \\
        Bright \\
    \end{tabular}
    &
    \begin{tabular}{@{}l}
        $2.40$ \\
        $2.95$ \\
         \\
        $3.15$ \\
        $2.41$ \\
        $2.91$ \\
        $2.32$ \\
        $3.16$ \\
    \end{tabular}
    &
    \begin{tabular}{@{}l}
        $\geq13.1$ \\
        $7.6$ \\
        $2.2$ \\
        $18.3$ \\
        $\geq2.0$ \\
        $\geq0.4$ \\
        $7.9$ \\
        $\geq12.2$ \\
    \end{tabular}
    &
    \begin{tabular}{@{}l}
        $205.1654$ \\
        $205.2539^{\dagger}$\\
        $205.2836^{\dagger}$\\
        $205.5001$ \\
         \\
        $205.5556$ \\
        $205.6505$ \\
    \end{tabular} 
    &
    \begin{tabular}{@{}l}
        $90\pm20$ \\
        $110\pm30^{\dagger}$\\
        $700\pm100^{\dagger}$\\
        $50\pm20$ \\
         \\
        $50\pm10$ \\
        $100\pm20$ \\
    \end{tabular} 
    \\ \hline \hline

\end{tabular}
\caption{List of flip-flop properties. Each horizontal line denotes a prolonged gap between observations, within which flip-flop properties could not be accurately identified. Two horizontal lines distinguish the early from the late flip-flops. Fluxes are determined in the $3\textup{--}79~\mathrm{keV}$ range from the NuSTAR FPMA spectra, and are expressed in units of $10^{-9}~\mathrm{erg}~\mathrm{cm}^{-2}~\mathrm{s}^{-1}$. State durations are calculated based on the first and last measurement performed by \textit{NuSTAR}, \textit{XMM-Newton}, or \textit{Astrosat}, that can be confidently associated with this state. The time of the centre of transition, and its duration were calculated using Equation \ref{transtimefit} on the \textit{NuSTAR} light curves, restricted to be within 100 s before the start of the transition, and 100 s after the end of the transition. For the late flip-flops, we restricted the region of the light curve to be fitted, to 200 s before the start, and 200 s after the end of the transition. $\dagger$ indicates transition times calculated from the \textit{XMM-Newton} Timing mode light curves, as they were not seen by \textit{NuSTAR}. $*$ denotes transitions only observed by \textit{XMM-Newton} in Burst mode, and calculated from those data. These are less reliable estimates and are therefore significantly larger. Blank cells indicate states or transitions for which certain properties could not be determined reliably.  \label{tabflux}}
\end{table}

In table \ref{tabflux}, we see that the greatest ratio of bright to dim state flux between neighbouring states that we observed, is 1.77, and occurred during the first observed flip-flop transition. Subsequent flux ratios indicated a decreasing amplitude of fractional flux variation between the two states over time. The second to last of the flip-flops that we observed directly, had a flux ratio of merely 1.49. This is reminiscent of a damped oscillation.

In Fig. \ref{LC1}, \ref{LC2}, and \ref{LC3}, we plotted the light curves and hardness ratios of individual \textit{NuSTAR} and simultaneous \textit{XMM-Newton} observations containing flip-flops. \ref{LC1} also features the \textit{ATCA} observation of a radio flare, and \ref{LC3} is complemented by an \textit{Astrosat} observation. These figures show that changes in the total flux occurred coincident to changes in the hardness. Higher flux corresponded to greater hardness, but the fractional increase in hardness was smaller than the fractional increase in flux. We can therefore deduce that a flip-flop transition from a dim to a bright state corresponds to a flux density increase for both hard and soft X-ray energies. But the hard X-ray flux must have a larger fractional increase than the soft X-ray flux, to generate the observed change in hardness. However, this trend does not extend to very high energies, as the $20\textup{--}200~\mathrm{keV}$ light curve of Fig. \ref{LCtot} does not feature as significant fractional flux variations within the flip-flop intervals, as the $5.5\textup{--}10~\mathrm{keV}$ light curve does.

The hardness ratios in Fig. \ref{LCNX} seem to be correlated with flux variations in the bright state, and anti-correlated with flux variations in the dim state. This is verified in the best fit lines of individual regions within the \textit{NuSTAR} HID of Fig. \ref{HIDN}. Remarkably, despite being separated by several days, all bright states, and all dim states lie in the same region of the HID. We combined all the bright state measurements to find a correlation coefficient of $r_{xy}=0.66$, and associated two-tailed p-value of $p=2.5\times 10^{-7}$, indicating a strong positive correlation between intensity and hardness. In contrast, the combined dim state measurements have correlation statistics of $r_{xy}=-0.17$, and $p=0.045$.

We observed six bright states, seven dim states, and nine transitions between the two, in observations by \textit{NuSTAR}, \textit{XMM-Newton}, and \textit{Astrosat}. In Fig. \ref{LC3}, we noticed one additional instance of a significant flux difference on either side of a gap within an observation by \textit{NuSTAR}, caused by its low Earth orbit, at about MJD 58181.84. The differences in the flux, timing, and spectral properties on either side of the gap, are however consistent with the flip-flop behaviour, and we consider the unobserved region to also feature a flip-flop. 

In Fig. \ref{LCff}, we zoom into the flip-flop region, to better illustrate what is happening within it. Thanks to frequent observations by \textit{Swift}, and \textit{Insight}-HXMT, we can narrow down the start of the flip-flop interval to MJD 58174. Interestingly, this implies that flip-flops started at the same time as the highest intermediate, and hard X-ray flux was reached, and the major radio flare was detected. 

On MJD 58184, following an extended bright state, which appeared to last for more than two days, we observed a final transition to a significantly lower flux. None of our observations over the subsequent 20 days showed any flip-flop activity. Therefore, it seems like this final drop down transition on MJD 58184--58185 marked the end of this interval of flip-flops, which had lasted for about 11 days. 

\begin{figure}[h]
\resizebox{\hsize}{!}{\includegraphics{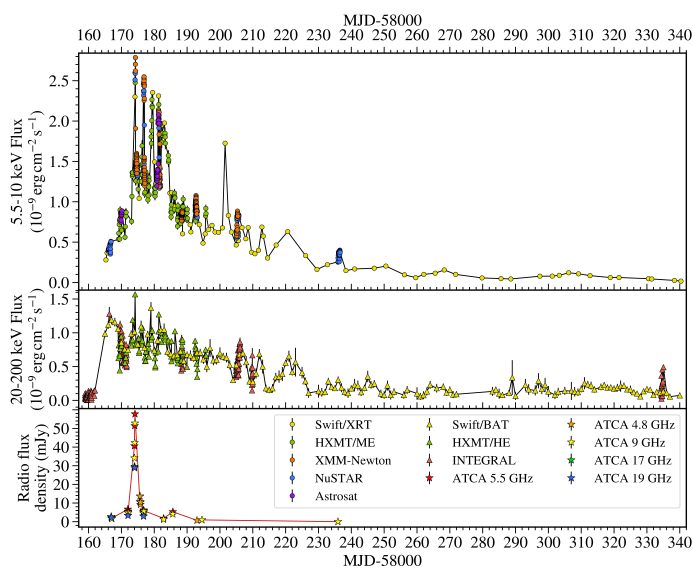}}
        \caption{Light curve of the entire outburst, for X-rays of  $5.5\textup{--}10~\mathrm{keV}$, $20\textup{--}200~\mathrm{keV}$, and at radio wavelengths.
        \label{LCtotout}}
\end{figure}

The \textit{Insight}-HXMT, and \textit{Swift}/XRT light curves during the flip-flop interval revealed the existence of seemingly long ($\sim1~\textrm{day}$) dim state periods. These were interspersed with shorter segments involving several flip-flop transitions, to and from bright states, which each lasted for $2.2\textup{--}19.0~\textrm{ks}$, or longer (see Table \ref{tabflux}). It is however likely that we missed several bright states between individual \textit{Insight}-HXMT, and \textit{Swift} observations. In the light curve of Figure \ref{LCff}, there appear to be four shorter segments within which bright states, and flip-flop transitions were observed. We obtained detailed measurements with \textit{NuSTAR}, \textit{XMM-Newton}, and \textit{Astrosat} of three of these segments, which can be examined in greater detail in Fig. \ref{LC1}, \ref{LC2}, and \ref{LC3}. The duration of these short segments in which one, or more bright states were observed, increases from one instance to the next (lasting approximately 0.15, 0.25, 0.5, and 3.3 days, respectively), whereas the time between them, in which no bright state was observed, appears to decrease (lasting approximately 2.5, 2.0, and 1.6 days, respectively).

In the $20\textup{--}200~\mathrm{keV}$ range, we detected similar up-and-down behaviour in the flux, which corresponds to, but is less pronounced than the light curve evolution observed in the $5.5\textup{--}10~\mathrm{keV}$ interval. Individual states are harder to identify in the hard X-ray light curve, so we focused on the $5.5\textup{--}10~\mathrm{keV}$ one. However, as the hardness ratios of Fig. \ref{LCNX} indicate, another way to single out flip-flop transitions is to monitor the ratio of the disk black body to power law emission. 

The flux change during transitions decreased gradually over time, as can be seen by the dotted lines in Fig. \ref{LCtot}, and \ref{LCff}. But there is only very little variation between the flux changes of adjacent transitions. This is illustrated by the blue line in Fig. \ref{ADPS}, which does not show any discontinuities at the transitions. The gradual flux difference decrease observed in the flip-flops is reminiscent of the amplitude decay of a damped oscillator.

We determined the time it takes to transition between the two states, by fitting this function: 
\begin{equation}\label{transtimefit}
y(t) = 
\begin{cases}
  y_0 + m_1 (t-t_0)& \text{if $t\leq t_0$} \\
  y_0+\frac{y_1-y_0}{\Delta t}(t-t_0) & \text{if $t_0< t \leq t_0+\Delta t$} \\
  y_1 + m_2 (t-t_0-\Delta t)& \text{if $t>t_0+\Delta t$}
\end{cases}
\end{equation}
to the region of the light curves from $100~\mathrm{s}$ before the start, to $100~\mathrm{s}$ after the end of each transition. $y_0$ is the flux just before the transition, and $y_1$ is the flux just after it. $t_0$ is the time when the transition starts, and the transition duration is $\Delta t$. $m_1$ and $m_2$ are the gradients of the light curve just before and after the transition. We found that the transition durations measured from observations by \textit{XMM-Newton} in Burst mode were substantially longer than other measurements, which is a consequence of its larger uncertainties. Transitions from bright to dim states were found to take longer than transitions from dim to bright. Bright to dim transitions lasted for about $42\textup{--}170~\mathrm{s}$, whereas dim to bright transitions only lasted for about $26\textup{--}35~\mathrm{s}$.

In the inset of Fig. \ref{LC3}, we indicate the presence of a short ($\sim 90~\mathrm{s}$) drop in the light curve, with a fractional flux decrease of $\sim 15\%$, just $300~\mathrm{s}$ before the transition from the bright to the dim state. We also see two long ($\sim 600\mathrm{s}$), almost triangular flux drops of $\sim18\%$ just before the first flip-flop transition, as can be seen in the inset of Fig. \ref{LC1}. These features might be interpreted as failed transitions back down to the dim state. 

After the end of the flip-flops, variability of the light curves on timescales of days was small (see Fig. \ref{LCtot}). We did not find any evidence of flip-flops in the two \textit{NuSTAR} and simultaneous \textit{XMM-Newton} observations made in this interval, and even frequent monitoring by \textit{Swift}, and \textit{Insight}-HXMT did not reveal any significant flux changes. The flux decreased in an exponential manner. 

Both \textit{NuSTAR} and \textit{XMM-Newton} observations in this interval lie in almost exactly the same region of the HID (See Fig. \ref{HIDN}), which is however very clearly distinguished from every other observed part of the outburst. We see a negative correlation between hardness and intensity when combining the two observations, with $r_{xy}=-0.75$, and $p=2.6\times10^{-22}$.

\begin{figure}[h]
\resizebox{\hsize}{!}{\includegraphics{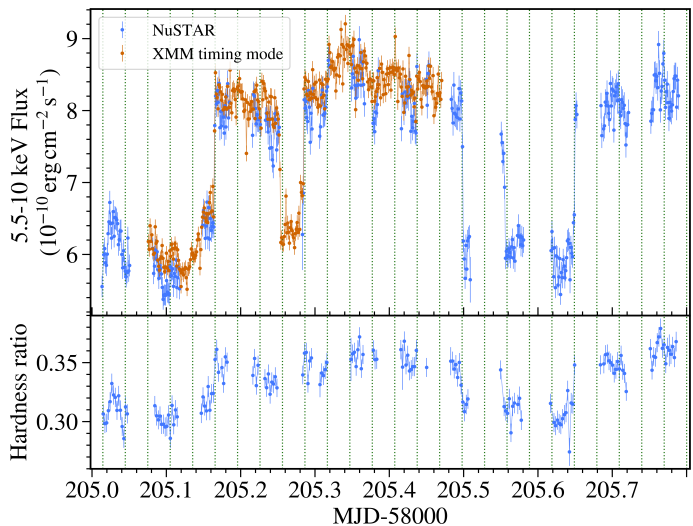}}
\caption{Light curves and hardness ratio of the \textit{NuSTAR} and \textit{XMM-Newton} observation of the late flip-flops. We used the same energy ranges and time bins as in Fig. \ref{LCNX}. The Burst mode is not included here, as it was strongly affected by soft proton flares. Vertical lines are drawn every 2610 s, showcasing that the late flip-flops also feature a period underlying the times of transition between the two flip-flop states. 
\label{LCNXlff}}
\end{figure}

On MJD 58201, \textit{Swift/XRT} detected an X-ray flare. Five days later, in the next observation by \textit{NuSTAR} and \textit{XMM-Newton}, we detected flip-flops again (See Fig. \ref{LCNXlff}). These however differ quite significantly from the ones seen previously. They have much smaller flux differences, with flux ratios of between 1.22 and 1.36 between the bright and dim states in the 3--79 keV energy range. 

Interestingly, the X-ray flare is placed in exactly the same region of the HID as the early flip-flops were observed in. The late flip-flops observed afterwards, however fall into a different region, as can be seen in Fig. \ref{HIDS}.

Unlike the early flip-flops, here we observe a positive correlation between hardness and intensity in both the bright, and dim states. The same correlation applies to both, and has a correlation coefficient and p-value of: $r_{xy}=0.95$, and $p=4.1\times10^{-20}$. 

Due to fewer observations of Swift J1658.2-4242 during the late flip-flop interval, we cannot place stringent bounds on its start and end times. It is possible that the X-ray flare marked the start of this interval. In the days that followed the flare, we detected a greater variability between subsequent observations by \textit{Swift}/XRT and BAT than in the 15 days that followed the end of the first flip-flop interval. This variability seemed to end by about MJD 58215, which is our estimate for the end of this late flip-flop interval.

We observed four bright states, four dim states, and six transitions between them. One additional transition is inferred, but was not observed. Contrary to the early flip-flops, the late flip-flop transitions from bright to dim states (50--110 s) were faster than the reverse transitions (90--700 s). 

\begin{figure}[h]
\resizebox{\hsize}{!}{\includegraphics{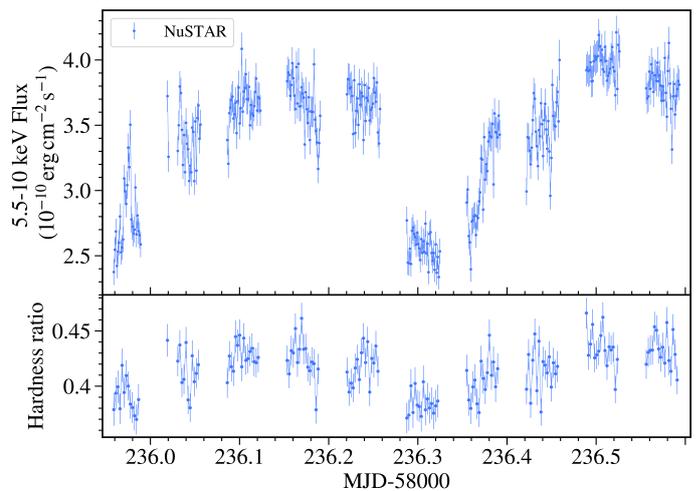}}
\caption{Light curve and hardness ratio of the final \textit{NuSTAR} observation, starting at MJD 58235.96. We used the same energy ranges and time bins as in Fig. \ref{LCNX}
\label{LCNSIMS}}
\end{figure}

The final \textit{NuSTAR} observation, on MJD 58235--58236 contained a drop in flux of about 27\% during a gap in the data, followed by a gradual flux rise up to the previous level (see Fig. \ref{LCNSIMS}). The hardness ratio correlates with the flux for this observation (with $r_{xy}=0.89$, and $p=4.1\times10^{-20}$), but the shape of the light curve is vastly different to what was seen during the early or late flip-flops. We therefore decided against labelling these flux changes as a flip-flop. 

In the first \textit{NuSTAR} observation, two absorption dips are seen (\citeads{2018ApJ...865...18X}), which are distinguished from flip-flops via their short duration, their shape in the light curve, and the increase of the hardness ratio. We did not observe dips in any of the other observation of Swift J1658.2-4242.

The major radio flare seen at the start of the early flip-flop interval died down quickly. During the next observation, $\sim1.5$ days after the detection of the radio flare, the flux had almost completely returned to its previous level again. We did not detect any additional radio flares at any other point during the outburst. 

In Fig. \ref{HIDS} we show all \textit{Swift}/XRT observations of the outburst, which continued until September 28, 2018 (MJD 58389). The rest of the outburst features the typical decrease in flux, followed by an increase in hardness, as can be seen in Fig. \ref{LCtotout}. As far as we can tell, Swift J1658.2-4242 never reached the HSS, but remained in the intermediate states for the rest of the Swift/XRT observations. 

The duration of the flip-flop states seems to decrease over time. By this we define the amount of time for which a particular bright, or dim state was maintained, from the end of the transition leading into it, to the start of the transition leading out of it. In the first observation by \textit{NuSTAR}, and \textit{XMM-Newton} within the early flip-flop interval, only one bright and one dim state was seen. In the subsequent two observations, two bright, and two dim states were detected. And within the late flip-flop interval, a single \textit{NuSTAR} observation spanned four bright, and four dim flip-flop states (See Fig. \ref{LCNX}). In this outburst alone, we recover some of the variety of flip-flops seen in the literature. We observe a range of flip-flop amplitudes, durations, and also detect different behaviour in the time domain.

\section{Discovery of an underlying clock in flip flop transitions}

Flip-flop states have vastly different durations. The time between one flip-flop transition and the next was observed to range from anywhere between $2.56~\mathrm{ks}$ to at least $65.2~\mathrm{ks}$. There does not seem to be a fixed time after which the light curve repeats itself in a periodic manner. However, in the \textit{XMM-Newton} observation 0811213401, we noticed that two of the observed flip-flop states during MJD 58176--58177 had times between transitions which had a ratio very close to 2:1 (See Fig. \ref{LC2}, and Table \ref{tabflux}). The time between the first and second, and between the second and third transition in this observation, is 5590 s and 2785 s, respectively. Based on this finding, we investigated the possibility of an underlying timescale manifesting itself throughout all flip-flops. Starting from one of the transitions of this observation, we extrapolated integer multiples of the shorter of the two durations throughout the early flip-flop interval via the equation:

\begin{equation}\label{compereq}
f(n) = nP+t
\end{equation}

Where $t$ is the centre time of one particular flip-flop transition. $P$ is a period, which we initially estimated to be $P=2785~\mathrm{s}$, and $n\in\mathbb{Z}$. We found that numerous transitions lie very close to one of the $f(n)$ times determined by this formula. Transitions do not happen at every integer $n$, but when they occur, they lie close to one of the $f(n)$. The light curve does not appear to be periodic, but changes in the light curve occur at seemingly random multiples of a fundamental period, $P$. Due to the lack of overall periodicity, the standard methods to search for a period over which a cycle repeats itself, are not applicable here. 

We used the centre times of each transition, which were determined by fitting the light curves around every transition with Equation \ref{transtimefit}. Of the nine transitions we observed, eight are fitted well by Equation \ref{compereq}, and a period of about $P=2785~\mathrm{s}$. Often, one of the $f(n)$ times passes through the width of a transition, or is just barely shy of it (see Fig. \ref{LCNX}). However, the first transition observed by \textit{XMM-Newton} in Burst mode on MJD 58181 (Fig. \ref{LC3}) is not close to any of the times calculated by Equation \ref{compereq}. 

\begin{figure}[h]
\resizebox{\hsize}{!}{\includegraphics{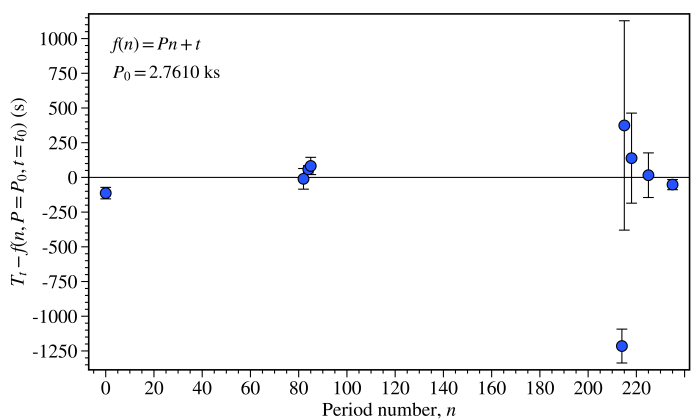}}
        \caption{Here we plotted the difference between the best fit period, and the observed time of the centre of each transition. We ignored the anomalous transition time in the fit, but show it on the graph nevertheless. \label{Comper_eff}}
\end{figure}

Excluding this anomalous transition time, and using $\sim2.78~\mathrm{ks}$ as our initial estimate of the period, we determined a best fit with $P=2761.0\pm0.4~\mathrm{s}$, and $t=58174.2955\pm0.0006~\mathrm{MJD}$, having a goodness of fit of $\chi^2=16.5$, for 6 degrees of freedom. The results of this fit are shown in Fig. \ref{Comper_eff}. The vertical lines drawn in Fig. \ref{LCNX} showcase the best fit results overplotted by the light curve. We also searched for a different period that could fit our observations better than this one, but did not find any other period larger than $2761.0~\mathrm{s}$, which fitted better than it. Shorter periods were found, but these turned out to be merely fractions of the $2761.0~\mathrm{s}$ period. 

We next considered the significance of this detection, by investigating the likelihood of obtaining at least as good of an agreement from a sample that was not based on a fundamental period. We therefore simulated sets of nine random transition times within the window of our observations of the early flip-flop interval, to which we randomly assigned our measured errors in the transition times. The shorter the period is, the greater the likelihood is of finding a good agreement between randomized transition times, as the maximum delay between a transition time and an extrapolated period decreases. We therefore considered the probability of being able to fit for a period in the way we did, which is at least as long as the one we determined. We fitted all simulated transition times with Equation \ref{compereq}, using a given input period, and having set $n=0$ for the chronologically first transition. For each set of nine simulated transitions, and each initial period estimate, we excluded the transition deviating the most from the best fit, leaving only eight transitions to be fitted. In doing so, we ensured that simulated and real data were treated in exactly the same way. We fitted the remaining eight transitions again, to determine the closest integer $n$ for each of them, to make sure that the ignored transition did not affect the result. Using the newly found best fitting integers $n$, we performed a third fit to find the optimal $P$ and $t$, and to minimize the $\chi^2$, which we finally compared with the best fit for the actual data. We provided a sample of initial period estimates, starting from 2761.0 s, up to the maximum period that would allow for nine transitions to occur within our set of observations. The increment in the initial period estimates, was chosen in such a way as to correspond to a decrement of one in the number of multiples of the period that could be placed between the start and the end of the early flip-flop interval. In doing so, we sampled the entire range of possible periods. Every set of simulated transition times was fit with all of these initial period estimates, and we investigated how many simulations had at least one instance where a better, or equally good fit was obtained as we had found for the actual data, using the same procedure.

When running $10^5$ iterations of this simulation, we found that only $3.38\%$ of randomly generated sets of transition times fitted with a $\chi^2$ of at most 16.5, to a fundamental period at least as large as $2761.0~\mathrm{s}$, when the worst fitting time was excluded from the fit.

We noticed that the single anomalous transition occurs very close to the half-way point between two of the $f(n)$. This raises the question of whether a period of about half of our first estimate is instead the actual recurrent timescale of the flip-flop transitions. Indeed, a period of $P=1380.5\pm0.2~\mathrm{s}$ agrees well with all the early flip-flop transition times we observed, fitting with a $\chi^2$ of 18.2 for 7 degrees of freedom. However, we found that in $10^4$ simulations, the probability of obtaining at least as good of an agreement between nine randomly generated transition times, for a period at least as long as this one, is 18.2\%. Despite the existence of one anomaly, there is a greater significance in the existence of an underlying period of $2761.0~\mathrm{s}$ in the early flip-flops. It is possible that $P=1380.5\pm0.2~\mathrm{s}$ is the actual underlying period defining the times of transition, but our data are insufficient to distinguish periodicity on these time scales from noise. We therefore rely on the agreement obtained for a period of $2761.0~\mathrm{s}$.

Next, we turned our attention to the late flip-flops, and the six transitions observed in them. A simple extrapolation of the best fit we had found for the early flip-flops, did not fit the transition times of the late flip-flops. Additionally, two transition times are separated by about 2560 s, which suggests that the late flip-flops cannot be fitted well with a period of $2761.0~\mathrm{s}$, and might require a different period. We investigated the possibility of a different linear relation linking the times of the late flip-flop transitions, via Equation \ref{compereq}, independent of our results for the early flip-flops. The best fit to these data was obtained with $P=2610\pm20~\mathrm{s}$ and $t=58205.165\pm0.003~\mathrm{MJD}$, and has a goodness of fit of $\chi^2=42.5$, for 4 degrees of freedom. These best fit parameters were used to plot the vertical lines in Fig. \ref{LCNXlff}. This best fit period underlying the late flip-flop transitions is about $5.5\%$ smaller than the one found for the early flip-flops.

Using a similar simulation as before, after updating the parameters to match the observations of the late flip-flops, we found that for $5\times 10^5$ iterations, the probability that six random transition times within our late flip-flop observations, fit with a period at least as large as 2610 s, and a goodness of fit of at most 42.5, is about $7.37\%$. This higher probability is a consequence of the large $\chi^2$ that we obtained for the best fit to the observed times of transition in this interval. The timing of transitions in the late flip-flops is less consistent than in the early flip-flops. 

The detection of similar underlying periods in both intervals is intriguing. We investigated whether all transitions, in both the early and the late flip-flop intervals, could be fitted with one universal equation, of the form: 

\begin{equation}\label{compereqdec}
f(n) = nP+An^2+t
\end{equation}

Where $P$ describes the initial period, at $n=0$. $t$ describes the reference time, corresponding to the time of the transition at $n=0$. $A$ describes the gradual change in the period, and is equal to $A=\frac{1}{2}\Pi\dot\Pi$, where $\Pi(n)=P+2nA$, is the period at $n$. We however do not measure the periods at any of the $n$, but rather the time between any two $n$, such as: $f(n)-f(n-1)=\int_{n-1}^n \Pi(n)dn = P+(2n-1)A$. 

\begin{figure}[h]
\resizebox{\hsize}{!}{\includegraphics{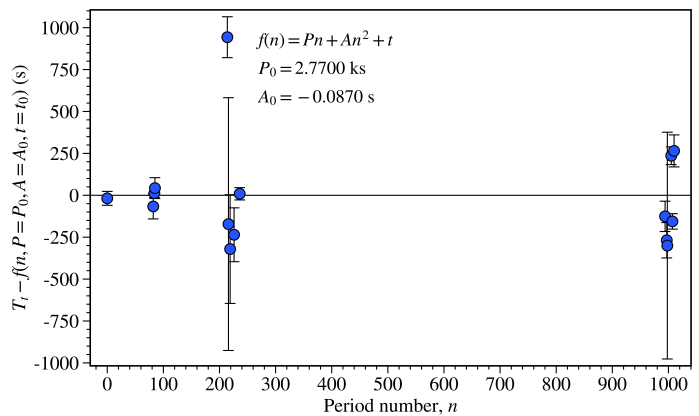}}
        \caption{Here we plot the best fit of Equation \ref{compereqdec} to the times of transition in both the early and late flip-flops. The transition times can not be fitted with Equation \ref{compereq}, assuming a constant, and not too short period. But by adding an extra term to allow for a change in the period, we are able to fit the times of transition reasonably well. \label{Comper_decper}}
\end{figure}

By setting appropriate initial estimates for $P$, of the best fitting period found for the early flip-flops, and for $A$, such that the $5.5\%$ difference in the periods of the early and late flip-flops could be generated within the 32 day range of early to late flip-flop observations, we discovered a fit that achieves a good agreement for all transition times in both intervals, except for the one anomalous transition in the early flip-flops. Excluding that one, the best fit parameters were found to be $P=2770.0\pm0.6~\mathrm{s}$, $t=58174.2963\pm0.0008~\mathrm{MJD}$, and $A=(-8.70\pm0.06)\times 10^{-2}~\mathrm{s}$, resulting in an initial rate of period decrease equal to $\dot\Pi(0)=(-6.28\pm0.04)\times10^{-5}$. The best fit has 11 degrees of freedom, and a goodness of fit of $\chi^2=52.6$. We indicate the difference between the measured transition times, and the closest value of Equation \ref{compereqdec} to each transition time, using these best fitting parameters, in Fig. \ref{Comper_decper}. The ability to fit both early and late flip-flop transitions could indicate a strong connection between them.

We once again investigated the significance of this detection, under the assumption that the early and late flip-flops should be considered as one phenomenon, rather than two. To do so, we modified the above simulations, to instead generate 15 random transition times within our observations of early and late flip-flops. We fitted these times with Equation \ref{compereqdec} instead, and once again removed the worst fitting instance. We also had to specify a range of input values for $A$, which we limited to have an absolute value of less than $0.0870$. After $10^4$ iterations, we found that the probability to obtain at least as good of an agreement in randomized data, as we had found for our observations, is $0.23\%$, or $2.8\sigma$. 

This analysis is based on the idea that the early and late flip-flops are the product of the same process, which evolves over time, but is not seen for about two weeks. A complementary possibility is that the early and late flip-flops are completely separate realisations of the same process, which are possibly a consequence of the major flares observed in X-ray and radio. 

We therefore also consider the probability of obtaining a linear fit from Equation \ref{compereq} to both the early and late flip-flop intervals individually, in both instances fitting with a period of at least as large as the one we found, for equally many transitions as we observed, and having a goodness of fit smaller than the one we detected, and also having periods in both intervals that are separated by less than $5.5\%$. Using the results of our simulations for a constant period in the early and late flip-flops described above, we calculated the probability that any randomly chosen combination of an early and late flip-flop period, which managed to meet our fitting criteria applied to simulated data, had a separation of at most $5.5\%$. This probability was found to be about $26\%$. Therefore, the probability of randomly generated, and independent early and late flip-flop transition times being able to be fit with a constant period in each interval, which is greater than or equal to the one we found in the data, which fit at least as well as our measurements did, and have periods separated by at most $5.5\%$, is $P=0.0338\times0.0737\times0.26=6.48\times10^{-2}\%$, or $3.2\sigma$. This result is based on $10^5$ simulations for the early flip-flops, and $5\times 10^5$ simulations for the late flip-flops. In these simulations, we excluded the worst fitting transition, to ensure that the data and simulations were handled in exactly the same way. We used the exact same procedure on both the real data, and all simulated data. These results therefore suggest that this underlying period is significant.

This entire analysis is based on the 15 flip-flop transitions that we observed. There were two more instances when a flip-flop transition was inferred to have occurred in a short gap in the observations, on MJD 58181.8, and MJD 58205.5. We note that in both instances, there was one $f(n)$ time that occurred within the gap and might have coincided with the unobserved transition. Neither of the two gaps contradict the detection of a period in either the early, or the late flip-flop interval. We did not use the gaps as additional measurements of times of flip-flop transition, as the associated error in their measurements is so large, that they did not aid the investigation of possible periods.

\section{Power density spectra and QPOs}

\begin{figure}[h]
\resizebox{\hsize}{!}{\includegraphics{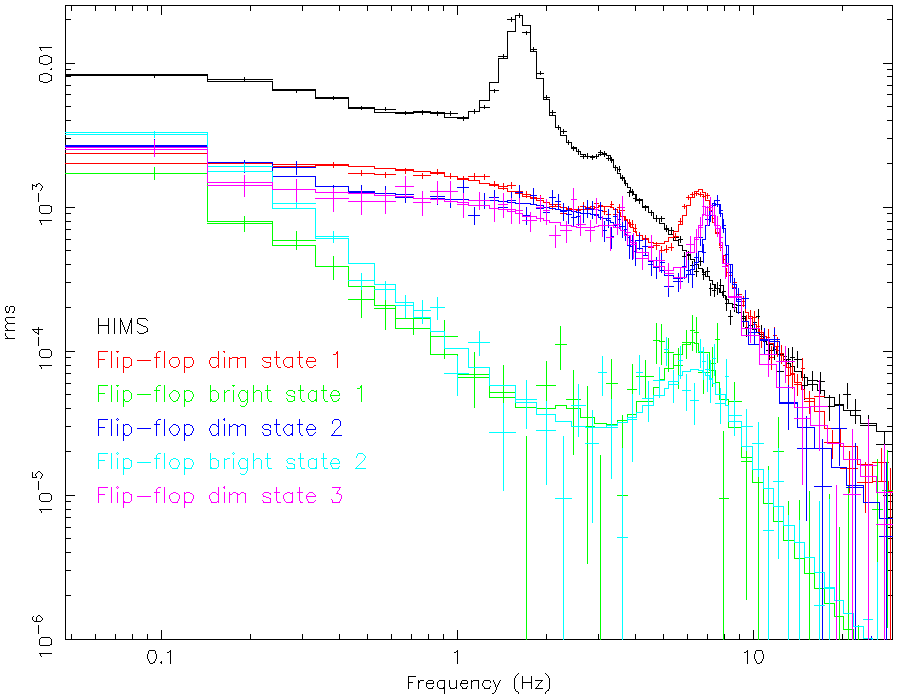}}
        \caption{Poisson noise subtracted, squared fractional rms normalized \textit{Astrosat} PDSs of different intervals of the outburst. The first \textit{Astrosat} observation in the LHS is marked in black, all other PDSs originate from different regions of the second \textit{Astrosat} observation, which contains flip-flops (See Fig. \ref{ADPS}). Both the LHS and Flip-flop dim state 1 PDSs have inflated widths due to being generated from a large time interval containing a variable type C QPO frequency. Producing the PDS for shorter intervals yields a narrower QPO. The same is not true for the type A QPOs seen in the bright states. \label{AsatPSDs}}
\end{figure}

\begin{figure*}[h]
\resizebox{\hsize}{!}{\includegraphics{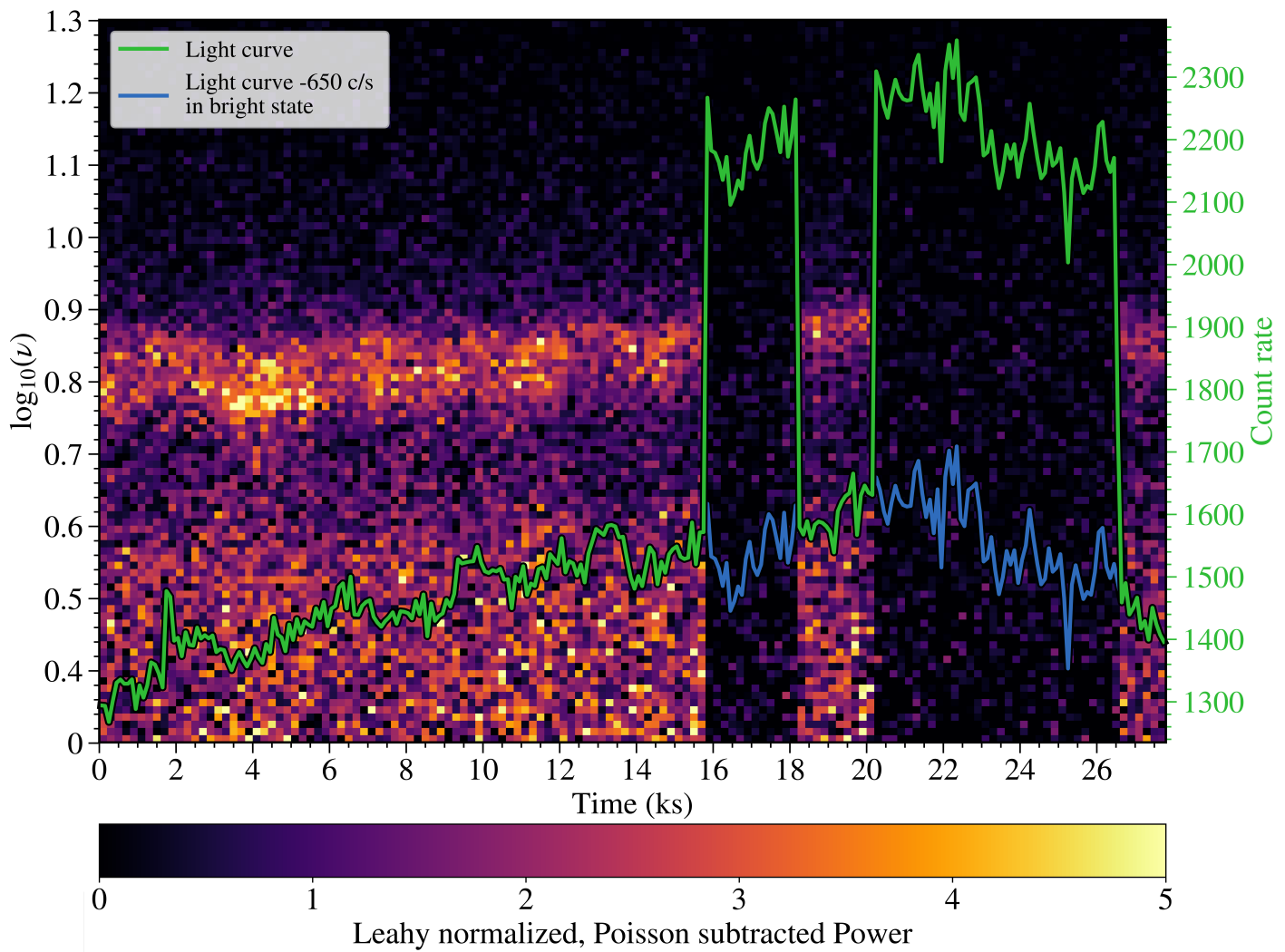}}
        \caption{Spectrogram of the second \textit{Astrosat} observation, starting on MJD 58180.786, with an overplotted light curve. Colours denote the Leahy normalized, Poisson noise subtracted power. The light curve is plotted in green, the blue curve represents the same light curve, but with a count rate reduced by $650~\mathrm{counts}/\mathrm{s}$ in the bright states. Gaps in the observations due to the low Earth orbit of \textit{Astrosat} have been left out for display clarity. The time is measured in seconds of observing time since the start of the observation. The light curve is split into $200~\mathrm{s}$ segments, for which the average count rate, and the PDS were computed. The PDS frequencies are rebinned on a logarithmic scale. The first three state transitions shown here all lie within gaps in the data. We do however see the entire fourth transition, near the end of the observation. \label{ADPS}}
\end{figure*}

\begin{figure}[h]
\resizebox{\hsize}{!}{\includegraphics{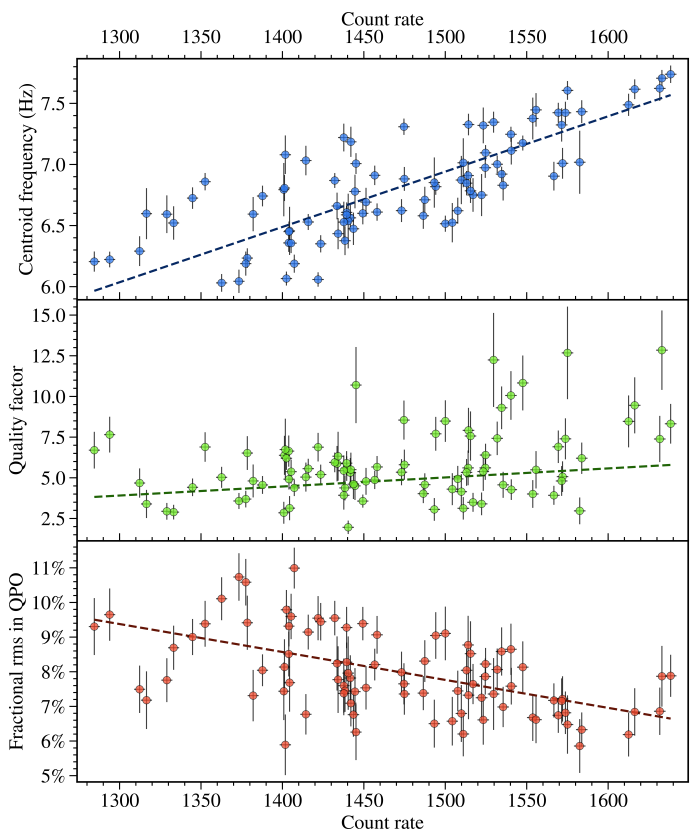}}
        \caption{Correlation between the best QPO fitting parameters and the count rate, in each of the $200~\mathrm{s}$ intervals of Figure \ref{ADPS}. We fitted each PDS between 4 and 10 Hz with a Lorentzian added to a straight line with a constant gradient. The properties of the Lorentzian were extracted to determine the centroid frequency, quality factor, and fractional rms in the QPO, as described in the section on \hyperref[sec:QPOIntro]{Quasi-periodic oscillations}. Different ranges of values for each of these parameters were found when binning over longer time periods. 
        \label{PSDfitpar}}
        
\end{figure}

\begin{figure}[h]
\resizebox{\hsize}{!}{\includegraphics{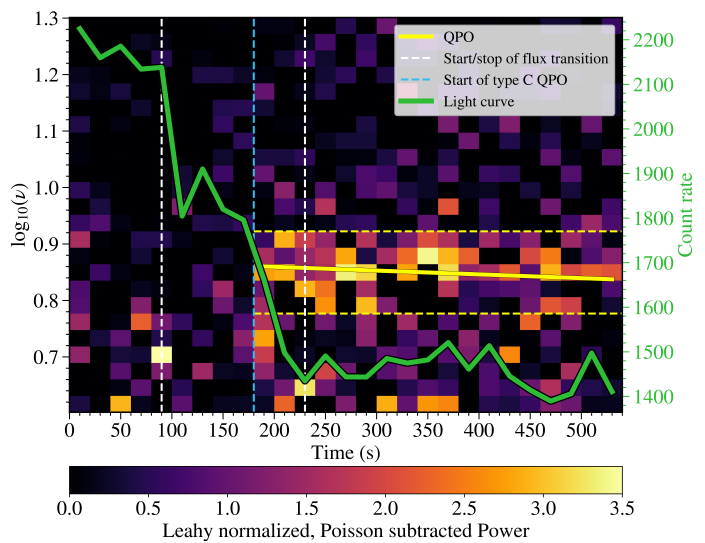}}
        \caption{Spectrogram of the last flip-flop transition shown in Fig. \ref{ADPS}. Colours denote the Leahy normalized, Poisson noise subtracted power. The light curve is plotted in green. The time is measured in seconds of observing time since the start of the orbit. We computed the PDS, and the average count rate, for each of the 20 s intervals into which we divided the data. We plotted a smaller frequency range for this figure, due to significant noise at low frequencies, a consequence of the short interval size. \label{ADPSt}}
\end{figure}

In order to learn more about the flip-flop transitions, we generated PDSs of individual flip-flop states, and of entire observations without flip-flop activity, from the light curves of \textit{Astrosat}, \textit{NuSTAR}, and \textit{XMM-Newton}. In Fig. \ref{AsatPSDs} we plotted the PDSs of all the flip-flop states observed by \textit{Astrosat}, as well as the PDS of the \textit{Astrosat} observation in the HIMS, for comparison.

During the initial rise in flux in the outburst, we found a narrow QPO, perched on a strong broad-band continuum, featuring a second harmonic, and no subharmonics. It has properties indicative of a type C QPO (\citeads{2018ApJ...865...18X}). We noticed that the frequency of this QPO increased alongside the flux, rising from $0.13~\mathrm{Hz}$ at the start of the first observation by \textit{NuSTAR}, to $1.9~\mathrm{Hz}$ by the end of the first \textit{Astrosat} observation. This occurred alongside a flux rise from $3.9\times10^{-10} ~\mathrm{erg}~\mathrm{cm}^{-2}~\mathrm{s}^{-1}$ to $1.1\times10^{-9} ~\mathrm{erg}~\mathrm{cm}^{-2}~\mathrm{s}^{-1}$ in the energy range $3\textup{--}79 ~\mathrm{keV}$. 

We analysed PDSs of the flip-flop states individually, and noticed a clear duality of properties, matching the duality of the flux levels. All bright states have similar PDSs, and so do all dim states. But the PDSs of the bright states differ remarkably from those of the dim states (see Fig. \ref{AsatPSDs}). 

The dim flip-flop states detected by \textit{Astrosat}, \textit{NuSTAR}, and \textit{XMM-Newton} all contained a reasonably narrow ($4.7\leq Q \leq17$) QPO at frequencies of $6\textup{--}7.5~\mathrm{Hz}$ (see the purple, red, and dark blue PDSs in Fig. \ref{AsatPSDs}). They feature a sub-harmonic, but no higher order harmonic. The rms in the QPO is $2.5\textup{--}8.2\%$. Together with a strong, flat, broad-band continuum with a break frequency at about $3~\mathrm{Hz}$, the total rms of the dim states between $0.5~\mathrm{Hz}$ and $50~\mathrm{Hz}$, was found to be $6.4\textup{--}10.3\%$. The large effective area of \textit{Astrosat} enabled a more detailed examination of the change in QPO properties on shorter timescales. We noticed that the centroid frequency of the QPO in the dim states of the second \textit{Astrosat} observation scales with the count rate (Fig. \ref{ADPS} and \ref{PSDfitpar}), with $r_{xy}=0.78$ and $p=7.3\times10^{-19}$. We additionally found that the quality factor increases slightly with increasing count rate ($r_{xy}=0.38$, $p=3.2\times10^{-4}$). And the rms in the QPO was found to decrease with increasing count rate ($r_{xy}=-0.51$, $p=3.8\times10^{-7}$), thereby indicating that the rms also correlates negatively with the centroid frequency ($r_{xy}=-0.64$, $p=2.0\times10^{-11}$).

By comparing these results with the distinguishing characteristics of the three different QPO types (\citeads{2016AN....337..398M}), we find that almost all of these properties suggest that this is a type C QPO. However, this QPO has a large range of quality factors ($4.7\leq Q \leq17$), and is usually found at $Q<10$, which is too wide to fit the standard definition of a type C QPO (\citeads{2016AN....337..398M}). Type C QPOs with a smaller quality factor than the standard definition have however been seen before, by \citeads{2012MNRAS.427..595M}, suggesting that the lower limit of the quality factor of type C QPOs is ill-defined. The shape of the continuum, the large rms, and the correlations between count rate, centroid frequency, and rms, are all inconsistent with type B and A QPOs. 

Whereas the type C QPO of the dim states can easily be seen in the spectrograms of Fig. \ref{ADPS} and \ref{ADPSt}, no QPO can be identified during the bright states. So the QPO seems to disappear entirely in dim to bright transitions. This is however a consequence of the low amplitude and broad profile of the QPO in the bright state, as well as the short time sampling used in these figures. Generating PDSs over longer time intervals instead revealed the existence of a very low amplitude, broad peak with a centroid frequency of about $5.0\textup{--}7.2~\mathrm{Hz}$, a quality factor of $2.0\textup{--}3.8$, and an rms of $1.4\textup{--}2.3\%$ (see Fig. \ref{AsatPSDs}). Fig. \ref{ADPS} also shows, that the broad-band continuum at low frequencies is strongly suppressed in bright states. Fig. \ref{AsatPSDs} however indicates that the continuum grows at low frequencies. It contributes slightly to the total integrated rms in bright states, which is about $2.0\textup{--}2.7\%$. These broad, low amplitude QPOs are very difficult to detect, and could only be seen in the \textit{Astrosat} PDSs. This QPO type is undoubtedly a type A QPO, because all of these fitting parameters, as well as the shape of the PDS, agree with the standard properties of a type A QPO (See e.g. \hyperlink{N03}{N03}). Even though we did not detect the type A QPO in the PDSs of \textit{NuSTAR} and \textit{XMM-Newton}, we found that the shape of their PDSs, and the low rms in the bright states, is consistent with the \textit{Astrosat} result, of the bright states featuring type A QPOs. 

Following \citeads{1988SSRv...46..273L}, we determined the detection significance of the type A QPOs. We split the data into individual orbits, and only kept data firmly belonging to the bright flip-flop state. For each of these intervals, we determined the QPO detection significance, and found it to be between $3.2\,\sigma$, for a very brief interval, and up to $11.6\,\sigma$. We therefore conclude that all bright states observed by \textit{Astrosat} featured a type A QPO. Through the similarity in their properties, we infer that a type A QPO was also present in all other bright states of the early flip-flops. 

Thanks to the large effective area and small time resolution of \textit{Astrosat}, we were able to compute accurate PDSs over short time intervals as well, enabling the creation of the spectrograms shown in Fig. \ref{ADPS}, and \ref{ADPSt}. In Fig. \ref{ADPSt}, we zoom into the first ever detection of a direct transition from a type A to a type C QPO. Other groups (\hyperlink{C04}{C04}, \citeads{2008mqw..confE..91D}) have detected a type C QPO in one observation, and a type A in the next. But they could not rule out the possibility that the interval between the observations contained a transitional type B QPO. We, however, see multiple direct transitions between types A and C, and can therefore determine whether the change between the two QPO types involved a type B QPO. If a type B QPO had occurred during the transition, it would have featured as a very bright spike in the spectrograms, with next to no broad-band continuum, and would presumably have been found at a slightly different frequency than the type C QPO. Visually, we do not see any features matching this description in Fig. \ref{ADPSt}. We also fitted the individual power spectra across the transition in the search for a type B QPO, but did not find any. We even used intervals as short as $10~\mathrm{s}$ without detecting any type B QPO. The search for QPOs in short time segments becomes increasingly difficult, because many cycles are needed to crystallize out a QPO from a strong continuum, and noise. So even though we cannot rule it out entirely, we can rule out the existence of a type B QPO on any timescale longer than about $10~\mathrm{s}$ during this transition.

We note that the change in the PDS is very sudden. In Fig. \ref{ADPSt}, we see that the type C QPO takes less than $20~\mathrm{s}$ to appear. Using smaller time intervals indicates that it takes at most $\sim 10 ~\mathrm{s}$ to get started. There was no indication before then, that a type C QPO was being established there. This is significantly shorter than the flux transition, which instead takes $\approx 140~\mathrm{s}$. Another interesting feature is that the QPO suddenly appears about $100~\mathrm{s}$ after the start of the transition in the light curve, and $\sim 40~\mathrm{s}$ before the flux has finished dropping down to the dim state level.

\begin{figure}[h]
\resizebox{\hsize}{!}{\includegraphics{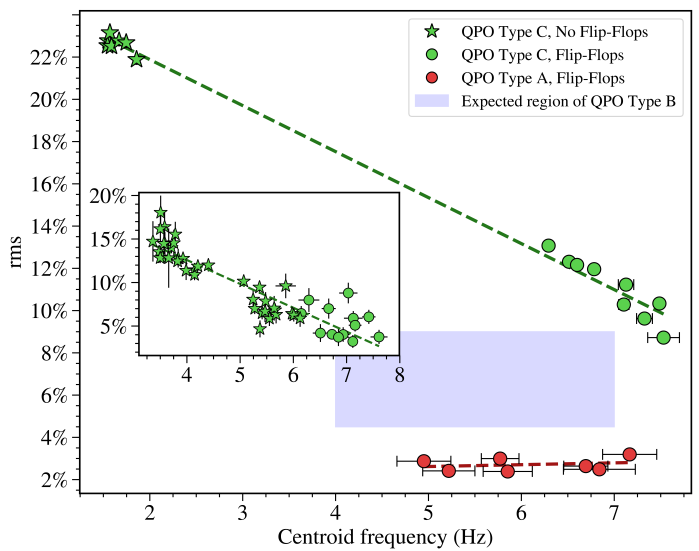}}
        \caption{Here we plot the total integrated rms and QPO centroid frequency of each orbit from the two \textit{Astrosat} observations. The inset shows the centroid frequency, and the corrected rms in the QPO, detected in individual orbits of \textit{NuSTAR}, and in $2~\mathrm{ks}$ regions of the \textit{XMM-Newton} light curves. Total rms values are computed in the $0.5\textup{--}50~\mathrm{Hz}$ range. We use colours and shapes to distinguish the type of QPO, and whether the measurement was made during the flip-flop period, or not. \label{ARMSnu}}
\end{figure}

In Fig. \ref{ARMSnu}, we plot the relation between the \textit{Astrosat} total integrated rms and the QPO frequency. In the inset we plot the dead time corrected rms in the QPO, in observations by \textit{NuSTAR}, and \textit{XMM-Newton}. Comparing this with the rms vs. $\nu$ plots of other transient BHTs (see e.g. \citeads{2011MNRAS.418.2292M}), we can distinguish QPO types, which fall into different regions of this graph. We see the typical negative gradient relation between the rms and centroid frequency of the type C QPOs ($r_{xy}=-0.997$, $p=1.0\times10^{-16}$ for the \textit{Astrosat} observations, and $r_{xy}=-0.91$, $p=1.0\times10^{-20}$ for the combined \textit{NuSTAR}, and \textit{XMM-Newton} observations), and the independence of the rms and centroid frequency of type A QPOs ($r_{xy}=0.23$, $p=0.61$). We indicate the region where we would expect the type B QPOs to appear, based on the observations of other BHTs. Compared to similar plots by e.g. \citeads{2011MNRAS.418.2292M}, we see the clear lack of type B QPOs in this outburst. This is yet another indication that we only observed QPO types A and C, but no type B QPO. 

We also tested whether the change in the timing properties could be explained by the addition of a separate flux component which dilutes the variation seen in the PDS. We added a component to the dim state \textit{Astrosat} light curve, whose value in every time bin is given by a Poisson distribution with a mean value set equal to the mean counts per bin of this \textit{Astrosat} dim state. This models a $100\%$ flux increase by a separate feature with purely Poissonian variability. As expected, this hardly affected the overall shape of the PDS, and QPO properties. We retained the same broad-band continuum, and comparatively narrow QPO. Hence, the increase in flux is not directly responsible for the change in the QPO. The source of the type C QPO, and of the broadband continuum must therefore either be disrupted or blocked from being viewed, to obtain the observed changes to the PDS. 

In the time between the early and late flip-flops, we once again detected type C QPOs. These were found at frequencies of $3.5\textup{--}6.1~\mathrm{Hz}$, but were stronger (with an rms in the QPO of $\sim 10\%$), and narrower ($Q\sim 10$) than those observed in the dim flip-flop states. 

We did not find any QPOs in the late flip-flops but do detect a change to the PSD in these transitions. NuSTAR suffers from a significant and variable deadtime of $\sim 2.5 ~\mathrm{ms}$, whose effect can however be mostly corrected for by cross correlating the two light curves to generate a cospectrum (\citeads{2015ApJ...800..109B}). The total integrated rms obtained by this method differs slightly from the one that would be obtained from an ideal non-dead time affected PDS. But substantial changes to the rms are nonetheless significant. Despite the non-detection of a QPO, we still determined the total integrated rms between 0.01 and 20 Hz, of the NuSTAR cospectra in the bright and dim states of the late flip-flops. It was found to be about 8-12\% in the dim states, and 6\% in the bright states. We therefore still observe a significant change in the PDS in late flip-flop transitions, which cannot be obtained merely by a dilution of variations due to the higher count rate of the bright states. The value of the rms in the two states, as well as their position in the HID, seems to suggest that the late flip-flops correspond to transitions between the HIMS and the SIMS. However, the bright state, with the lower rms was found at a greater hardness than the dim state. 

The final observation by \textit{NuSTAR} in the SIMS (see Fig. \ref{LCNSIMS}) also featured some changes in flux. We also did not detect any QPO in this observation. The orbit with the lowest flux level had an rms of 7\%, and the orbit with the highest flux had an rms of 10\%. So there is some indication of changes to the PDS in this observation as well, though the difference is smaller, and less significant than the rms differences found in any of the early or late flip-flops. 

No HFQPOs were detected at any point in the outburst. 

\section{Energy spectra}

\subsection{Dust Scattering Halo and Energy Shifts}

\begin{figure}[H]
\begin{subfigure}{0.5\textwidth}
    \resizebox{\hsize}{!}{\includegraphics{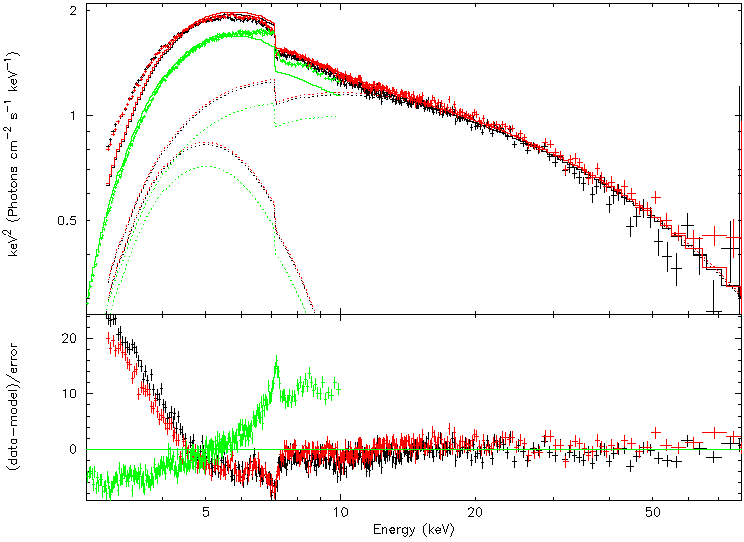}}
\caption{Spectra without a DSH, and without an energy scale correction.}
    \label{SpecNX1}
\end{subfigure}
\begin{subfigure}{0.5\textwidth}
    \resizebox{\hsize}{!}{\includegraphics{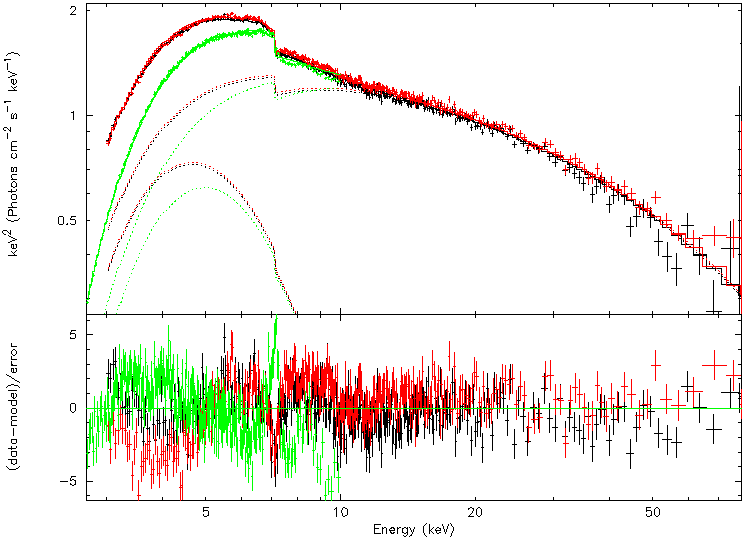}}
    \caption{Spectra with a DSH, but without an energy scale correction.}
    \label{SpecNX2}
\end{subfigure}
\begin{subfigure}{0.5\textwidth}
    \resizebox{\hsize}{!}{\includegraphics{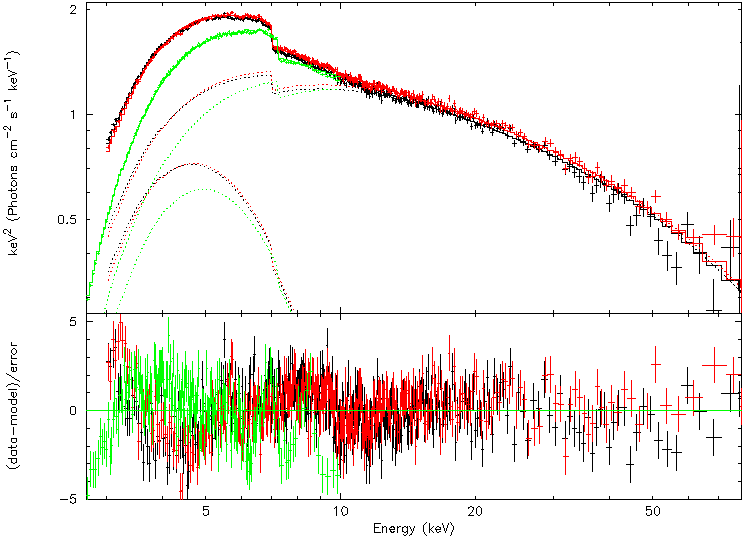}}
    \caption{Spectra with both a DSH, and an energy scale correction.}
    \label{SpecNX3}
\end{subfigure}
\caption{Absolutely simultaneous \textit{NuSTAR} (black for FPMA, red for FPMB) and \textit{XMM-Newton} (green) spectra from the long dim state observed on MJD 58174, showcasing the necessity of including both a DSH correction, and an energy calibration correction. Only when both corrections are applied do the spectra agree reasonably well.}
\label{SpecNXcomp}
\end{figure}

In Fig. \ref{SpecNXcomp} we plot the \textit{NuSTAR} and \textit{XMM-Newton} spectra of an absolutely simultaneous set of intervals from the long dim state observed on MJD 58174. We only include times when both instruments were observing Swift J1658.2-4242. As Fig. \ref{SpecNX1} shows, these spectra differ quite a lot, especially at low energies. It is impossible to fit them both using exactly the same parameters, unless only data at energies exceeding $5.5~\mathrm{keV}$ is considered. This is why we restricted the medium energy X-ray light curve of Fig. \ref{LCtot}, \ref{LCNX}, \ref{LCff}, and \ref{LCtotout} to the range of $5.5\textup{--}10~\mathrm{keV}$. The difference at low energies is caused by the DSH, combined with distinct source extraction methods and regions. 

Using the DSH model we developed for the \textit{XMM-Newton} pn CCD in Timing mode, and the standard DSH model by \citeads{2017MNRAS.468.2532J} and \citeads{2019ApJ...875..157J} for \textit{NuSTAR}, we managed to remove almost the entire difference between the \textit{NuSTAR} and \textit{XMM-Newton} spectra at low energies, as is shown in Fig. \ref{SpecNX2}. The remaining difference between the two spectra at low energies could be due to background contamination or indicate the limit of the usefulness of our DSH model at large hydrogen column densities. 

\citeads{2018ApJ...865...18X} and \citeads{2019ApJ...879...93X} describe an absorption line at an energy of $7.03~\mathrm{keV}$, very close to the iron K-edge at $7.112 ~\mathrm{keV}$. This line can be seen in Fig. \ref{SpecNX2}, and might be the signature of an accretion disk wind (\citeads{2012MNRAS.422L..11P}, \citeads{2016AN....337..512P}, and \citeads{2016AN....337..368D}). The simultaneous \textit{XMM-Newton} spectra however feature an emission, rather than an absorption line at this energy. Clearly there is a contradiction here which needs to be resolved. As the absorption and emission lines are very close to the iron K-edge, we investigated the possibility that a wrong energy calibration at $\sim 7~\mathrm{keV}$ could produce a fake emission or absorption line. By adding a non-zero redshift component to describe a slight shift in energy, having different values for \textit{NuSTAR} and for \textit{XMM-Newton}, we achieved a good agreement between the two spectra at these energies. 

Therefore, either the \textit{NuSTAR}, or the \textit{XMM-Newton} spectra, or both, require an energy recalibration. We examined the \textit{Chandra} HETG spectrum, which was obtained simultaneous to the last \textit{NuSTAR} observation, after the end of the late flip-flop region. As the \textit{Chandra} spectrum is also strongly affected by the DSH at low energies, we restricted the fit to $5.5\textup{--}10~\mathrm{keV}$, to ensure that the DSH correction would not bias our results. The \textit{NuSTAR} spectrum contains a strong absorption line at $\sim7~\mathrm{keV}$, as in all other observations. We fitted the possible absorption or emission line by including the additive \texttt{gauss} model in the spectral fit. For \textit{NuSTAR}, this possible absorption line was best fitted with with a centroid energy of $7.224^{+0.099}_{-0.056}~\mathrm{keV}$, a variance of $(1.47\pm0.85)\times10^{-1}~\mathrm{keV}$, and a normalization of $\left(-2.23^{+0.75}_{-0.53}\right)\times 10^{-4}~\mathrm{photons}~\mathrm{cm}^{-2}~\mathrm{s}^{-1}$, with the negative sign indicating that this is an absorption line. This line has an associated equivalent width of $-28.5^{+9.6}_{-6.8}~\mathrm{eV}$. However, the \textit{Chandra} spectrum was best fitted without the inclusion of an emission or absorption line at this energy. When the additive gaussian component was included, and the centroid energy allowed to vary, it did not fit to anywhere near to the value found by \textit{NuSTAR}. When forcing the centroid energy and variance of the gaussian included in the \textit{Chandra} fit to be equal to their best fitted values in the \textit{NuSTAR} spectrum, we found the best fit Gaussian normalization to be: $\left(1.9^{+3.2}_{-2.9}\right)\times 10^{-4}~\mathrm{photons}~\mathrm{cm}^{-2}~\mathrm{s}^{-1}$, corresponding to an equivalent width of $\left(2.4^{+4.0}_{-3.7}\right) \times10^{-2}~\mathrm{keV}$. This does not support the existence of either an absorption or an emission line. Due to the higher resolution and better calibration of \textit{Chandra} spectra, we therefore decided to use an energy shift in the \textit{NuSTAR} and \textit{XMM-Newton} spectra to ensure their consistency with the non-detection of an emission or absorption line at $\sim7~\mathrm{keV}$ by \textit{Chandra}. 

We obtained further support for this idea, when fitting for the energy of the iron K-edge. We not only find that the best fit energy of the edge differs from the value it should have, of $7.112 ~\mathrm{keV}$, but that in using the best fit edge energy, we no longer see absorption or emission lines at these energies.  

To measure the energy shift of the iron K-edge, we fitted \textit{NuSTAR} and \textit{XMM-Newton} spectra between $5.5\textup{--}9~\mathrm{keV}$ using the combination of XSPEC models: \texttt{edge*diskbb}. By comparing the fitted edge energy with the value it should have, of $7.112 ~\mathrm{keV}$, we can determine a redshift which can be implemented into a model for the entire spectrum, to correct for the energy offset observed at the iron K-edge, and ensure consistency in the spectra of \textit{NuSTAR}, \textit{XMM-Newton}, and \textit{Chandra}.

The best fit results of the shift in iron K-edge energy from its expected value, in the \textit{NuSTAR} and \textit{XMM-Newton} spectra, are shown in Fig. \ref{FeKedge}. As expected, \textit{NuSTAR} spectra require a negative redshift energy correction, and \textit{XMM-Newton} spectra require a positive redshift energy correction. When applying these energy corrections via a redshift component, we obtain the spectra shown in Figure \ref{SpecNX3}. We still retain slight differences between the spectra, but they are now undoubtedly more consistent than they were without the DSH correction, and without the energy calibration correction. 

\subsection{Spectral Analysis}

Comparing the spectra of the bright, dim, and late flip-flop states, we find that they are remarkably similar (see Fig. \ref{Speccomp}). Visually, besides having a different normalization, these spectra are more similar to each other than to either of the three non flip-flop spectra shown in this figure. The difference between the bright and dim flip-flop spectra is larger at higher energies. This results in the increase in hardness ratio simultaneous to flux increases, as seen in Fig. \ref{LCNX}. 

The initial LHS spectrum differs noticeably from the other spectra we measured (see Fig. \ref{Speccomp}). We also find that the spectrum in the interval between the two sets of flip-flops, and the spectrum after the completion of the late flip-flops, are visually rather similar, besides having a different normalization. This would indicate that the spectra themselves distinguish the flip-flop intervals from other parts of the outburst. This behaviour is reminiscent of state transitions, which typically are observed to occur at a very similar hardness.

\begin{figure}[h]
    \resizebox{\hsize}{!}{\includegraphics{deltaEFeKedge4.pdf}}
    \caption{Graph of the difference between the energy at the expected iron K-edge, at $7.112~\mathrm{keV}$, and the best fitted edge energy, in \textit{NuSTAR} and \textit{XMM-Newton} spectra. The energy shift was determined by fitting a simple \texttt{zedge*diskbb} model to spectra confined to the energy range of $5.5\textup{--}9~\mathrm{keV}$. \label{FeKedge}}
\end{figure}

To obtain a better understanding of the differences between the bright and dim states of the flip-flop, and what distinguishes the flip-flop intervals from other parts of the outburst, we fit the spectra of individual flip-flop states, and entire observations which did not feature any flip-flops, using two sets of XSPEC models. 

In Model 1, we fit the spectral energy contribution from the multicolour black body component of the accretion disk, using the model \texttt{diskbb}. The high energy power law component is represented by the \texttt{cutoffpl} model, which includes a cutoff energy, beyond which the power law becomes steeper. We also add a \texttt{diskline} component to describe the iron $K_{\alpha}$ line at $6.4~\mathrm{keV}$. The effect of absorption by interstellar dust is modelled through the multiplicative \texttt{ztbabs} component. We shift the energy of the spectrum, to ensure the iron K-edge is located at $7.112~\mathrm{keV}$, through freezing the redshift at the energy shift found previously for each individual spectrum. We correct for the effects of the DSH through the multiplicative model \texttt{dscor}. We also used the \texttt{constant} multiplicative model to apply a cross-calibration correction constant between the two focal plane modules of \textit{NuSTAR}. Combining all these elements, our \textbf{Model 1} in XSPEC jargon is: \texttt{dscor*ztbabs*constant*(diskbb+diskline+}\newline\texttt{cutoffpl)}.

The inclination of the accretion disk around the black hole is unknown, but the dips seen at the start of the outburst, and the shape of the HID (compare Fig. \ref{HIDS} with the HIDs of \citeads{2013MNRAS.432.1330M}), suggest that the accretion disk around the black hole has a high inclination relative to the line of sight. To maintain consistency in the spectral fits, we set the inclination equal to $70^{\circ}$ for all spectra. The choice of inclination does not affect the other spectral fitting parameters significantly. Unless a close to face-on inclination is chosen, the best fit parameter values agree within their respective errors. The energy of the added \texttt{diskline} was set equal to $6.4~\mathrm{keV}$. We were unable to fit for the inner and outer disk radii using this model, so we set these parameters equal to $6\,GM/c^2$, and $1000\,GM/c^2$, respectively. The remaining spectral parameters were left free, and their values determined when fitting the data using this Model 1. These fitting parameters are: the hydrogen column density ($N_H$), the inner accretion disk temperature ($T_{in}$), the emissivity power law index of the iron line ($B_{10}$), the power law photon index ($\Gamma$), the high energy cutoff ($E_{cut}$), as well as the normalization of the three additive spectral components, \texttt{diskbb}, \texttt{diskline}, and \texttt{cutoffpl}. 

We also fitted the energy spectra using a more physical comptonization model by \citeads{1996ApJ...470..249P}, \texttt{compPS}. As this model describes both the disk black body, and the power law component of the spectrum, we replaced both these components of Model 1 by \texttt{compPS}, unlike \citeads{2019ApJ...879...93X}. Therefore, \textbf{Model 2} is: \texttt{dscor*ztbabs*constant*(diskline+compPS)}. We used the same fixed values for the iron $K_{\alpha}$ line energy, the inner and outer accretion disk radii, and the inclination of the accretion disk, which were already used in Model 1. We used a spherical corona geometry and specified a multicolour disk black body component by confining the disk temperatures specified in the model to have negative values. The absolute magnitudes of those values then represent the temperatures at the inner edge of the geometrically thin, optically thick accretion disk. We fit the spectra with a free hydrogen column density ($N_H$), emissivity power law index ($B_{10}$), coronal electron temperature ($T_e$), inner accretion disk temperature ($T_{in}$), coronal optical depth ($\tau_y$), relativistic reflection normalization ($R_r$), and normalization of the \texttt{compPS}, and \texttt{diskline} components. Spectra fitted with both Models 1 and 2 always had a lower $\chi^2$ at the same number of degrees of freedom, when fitted with Model 2. This indicates that the more physical model provides a better description of the data.

\begin{figure}[h]
    \resizebox{\hsize}{!}{\includegraphics{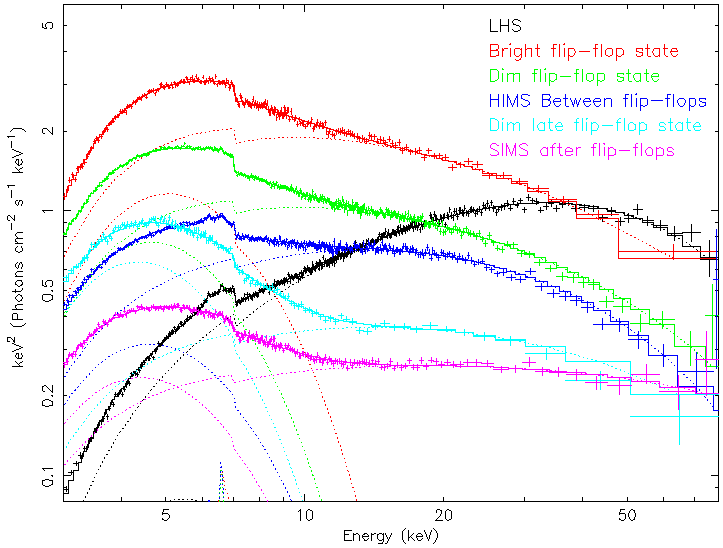}}
    \caption{Comparison of the \textit{NuSTAR} FPMA spectra and the varying strengths of the individual components they are fit with, at different times during the outburst. The spectra here are all fitted with Model 1. \label{Speccomp}}
\end{figure}

We fitted these two models to the \textit{NuSTAR} FPMA and FPMB spectra on their own, as well as to the \textit{NuSTAR} FPMA, FPMB, and \textit{XMM-Newton} spectra in absolutely simultaneous intervals of each observation, or flip-flop state. After correcting for the effects of the DSH and applying an energy correction, there was a good agreement in the best fit parameter values of both sets of fits (see Fig. \ref{SpecNXcomp}). Due to the shorter \textit{XMM-Newton} exposure times, the latter selection contained a smaller number of flip-flop states, and shorter intervals of simultaneous spectra. To remain consistent, and describe all the observed flip-flops seen by either of the two instruments, we only plot (Fig. \ref{Specparcpl}, and \ref{Specparcps}) and list (Tables \ref{Tabspecfitcpl}, and \ref{Tabspecfitcps}) the best fit parameter values found when fitting the \textit{NuSTAR} FPMA and FPMB spectra together. In such fits, we only allow the cross-calibration constant to differ between the spectra from the two focal plane modules, along with the difference in the source extractions regions used, an important element of the DSH model. Errors are quoted at a $90\%$ confidence interval for each parameter. 

In Fig. \ref{specdiffT}, we analyse the differences between the bright and dim state spectra. All the spectra in this figure are fitted with Model 2. Initially, the parameters of both the bright and dim state spectra were set equal to the best fit values of the dim state spectrum, in (a). The dim state spectrum is fitted very well, but the bright state spectrum has huge residuals. Next, we changed the inner disk temperature of the spectral fit describing the bright state, to the best fit value that had been found when this spectrum was fitted on its own, with all parameters described above being set free to vary, in (b). This one change on its own already accounts for most of the difference between the bright and dim state spectra. So, the dominant cause of the change in flux between the dim and bright states, is an increase in disk temperature, and the effect this has on the power law component. 

In plot (c), we also change the \texttt{compPS} normalization of the bright state spectral fit to its best fit value. This has now removed most of the remaining residuals of the fit, except at low energy, where the greater $N_H$ in the bright state has a noticeable impact. When the $N_H$ itself is changed from the dim to the bright state best fit value, in (d), the bright state spectrum is fitted very well, with only minor differences remaining. By changing all the other remaining parameters, we obtain the final graph, (e). 

In Fig. \ref{Specparcpl} and \ref{Specparcps}, we distinguished the results of the spectral fits, by whether flip-flops were detected in the observations, and what QPO type was observed, if any. Interestingly, the observations without a detectable QPO are the least bright, and the bright flip-flop states with a type A QPO are the brightest. All regions in which a type C QPO was observed lie between these two extremes. All identical or comparable spectral parameters involved in the two different fitting models showed similar, though not identical values and flux dependencies, when applied to the same set of data.

In the results of the spectral fits, we found $N_H$ to correlate with flux. When using all the observations shown in Fig. \ref{Specparcpl}, and \ref{Specparcps}, we find $r_{xy,1} = 0.93$, $p_1=1.0\times 10^{-8}$, and $r_{xy,2} = 0.83$, $p_2=1.2\times10^{-5}$ for the two spectral fitting models, respectively. There is however a partial degeneracy between $N_H$ and $T_{in}$, as both parameters push the \texttt{diskbb} spectral component to higher energies. We therefore analysed whether the observed change in $N_H$ could also be reproduced by a larger change in $T_{in}$. To do so, we fit all the spectra together, each spectrum with independent fitting parameters as described above, except for the hydrogen column density, which was tied together for all spectral fits. Then for comparison, we fit all spectra again, but this time with each spectrum having its own independent hydrogen column density. The second set of fits had 18 fewer degrees of freedom, but also had a $\chi^2$ which was smaller by $693$. We therefore conclude that the observed correlation between hydrogen absorption and flux is a real effect. This implies that a significant fraction of the $N_H$ must be local to the system, and able to vary within $\sim2~\mathrm{ks}$. However, this goes against the assumption of \citeads{2019ApJ...875..157J}, that all the absorption occurred in the DSH. 

We measured a lower average hydrogen column density for the dim flip-flop states than for the bright states. The effect of the different $N_H$ measured in the bright and dim states can be seen in Fig. \ref{specdiffT} (c), and (d). The low p-value for the null hypothesis which assumes a constant $N_H$ in both states, of $p=6.4\times10^{-4}$, and $p=1.9\times10^{-4}$ for the two models, indicates that the hydrogen column density changes during the flip-flops. Both \citeads{2019ApJ...879...93X} and \citeads{2019ApJ...887..101J} also detected a greater $N_H$ in the bright flip-flop states, though their values differ from the ones we found, as they did not include the DSH model in their spectral fits. As another test of whether the $N_H$ changes during the flip-flops, we repeated the above test, but this time only for the flip-flop spectra. We found that the decrease of 9 degrees of freedom was accompanied by a drop of the total $\chi^2$, by $103$. This also supports the notion that the $N_H$ changes during flip-flop transitions. 

In both models we find an increase in the inner disk temperature with X-ray flux, most notably between the bright and dim flip-flop states. As Fig. \ref{specdiffT} indicates, this change in temperature is the cause of most of the difference between the bright and dim state spectra. This increase is enshrined in the detected correlation between temperature and flux of the early flip-flop bright and dim states, with $r_{xy,1}=0.92$, $p_1=1.6\times10^{-4}$, and $r_{xy,2}=0.91$, $p_2=2.4\times10^{-4}$ for the two models. Interestingly, the two states by themselves do not show any significant correlation at all, with $p_1=0.51$, $p_2=0.81$ for the bright states, and $p_1=0.98$, $p_2=0.59$ for the dim states. At comparable fluxes, the disk has a lower inner temperature in both of the flip-flop intervals, compared to other parts of the outburst. This is a consequence of the power law component being stronger during the flip-flops, than in other regions of the outburst whose spectra we obtained, and can be seen in Fig. \ref{Speccomp}.

\begin{figure*}[h]
\begin{subfigure}{0.49\textwidth}
    \resizebox{\hsize}{!}{\includegraphics{specparcompfinal_cpl2_fig1a.pdf}}
\end{subfigure}
\begin{subfigure}{0.49\textwidth}
    \resizebox{\hsize}{!}{\includegraphics{specparcompfinal_cpl2_fig2a.pdf}}
\end{subfigure}
\caption{Comparison of the best fit \textit{NuSTAR} spectral parameters of individual observations and flip-flop states fitted with Model 1, \newline \texttt{dscor*tbabs*constant*(diskline+diskbb+cutoffpl)}. The spectra are distinguished by their corresponding fluxes, timing properties, and their flip-flop state classifications. The first \textit{NuSTAR} observation has been excluded, as its spectrum had a negligible black body component, and therefore differed too greatly from the spectra shown here. The blue dashed line in the graph of \texttt{diskline} normalization as a function of flux, depicts a line of constant equivalent width, assuming no change in spectral shape at an energy of 6.4 keV. This is a reasonably good assumption, as is demonstrated by Fig. \ref{Speccomp}. The two blue dashed lines in the \texttt{diskbb} normalization graph indicate the average value of this parameter for flip-flop and non flip-flop spectra. One dim flip-flop observation featuring a type C QPO has been omitted from the graph of the high energy cutoff for display clarity, as this parameter was not well constrained in the spectral fit of this particular observation, having a best fit of $110^{+220}_{-50} ~\mathrm{keV}$. We also excluded one bright state, and three late flip-flop data points from the graph of $\mathrm{Betor}_{10}$, as the fits were insensitive to this parameter, and the errors in the measurements could not be determined. \label{Specparcpl}}
\end{figure*}

\begin{figure*}[h]
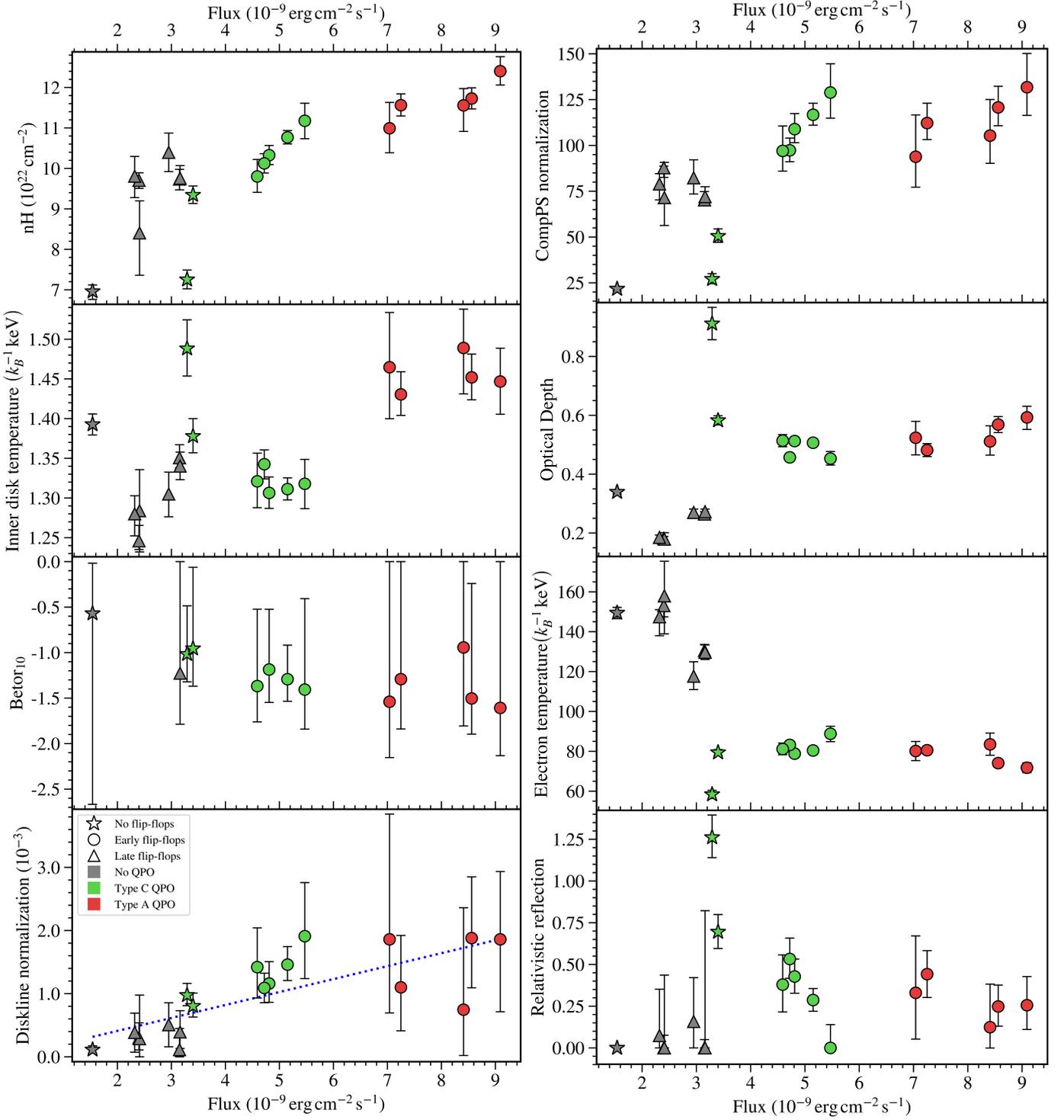

\begin{subfigure}{0.49\textwidth}
    \resizebox{\hsize}{!}{\includegraphics{specparcompfinal_cps2_fig1a.pdf}}
\end{subfigure}
\begin{subfigure}{0.49\textwidth}
    \resizebox{\hsize}{!}{\includegraphics{specparcompfinal_cps2_fig2a.pdf}}
\end{subfigure}
\caption{Comparison of the best fit \textit{NuSTAR} spectral parameters of individual observations and flip-flop states fitted with Model 2, \texttt{dscor*tbabs*constant*(diskline+compPS)}. We excluded three late flip-flop data points from the graph of $\mathrm{Betor}_{10}$, because the fits were insensitive to this parameter, and the errors in the measurements could not be determined. \label{Specparcps}}
 \end{figure*}

On the two graphs of the dependence of the iron $K_{\alpha}$ \texttt{diskline} normalization on X-ray flux, we plotted a straight line depicting a constant equivalent width, under the assumption of a uniform spectral shape. Fig. \ref{Speccomp} indicates that even though the spectra do have some differences in their spectral shape, this is still a reasonably good assumption. We notice that the equivalent width of the bright and dim flip-flop states is very similar ($46\pm12~\mathrm{eV}$). This suggests that the disk intercepts a similar solid angle of light emitted by the primary X-ray source in both states.  The interval between the early and late flip-flops features a greater equivalent width ($83\pm20~\mathrm{eV}$), and both the late flip-flops ($20\pm5~\mathrm{eV}$), and the final NuSTAR observation in the SIMS ($13\pm7~\mathrm{eV}$) have lower equivalent widths than the early flip-flops. the largest equivalent width ($166\pm8~\mathrm{eV}$) was measured in the first \textit{NuSTAR} observation, when the BHT was in the LHS. 

\begin{figure}[h]
    \resizebox{\hsize}{!}{\includegraphics{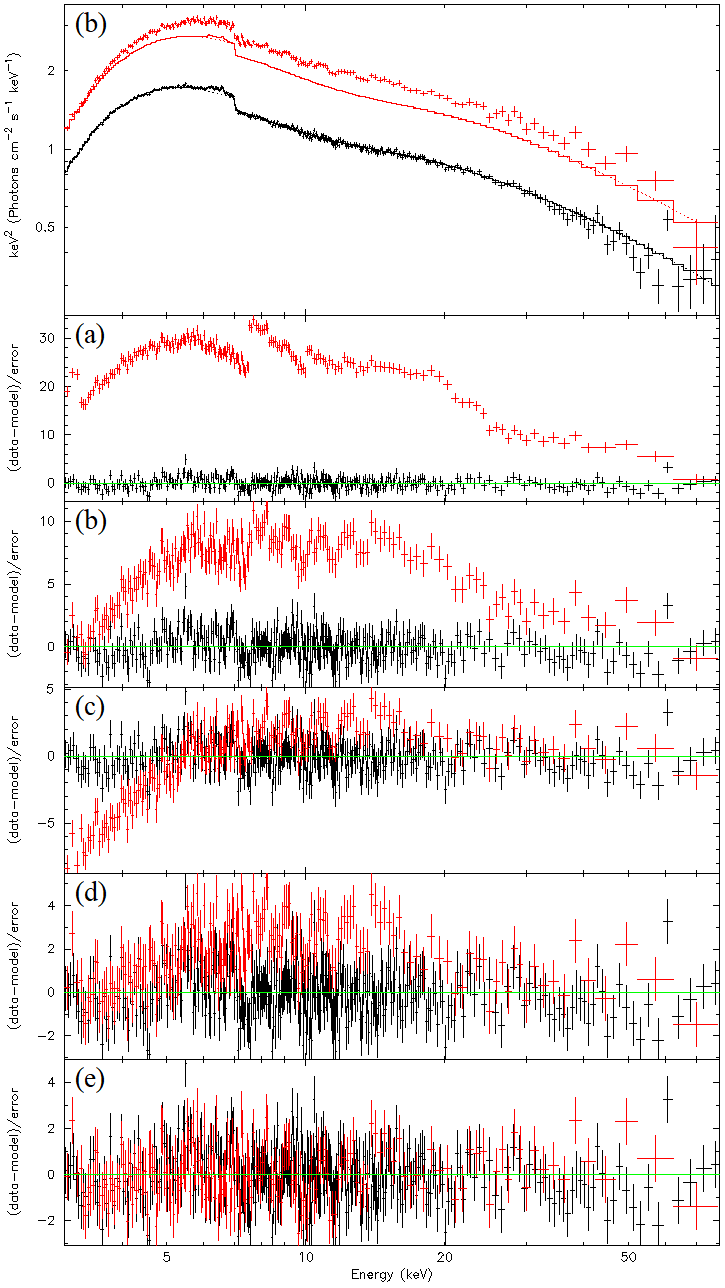}}
    \caption{Comparison of the bright (red) and dim (black) flip-flop spectra fitted with Model 2, with the spectral parameters initially set to the best fit values of the dim state (a). By setting the inner disk temperature of the bright flip-flop spectral fit to its best fit value, we obtain fit (b), and close most of the gap between the two spectra. In plot (c) we additionally modify the \texttt{compPS} normalization of the bright state spectrum to its best fitting value. Most of the remaining difference between the two spectra is corrected for by the change in $N_H$ (d). Finally, in plot (e) we show the residuals of the best fits to both bright and dim spectra, with completely independent parameters. \label{specdiffT}}

\end{figure}

The emissivity power law index, $B_{10}$, does not change significantly throughout the range of observations we analysed, in either of the two models.  

The normalization of the \texttt{diskbb} component, is given by the equation: $N_{dbb}=\left(\frac{R_{in}}{D_{10}}\right)^2 \cos{i}$. It is proportional to the square of the accretion disk inner truncation radius ($R_{in}$), and the cosine of the disk inclination ($i$). The inclination of the accretion disk is not known, and the distance to the binary ($D_{10}$, measured in units of $10\, \mathrm{kpc}$) has a significant uncertainty, so absolute measurements of the truncation radius from the best fitted values of $N_{dbb}$ are very uncertain. We can however determine fractional changes in the inner radius, assuming a constant inclination. In Fig. \ref{Specparcpl} we observe that the \texttt{diskbb} normalization is almost constant in all flip-flop and late flip-flop states, and significantly larger (by a factor of $2\textup{--}4$) than in observations not involving any flip-flops. All the flip-flop observations are consistent with variance around a common mean of $67.3\pm7.1$, having $r_{xy,1}=-0.35$, and $p_1=0.18$. In contrast, the observations without flip-flops differ too much to be considered as having the same normalization. This therefore suggests that all flip-flops correspond to a similar geometry, which however differs significantly from the geometry responsible for the spectra of other observations, which also differ a lot amongst themselves. 

$\Gamma$, the photon index of the power law spectral component, does not change a lot within the spectra we considered. The early flip-flops in particular have an almost constant $\Gamma\approx2.05\pm0.12$, with $p=0.39$.

The high energy cutoff increases slightly with increasing flux during the early flip-flops, having $p_1=0.02$, but there is not sufficient evidence to support a change in the cutoff within the range of bright or dim state observations individually. In the last two \textit{NuSTAR} observations, in which we did not detect any QPO, we found that their spectra have a higher average cutoff ($83\pm40~\mathrm{keV}$), than any of the spectra of intervals with a measurable type C or A QPO ($52\pm21~\mathrm{keV}$). 

The normalization of the $\texttt{cutoffpl}$ component increases almost linearly for all observations, with no clear distinction between the flip-flop and non flip-flop states, or the nature of the QPO. When using the results in all spectra, we obtain a correlation coefficient of $r_{xy}=0.94$, and a p-value of $p=3.6\times10^{-9}$.

We notice a near constant optical depth for the bright and dim flip-flop states of $\tau_y\sim0.512\pm0.042$. But it nonetheless features a very gradual rise from dim to bright state, with $r_{xy}=0.67$, and $p=0.032$. A greater optical depth of $\tau_y\approx0.74\pm0.16$ was found in the interval between the two sets of flip-flops. The remaining observations of the late flip-flops and the SIMS, for which we did not manage to identify any QPO, had a lower average optical depth of $\tau_y\approx0.242\pm0.057$. 

We see a similar trend in the electron temperature as a function of flux, of the \texttt{compps} model, as in the high energy cutoff of the \texttt{cutoffpl} model. This was expected, as these two parameters are related. The flip-flops states have $T_e\approx80.2\pm{4.5}~\mathrm{k_B}^{-1}~\mathrm{keV}$, and only feature a very small negative dependence of electron temperature on flux, with $r_{xy}=-0.55$, and $p=0.097$. A lower electron temperature is found between the two sets of flip-flops, with an average of $T_e\approx68\pm{11}~\mathrm{k_B}^{-1}~\mathrm{keV}$. Observations of the late flip-flops, and the SIMS have a higher average electron temperature of $T_e\approx140 \pm{14}~\mathrm{k_B}^{-1}~\mathrm{keV}$.

Finally, the relativistic reflection component of the \texttt{compps} model does not indicate a dependence on flux for the early flip-flop states, which have an average of $R_r\approx0.30\pm0.15$. This also supports the idea that the area of the disk, as seen from the central X-ray source, remains constant during flip-flop transitions. Between the early and late flip-flop intervals, we detect an increase in this component, up to $R_r\approx0.98\pm0.28$. This is followed by a drop down to close to 0 for the late flip-flops, and the SIMS. 

To study the temporal evolution of the spectral features and to determine whether these change immediately at the transition in the light curve and spectrogram, we fitted the spectra of $2 ~\mathrm{ks}$ segments on either side of every observed flip-flop. The resulting fits indicated an immediate change of the inner disk temperature, and the power law normalization. There was a slight suggestion of delayed hydrogen column density changes in these spectra, but changes were small, and occurred within the error margin, so a possible delayed change in $N_H$ could not be corroborated. 

The final \textit{NuSTAR} observation still had both a strong multicolour disk black body component, in addition to a strong power law component. Neither of them could be excluded from the fit or was significantly dimmer than the other. This is indicative of the intermediate states. It does not agree with the HSS definition, in which the disk black body component dominates. The remaining observations by \textit{Swift/XRT} needed to be fit with both a black body, and a power law component. Therefore, it seems like Swift J1658.2-4242 never reached the HSS. 

\section{Discussion}

We have investigated the differences between the bright and dim flip-flop states, as well as the distinctions between the flip-flop intervals, and other parts of the outburst, to enhance our knowledge of the changes during, and the causes of the extreme flip-flops we observed. Unfortunately, there is a distinct lack of models describing these phenomena. 

The flip-flops we observe seem to be of an extreme and rare variety, which has never been seen before. Previously observed flip-flops had flux ratios of between 1.03 and 1.33 between the bright and dim states, and corresponded to transitions between QPO types A and B, or between types B and C. The flip-flops of Swift J1658.2-4242 however had flux ratios of up to 1.77, and corresponded to transitions between QPO types A and C. 

Flip-flops have been observed a few times before, but they are not very common, having only been seen in seven other systems so far, out of $\sim 60$ known transient black hole binaries (\citeads{2016A&A...587A..61C}). However, it is possible that a lot of flip-flops were missed, due to short exposure times, sparse monitoring, long flip-flop state durations, short intervals of flip-flop activity, or small flip-flop amplitudes.

The duration, and frequency of \textit{XMM-Newton}, and \textit{NuSTAR} observations in the critical period of transitions between intermediate states in the outburst of a BHT is unusual, and what enabled the discovery of 17 flip-flop transitions, 15 of which were directly observed, and 2 of which were inferred. This is more than have ever been seen before within a single outburst of a BHT, of a comparable flip-flop duration. The collection of these observations are therefore ideal for examining these phenomena within one outburst, without worrying about the differences between different outbursts, and different sources. As flip-flop states of the kind we observed for Swift J1658.2-4242 last for at least $2.7~\mathrm{ks}$, but can also last for tens of ks, or much longer, short observations risk missing flip-flops entirely. It is possible that part of the apparent scarcity of flip-flops in the literature is merely the product of shorter, and fewer observations than would be required to detect flip-flops. Daily averaged count rates seem to indicate the presence of flip-flops in several other outbursts but could not be verified without direct detection through long continuous observations. Flip-flops are possibly much more common than they appear to be. 

In the standard definition of spectral-timing states used to describe the outburst of a BHT (\citeads{2011BASI...39..409B}), the dim flip-flop states we observed are classified as HIMS, as they have both strong black body, and strong power law spectral components, feature a type C QPO, and have an rms in the range typical of the HIMS. The bright flip-flop states are classified into the AS (\citeads{2010LNP...794.....B}, \citeads{2012MNRAS.427..595M}), due to their location in the HID. This AS has similar timing properties as the SIMS, is however found at a greater hardness, and a significantly higher flux, than the dim state HIMS. 

The late flip-flops are classified into the HIMS (dim state), and SIMS (bright state), due to their location in the HID, and their rms, and PDS properties. 

Interestingly, we always found a negative correlation between count rate and hardness in all observations with a type C QPO (see Fig. \ref{HIDN}). In contrast, the bright flip-flop states, which feature a type A QPO, as well as later observations in which no QPO was detected, have a positive correlation between count rate and hardness. This is predominantly a result of the relation between the path traced by a BHT in outburst in the HID, and the QPO types observed at particular points of the outburst. 

Compared to previously observed flip-flops, those in Swift J1658.2-4242 fit into the subsample with long duration states (of order $\sim0.1\textup{--}10~\mathrm{ks}$), and long transition times (of order $\sim10\textup{--}100~\mathrm{s}$). We did not observe any of the short duration flip-flops, which have durations of order $\sim10~\mathrm{s}$, and transition within fractions of a second. This supports the idea that although the duration and transition time of a flip-flop can change quite a lot within one outburst of a system, they remain within about 1 order of magnitude of each other. No system has been found yet, in which both long and short duration flip-flips were found. This could however be the result of an insufficiently large sample of flip-flops. 

The non-appearance of a type B QPO is at odds with previous findings by \hyperlink{C04}{C04} according to which a flip-flop transition always involve a type B QPO. They plot a hierarchy of QPO types as a function of flux, with type A occurring at the greatest fluxes, type C at the lowest fluxes, and type B in between the two. The flux limits separating the three QPO types decayed exponentially with time. Although all flip-flops between types B and C follow this hierarchy, several A to B flip-flops contradict it. It was thought that type B QPOs are essential for flip-flops and have to appear in either the bright or the dim state of every flip-flop. We however detected the first instance of flip-flops not involving a type B in either of the two states and conclude that flip-flops do not have to involve a type B QPO. 

Indeed, all hard to soft state transitions were thought to require a type B QPO. Although the bright flip-flop states fits into the properties of the AS rather than the SIMS, the observed change in rms is equivalent to the standard distinction between the HIMS and SIMS. We however find that these kinds of state transitions do not require a type B QPO.

\begin{figure}[h]
\resizebox{\hsize}{!}{\includegraphics{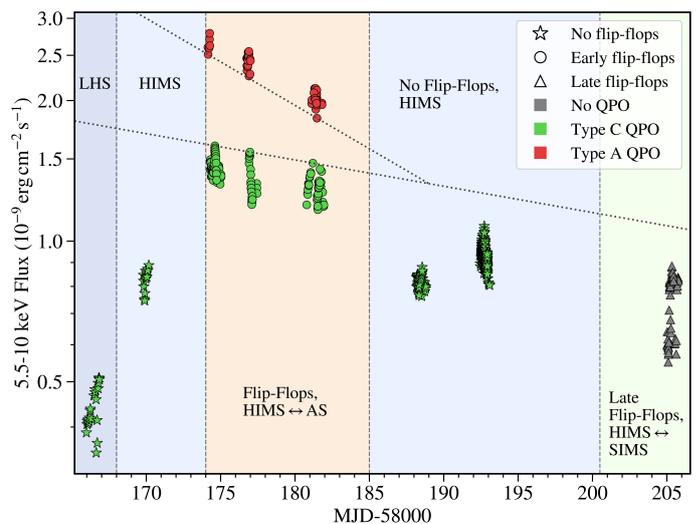}}
    \caption{Swift J1658.2-4242 light curve, with flux plotted on a logarithmic scale, analogous to Fig. 14 of \protect\hyperlink{C04}{C04}. We also plotted exponentials describing the minimum type A flux, and the maximum type C flux, as a function of time. We only included \textit{NuSTAR}, and \textit{XMM-Newton} data, as their spectra enable an accurate flux determination, and as the QPO type can be unambiguously identified from their light curves. We removed datapoints during transitions between the two states to show the limiting fluxes of each states when there is no transition.}
    \label{Caslines}
\end{figure}

In Fig. \ref{Caslines}, we plot a graph distinguishing QPO types in flux and time, analogous to fig. 14 of \hyperlink{C04}{C04}. Plotting the temporal evolution of the maximum type C flux, and the minimum type A flux as decaying exponentials, we also find a clear gap between the two. A noticeable difference is however that we detected different decay rates of these two exponentials, whereas \hyperlink{C04}{C04} found parallel lines in this log-linear light curve. 

\hyperlink{C04}{C04} observed type B QPOs in the gap between types A and C, but we did not. It should however be noted that they detected long-lived bright, or dim flip-flop states in this flux interval, rather than short transition periods, as we did. Perhaps the reason we did not detect any type Bs in this interval is that they take longer to switch on than the transition lasts for. The length of the transitions we found, is however longer than the observed time it took other type B QPOs in similar systems to appear or disappear. It therefore seems as though longer transitions would also not have involved type B QPOs.

It seems possible that there is a QPO phase diagram of parameter values, involving more than just the luminosity, which separate the different QPO states of a transient black hole, with transitions between different states happening at specific thresholds. So far, we have known of thresholds separating types A and B, and types B and C. Now we however know of a threshold separating types A and C directly, without a type B in between, in analogy to the solid to gaseous transition in a phase diagram.

The direct transition between QPO types A and C also supports the argument of \citeads{2012MNRAS.427..595M}, that types A and C originate from the same physical process, but type Bs have an entirely different cause.

We note that it is possible for the type A QPOs to be present in the PDSs at all times, but remain undetected whenever a type C appears, due to its very shallow amplitude and low power compared to the type C QPO continuum, and background. Either the type C replaces the type A, or the type C appears, and disappears alongside changes in flux, but the type A always remains in the PDS. We could not rule out either possibility but note that a validation of either scenario could assist the interpretation of the origin of both flip-flops and QPOs.

We classify the early flip-flops of Swift J1658.2-4242 as transitions between the HIMS and the AS. Transitions between these states have not been described before, and most flip-flops observed so far would be associated with transitions between the HIMS and the SIMS. The extreme flip-flop properties observed in this system are possibly the result of transitions from the HIMS to the AS, rather than to the SIMS.

The state of the system at the time of jet ejection, and the start of the radio flare, is unknown. The last observation with \textit{Astrosat}, four days before the start of the flare was observed, found the source to be in the HIMS. The first observation after the start of the flare, by \textit{NuSTAR}, was the first detected bright flip-flop state, which is classified into the AS. Even though we did not observe a peak of the radio flux, we can infer that it almost certainly reached its peak during the first flip-flop observation. 

We note that the power law involved in the flip-flops differs from the standard power law of the LHS. The dominant power law of the LHS is the source of its high variability, with rms values of $10\textup{--}40\%$ (\citeads{2016AN....337..398M}). But during the flip-flops, we see a relative increase in the strength of the power law, and an associated sharp drop in the rms variability. This suggests that the corona generating the comptonized power law spectral feature has different properties during the flip-flop intervals, than at the start or end of an outburst. This idea is supported by the existence of the high energy cutoff of the power law component in the spectra. 

Within the sole detailed observation of a flip-flop transition, we found flux variation to precede changes in the PDS, but take a longer time to transition. This suggests the existence of two different, yet interrelated mechanisms: one for changes in the averaged flux, and another for differences in the PDS.

The simultaneity of the radio flare with the start of the flip-flop interval implies a possible causal connection between the two effects. Jet ejections, and subsequent radio flares have been linked with state transitions before (see e.g. \citeads{2004MNRAS.355.1105F}), and are characteristic features of X-ray binaries in outburst. However, we see no other radio flare at any later time, despite observing numerous flip-flops. We did not even see a flare on MJD 58175, when an \textit{ATCA} observation coincided with a \textit{NuSTAR} and \textit{XMM-Newton} measurement of a bright flip-flop state. If all flip-flop transitions are related to jet ejections, we should have detected a noticeably higher radio flux in this observation.

Due to the lack of additional radio flares, our observations suggest that they are only associated with the very first flip-flop transition. This view is supported by the observations described by \hyperlink{C04}{C04}. They detect an ejection episode just before the start of the first X-ray flip-flop interval, and another one within a second flip-flop interval, 11 days later. \hyperlink{K11}{K11} also found a rapid drop in radio emission, possibly belonging to the end of a radio flare, just before the start of the flip-flops. Therefore, the entire flip-flop behaviour might be the settling of the system back to an equilibrium state following the disruption caused by the initial jet ejection.  

One might expect that a radio flare should also have occurred at the start of the late flip-flop interval, but due to a lack of radio observations in this epoch, we cannot investigate this possibility. We encourage further analysis of similar events, to determine the connection between flip-flops observed in X-rays, and radio flares, or the lack thereof. 

We pose the question of whether the early and late flip-flops are two different instances of the same phenomena, or governed by entirely distinct mechanisms. The drop in the early flip-flop amplitude and state duration over time seems to continue into the late flip-flops as well. This is reminiscent of a damped oscillation, which encompasses both the early and the late flip-flops. There is a greater similarity between the spectra of the early and late flip-flops, than with the spectra obtained in the interval between the two sets of flip-flops, or after the late flip-flops have completed. For instance, the \texttt{diskbb} normalization implies that the flip-flops have very similar values of $R_{in}^2 \cos{i}$, which are significantly larger than corresponding values in observations outside of either of the two flip-flop intervals. 

However, there are some very clear distinctions between the two sets of flip-flops as well. The late flip-flops have a smaller amplitude, a greater frequency of transitions, and a shorter fundamental period underlying the timing of transitions, than the early flip-flops do. All early flip-flops are found at almost exactly the same locations on the HID, with only minor variations over several days. The late flip-flops however lie in an entirely different region of the HID and feature different correlations between hardness and intensity. QPOs were not detected in any of the late flip-flop states, but a noticeable difference in the rms was nonetheless observed. Transition times were longer on average than their earlier equivalents, but the duration of individual flip-flop states was shorter. The late flip-flop spectra also contain a noticeably smaller iron $K_{\alpha}$ equivalent width than the early flip-flops, suggesting a difference in geometry. 

In both the early and the late flip-flop intervals we observed a period relating the times of transition from one state to the other. The period found in the late flip-flops is slightly smaller than the one detected in the early flip-flops. We showed that it is possible both to fit the early and late flip-flop transition times under the assumption of a decreasing period, as well as to consider them separately. The similarity in the periods found in the two intervals suggests that either the late flip-flops are a continuation of the early flip-flops, or alternatively that they are separate events, but that flip-flops within the same outburst only allow for a limited variation of the transition period. The existence of this period implies a semi-periodic mechanism, by which the system can change from one to the other state only at specific points in time, but does not have to change at every multiple of the period. The time between one transition and the next appears to be a random integer multiple of the underlying period. 

The interval between the early and late flip-flops is clearly distinguished from them via its spectral, variability, and timing properties. This indicates that there is a different physical mechanism at play during it. These different properties also suggest that this interval really did not feature any unobserved flip-flops. However, of the two \textit{NuSTAR} and \textit{XMM} observations in this interval, the latter observations shows a significantly greater similarity to the flip-flop spectral and timing properties. The QPO in the first of these intra flip-flop observations has a frequency of 3.9 Hz. But the QPO in the second of these observations has a frequency of 5.7 Hz, much closer to the QPO frequency in the dim flip-flop states observed previously, of 6.2--8.3 Hz. The fitted spectral properties also support this interpretation, with the latter intra flip-flop observation having similar parameter values as the late flip-flops, as can be seen in Fig. \ref{Specparcpl}, and \ref{Specparcps}. This possibly indicates a gradual return to the set of conditions enabling flip-flops to occur. 

We notice similarities in the light curves of some, but not all, black hole transients featuring flip-flops. Swift J1658.2-4242, XTE J1859+226 (\hyperlink{C04}{C04}, and \citeads{2013ApJ...775...28S}), H1743-322 (\hyperlink{H05}{H05}), possibly also the 2002-2003 outburst of GX 339-4 (\hyperlink{N03}{N03}), all feature an initial flip-flop region at the brightest part of the outburst, which is followed by several days of almost constant flux, before a late flip-flop interval is initiated. Flip-flops in this interval were only detected and described for Swift J1658.2-4242, and XTE J1859+226, but comparable regions in the light curves of other black hole transients appear very similar and might also contain flip-flops. None of these systems exhibits a third region that appears to contain flip-flops. These similarities suggest that flip-flops could be linked to a particular type of evolution of the outburst, which only occurs for some BHTs. 

An intriguing distinction between the times when flip-flops occurred, and when they did not, is seen in the fitted truncation radius of the accretion disk. Fig. \ref{Specparcpl} points to a consistent value of the \texttt{diskbb} normalization for all flip-flop states, both within the early, and the late flip-flop interval. Other observations within a similar part of the outburst, which do not feature flip-flops, have a range of different values for this normalization, but these are all a factor of 2--4 smaller than what is found within the flip-flop intervals. The \texttt{diskbb} normalization is defined as:  \newline$N_{dbb} = \left(\frac{R_{in}}{D_{10}}\right)^2 \cos{i}$, where $R_{in}$ is the inner radius of the accretion disk, $D_{10}$ is the distance to the source, in units of 10 kpc, and $i$ is the inclination. The distance to the system can be assumed to be constant, so as long as the inclination does not vary a lot, the distinction of the \texttt{diskbb} normalizations is due to a difference in the truncation radii. This result would therefore indicate that the disk is truncated at a very specific radius whenever flip-flops occur. Along the same reasoning, other times of the outburst, which do not feature flip-flops, have no consistency in their truncation radii. 

The fitted \texttt{diskbb} normalization falls into the range of $17-74$ within the NuSTAR spectra of observations after the initial flux rise. Even though we do not know the exact inclination of the accretion disk relative to the line of sight, we assumed it to be $\sim 70^{\circ}$. Along with the estimate of the distance to the system, of $\sim10~\mathrm{kpc}$ (\citeads{2019ApJ...875..157J}), we can approximate the range of \texttt{diskbb} normalizations to correspond to accretion disk truncation radii in the range of $12\textup{--}25~\mathrm{km}$. We used the conversion by \citeads{1998PASJ...50..667K}, to obtain the true truncation radius from the spectrally fitted one. This correction increases the radius by $19\textup{--}65\%$, depending on the choice of $\kappa$, and we chose the maximum of $\kappa=2.0$. The resulting radii are still too small compared with the gravitational radius of a $10\,M_\odot$ black hole, of $14.9~\mathrm{km}$. The small spectrally fitted inner truncation radii could be a consequence of an overestimated black hole mass, or an underestimated inclination, or distance to the black hole. But all of these quantities cannot differ a lot from their estimated values. \citeads{2009ApJ...707L..87T} however points out that mixed results were obtained from using the spectrally fitted normalization of the disk black body component to estimate the inner radius. Additionally, in the relativistic rigid body precession model of the QPO, its frequency is determined by the truncation radius of the accretion disk. A larger truncation radius should yield a lower QPO frequency. However, the highest QPO frequencies were observed in the dim flip-flop states, when the spectrally fitted truncation radius was at its largest. The accretion disk inner radius as a function of time remains uncertain, but is nonetheless seemingly very close to the black hole in the brightest phase of the outburst, particularly whenever there is no indication of flip-flop activity. 

We found that the major spectral difference between the dim and bright flip-flop states is a result of an increase of the inner disk temperature, at a constant inner radius. An additional increase of the power law normalization, and the hydrogen column density are required to explain most of the remaining change in flux and spectral shape between the dim and bright flip-flop states. 

The flip-flop interval features a decrease in flip-flop amplitude over time, and this decrease seems to continue on into the late flip-flop phase, for which both the absolute, and fractional flux difference is substantially smaller than it was in the early flip-flop interval. This suggests an association of flip-flops with a damped oscillation with an irregular period. In this picture, the mechanism responsible gradually loses its amplitude, and approaches an equilibrium state. Perhaps the jet ejection triggered an imbalance in the system which initiates an instability that pushes the system from one unstable state to another one, resulting in an oscillation between the two unstable states.  

The physical mechanism causing flip-flop to occur, remains unknown. The association of flip-flops with alterations of QPOs, suggest that an understanding of the former might require a knowledge of the latter. The search for a model describing the cause of flip-flops is complicated by the ongoing discussion about the origin of QPOs. However, as our findings of transitions between QPO types A and C demonstrated, an understanding of flip-flops could aid in the search for the origin of QPOs as well. Below, we present some ideas to provide the reader a flavour of some of the ingredients that could play a role in shaping the flip-flop phenomenology.

Flip-flops are described by two timescales; the duration of a state before the next transition, and the time the BHT takes to change from one state to the other. For Swift J1658.2-4242, these timescales are of order $\sim 10~\mathrm{ks}$, and $\sim 100~\mathrm{s}$, respectively. Other BHTs feature flip-flops with a wide variety of different timescales, but the ratio between the two seems to be reasonably similar. This could suggest that the two timescales are related to each other, and the physical parameters of the disk, and the black hole. Indeed, this ratio is consistent with the expected ratio of viscous to thermal timescales within the inner region of a standard accretion disk.

A thermal instability could also account for the observed temperature variation observed during flip-flop transitions. On the other hand, we note that thermal-viscous instabilities typically result in much larger temperature variations (see e.g. \citeads{2018ApJ...857....1F}) than we observed in the flip-flops (of at most 20\%). It also remains unclear how such an instability could generate the observed change of the QPO, or the underlying clock in the times of transition.

A completely different framework posits the existence of an ionized, relativistic, beamed outflow misaligned with the spin of the black hole, therefore precessing with a complex pattern. Our discovery of an underlying clock governing the times of flip-flop transitions indicates the existence of a fundamental periodicity in the system, which would be achieved by this model. An increase in flux from dim to bright states would be the result of bulk comptonization of disk and corona emission whenever the outflow points into the line of sight. The bulk comptonization of this outflow might explain the observed spectral changes between dim and bright states, namely the increase in temperature and power law normalisation (eventually also the increased hydrogen column density, if the outflow is multi-phase) in the bright states. To explain the variation in the PSD we would require this outflow to have a significant radial extent, opening angle, and optical depth, to ensure that a narrow type C QPO would be broadened and weakened into a type A QPO due to scattering of radiation, and the corresponding time lags across this outflow.

This model provides a straightforward explanation for the near-consistency of the flux differences of adjacent flip-flop transitions. It also immediately justifies why the type C QPO that appears after the end of a bright state is so similar to the type C QPO observed just before the start of the bright state; it was there throughout, but was scattered to a type A QPO in the bright state. The pause in flip-flop activity would be explained by the outflow never pointing into the line of sight for a prolonged period of time. Eventually, the complex system of rotations would point the outflow back into the line of sight once in a while, and the flip-flops would be seen yet again, but this time with a lower amplitude of flip-flop flux variation. Alternatively, the pause in flip-flop activity could be the result of the outflow being quenched, and subsequently re-established.

We tried to model this hypothesis with a precessing conical outflow having a constant opening angle. We defined the system to be in a bright flip-flop state, whenever the angle between the line of sight and the outflow centre was less than the opening angle. Larger angles of separation were instead assigned to the dim state. We used this binary model to fit the occurrence of bright and dim states in our observations and compared our results with those found in randomly generated flip-flop states, which had the same number of transitions within our observing times. We were unable to fit our observations well without adding even more complexity, suggesting that this simplistic scenario can also not explain the wealth of observational constraints. Using this model, it was possible to fit randomly generated flip-flop states equally well in $45\%$ of our simulations. Our simulations therefore do not support this particular interpretation with these specific assumptions. It is however possible that a more complex system of rotations is required, or that the assumptions we used, in particular of a constant opening angle, are too simplistic. We also note that this model cannot account for the flip-flops observed in other systems, with transitions between different sets of QPO types.

In conclusion, some of the requirements of this model seem to be rather extreme, even though they are still within the realms of physical possibility. A clear verifiable prediction of this model, is a high degree of polarization in the bright flip-flop states, which could be investigated by future X-ray polarimetric observations by \textit{IXPE} or \textit{eXTP}. This model should be regarded as tentative, and incomplete. 

Finally, we note that it is possible for flip-flops to be caused by a combination of several different effects, rather than one single effect, as we have considered here.

\section{Conclusions}

We observed the black hole candidate Swift J1658.2-4242 throughout its outburst from February to September 2018, using a suite of X-ray and radio instruments. Swift J1658.2-4242 underwent some extreme changes in its light curve and PDS, which have never been seen before. 

Flip-flops with flux ratios of up to $1.77$ were observed, which occurred simultaneous to changes between QPO types A and C. We report the first detection of a direct transition between these two QPO types. The presence of a type B QPO with a duration of longer than $\sim 10~\mathrm{s}$ is ruled out in the transitions from QPO types A to C. The start of a type C QPO during a bright to dim transition was observed to be delayed relative to the start of the transition in the light curve, but the QPO changed significantly faster.

A major radio flare was detected at the same time as the greatest X-ray flux was reached, and the flip-flop interval started, suggesting that the two events are related. We found a second interval of flip-flop activity, which featured smaller flux ratios between the two states, a greater frequency of transitions, no identifiable QPO, but a significant change in total rms between the bright and dim flip-flop states. It started about 16 days after the end of the first flip-flop interval. But only one radio flare was observed throughout the entire outburst.

We found flip-flop transitions at seemingly random integer multiples of $2.761~\mathrm{ks}$ after the time of the first flip-flop. A slightly lower period of $2.61~\mathrm{ks}$ was observed in the late flip-flops. The existence of an underlying period in both intervals, which are separated by no more than 5.5\%, was detected with a significance of $3.2\sigma$. This suggests an inherent base timescale, which defines when flip-flop transitions can occur. 

Spectral fitting of flip-flop states revealed that the major increase in flux, and the slight change in their spectral shape was primarily the result of an increase in the inner accretion disk temperature. The remaining disparity was caused by an additional increase of the power law normalization and hydrogen column density. We found that the inner accretion disk radius remained almost constant throughout all flip-flop states and was considerably larger than at times without any flip-flops.

We demonstrated the necessity of using a DSH model to describe the spectrum of Swift J1658.2-4242, and other highly obscured sources. Energy corrections were applied to \textit{NuSTAR} and \textit{XMM-Newton} spectra to ensure consistency between them, and the \textit{Chandra} spectrum. 

We consider a misaligned, precessing, and beamed outflow as a tentative explanation for the phenomenology we observed. In this model, flip-flops would be generated at semi-periodic times whenever the outflow passes into the line of sight. 

Flip-flops have rarely been observed so far, but this might reflect more on the nature of past observations, rather than their fundamental rarity, as long continuous observations are required to detect them. We encourage further multiwavelength analysis of similar phenomena in other systems, and a continued search for the physical nature of flip-flops. 

\section*{Acknowledgements}

DB acknowledges financial support from the Bundesministerium f{\"u}r Wirtschaft und Energie (BMWi) / Deutsches Zentrum f{\"u}r Luft- und Raumfahrt (DLR) grant FKZ 50 OR 1812. TMB and GP acknowledge financial contribution from the agreement ASI-INAF n.2017-14-H.0. GP acknowledges funding from the European Research Council (ERC) under the European Union's Horizon 2020 research and innovation programme (grant agreement No. [865637]). CJ acknowledges the National Natural Science Foundation of China through grant 11873054, and the support by the Strategic Pioneer Program on Space Science, Chinese Academy of Sciences through grant XDA15052100. HP acknowledges the National Natural Science Foundation of China through grant 11803047 and U1838202. TDR is supported by a Netherlands Organisation for Scientific Research Veni award. TMD acknowledges support from the Spanish grants AYA2017-83216-P and RYC-2015-18148. PV acknowledges partial funding support from the DST-INSPIRE fellowship from the Government of India. FV acknowledges support from STFC under grant ST/R000638/1

This research has made use of data and/or software provided by the High Energy Astrophysics Science Archive Research Center (HEASARC), which is a service of the Astrophysics Science Division at NASA/GSFC and the High Energy Astrophysics Division of the Smithsonian Astrophysical Observatory. This research has made use of the \textit{NuSTAR} Data Analysis Software (NuSTARDAS) jointly developed by the ASI Science Data Center (ASDC, Italy) and the California Institute of Technology (Caltech, USA). This publication uses the data from the AstroSat mission of the Indian Space Research Organisation (ISRO), archived at the Indian Space Science Data Centre (ISSDC). This work made use of data supplied by the UK Swift Science Data Centre at the University of Leicester. This work made use of the data from the Insight-HXMT mission, a project funded by China National Space Administration (CNSA) and the Chinese Academy of Sciences (CAS). This research has made use of data obtained from the Chandra Data Archive and the Chandra Source Catalog, and software provided by the Chandra X-ray Center (CXC) in the application packages CIAO, ChIPS, and Sherpa. This research is based on observations with INTEGRAL, an ESA project with instruments and science data centre funded by ESA member states (especially the PI countries: Denmark, France, Germany, Italy, Switzerland, Spain), and with the participation of the Russian Federation and the USA. This paper includes archived data obtained through the Australia Telescope Online Archive (http://atoa.atnf.csiro.au). The Australia Telescope Compact Array is part of the Australia Telescope National Facility which is funded by the Australian Government for operation as a National Facility managed by CSIRO.

The authors thank the anonymous referee for providing useful comments which improved the paper.

\bibliographystyle{aa} 
\bibliography{bibliography.bib}      

\appendix

\section{Table of ATCA radio observations}

\begin{table}[h]
\centering
\setlength{\tabcolsep}{2pt}
\def\arraystretch{1.1}
\begin{tabular}{l l l l l}
    \begin{tabular}{@{}l}
         \textbf{Start time}\\
         \textbf{(MJD-5800)}\\
    \end{tabular} & 
    \begin{tabular}{@{}l}
         \textbf{Exposure}\\
         \textbf{(ks)}\\
    \end{tabular} &
    \begin{tabular}{@{}l}
         \textbf{Config}\\
         \\
    \end{tabular} &
    \begin{tabular}{@{}l}
         \textbf{Frequency}\\
         \textbf{(GHz)}\\
    \end{tabular} &
    \begin{tabular}{@{}l}
         \textbf{Flux density}\\
         \textbf{(mJy)}\\
    \end{tabular} \\ \hline \hline 
    
    166.83680 & 23.05 & 750B &
    \begin{tabular}{@{}l}
         5.50 \\
         9.00 \\
    \end{tabular} &
    \begin{tabular}{@{}l}
         $2.35\pm0.17$\\
         $2.17\pm0.15$ \\
    \end{tabular}  \\
    
    166.82534 & 22.36 & 750B &
    \begin{tabular}{@{}l}
         17.0 \\
         19.0 \\
    \end{tabular} &
    \begin{tabular}{@{}l}
         $2.27\pm0.09$\\
         $2.30\pm0.15$ \\
    \end{tabular}  \\ \hline
    
    171.93565 & 14.32 & 750B &
    \begin{tabular}{@{}l}
         5.50 \\
         9.00 \\
    \end{tabular} &
    \begin{tabular}{@{}l}
         $5.95\pm0.10$\\
         $4.80\pm0.07$ \\
    \end{tabular}  \\
    
    171.92430 & 14.33 & 750B &
    \begin{tabular}{@{}l}
         17.0 \\
         19.0 \\
    \end{tabular} &
    \begin{tabular}{@{}l}
         $3.35\pm0.12$\\
         $3.30\pm0.20$ \\
    \end{tabular}  \\ \hline  
    
    173.87801 & 23.01 & 750B &
    \begin{tabular}{@{}l}
         5.50 \\
         9.00 \\
    \end{tabular} &
    \begin{tabular}{@{}l}
         $50.70\pm0.60$\\
         $39.50\pm0.90$ \\
    \end{tabular}  \\
    
    173.89942 & 19.80 & 750B &
    \begin{tabular}{@{}l}
         17.0 \\
         19.0 \\
    \end{tabular} &
    \begin{tabular}{@{}l}
         $29.60\pm0.50$\\
         $28.85\pm0.50$ \\
    \end{tabular}  \\ \hline  
    
    175.62030 & 16.2 & 750B & 8.40 & $11.27\pm0.29$ \\ \hline
 
    176.59271 & 24.00 & 750B &
    \begin{tabular}{@{}l}
         5.50 \\
         9.00 \\
    \end{tabular} &
    \begin{tabular}{@{}l}
         $6.20\pm0.20$\\
         $4.75\pm0.20$ \\
    \end{tabular}  \\
    
    176.61366 & 19.77 & 750B &
    \begin{tabular}{@{}l}
         17.0 \\
         19.0 \\
    \end{tabular} &
    \begin{tabular}{@{}l}
         $3.00\pm0.15$\\
         $2.90\pm0.15$ \\
    \end{tabular}  \\ \hline  
    
    182.83993 & 2.81 & EW352 &
    \begin{tabular}{@{}l}
         5.50 \\
         9.00 \\
    \end{tabular} &
    \begin{tabular}{@{}l}
         $1.76\pm0.30$\\
         $1.24\pm0.10$ \\
    \end{tabular}  \\ \hline
    
    185.62130 & 7.43 & EW352 &
    \begin{tabular}{@{}l}
         5.50 \\
         9.00 \\
    \end{tabular} &
    \begin{tabular}{@{}l}
         $5.30\pm0.35$\\
         $4.10\pm0.35$ \\
    \end{tabular}  \\ \hline 
    
    192.93941 & 11.84 & EW352 & 8.40 & $0.70\pm0.20$ \\ \hline
    
    194.52511 & 1.21 & EW352 &
    \begin{tabular}{@{}l}
         5.50 \\
         9.00 \\
    \end{tabular} &
    \begin{tabular}{@{}l}
         $0.95\pm0.25$\\
         $0.98\pm0.25$ \\
    \end{tabular}  \\ \hline
    
    235.88102 & 8.02 & H168 &
    \begin{tabular}{@{}l}
         5.50 \\
         9.00 \\
    \end{tabular} &
    \begin{tabular}{@{}l}
         $<0.3$\\
         $<0.3$ \\
    \end{tabular}  \\ \hline  

\end{tabular}
\vspace{1 mm}
\caption{List of radio observations of Swift J1658.2-4242 by \textit{ATCA}. No radio emision was detected in the last observation, so the $3\sigma$ upper limit is quoted instead. \label{listobsR}}
\end{table}

\section{Spectral fitting results}

\begin{table*}[h]
\raggedright
\setlength{\tabcolsep}{2pt}
\def\arraystretch{1.25}
\begin{tabular}{ll||ll|llll}
    \textbf{Model 1} & & \textbf{80301301002}& & \textbf{80302302002}& & &   \\ 
    & & bright state & dim state & bright state 1 & dim state 1 & bright state 2 & dim state 2 \\  \hline \hline
    \texttt{ztbabs} & $N_H \,(10^{23}~\mathrm{cm}^{-2})$ & $14.5\pm0.6$& $12.3\pm0.2$& $13.8\pm0.4$& $13.2\pm0.7$& $13.2^{+0.7}_{-0.9}$& $11.9\pm0.3$\\ 
    & $z$ & $0.012$& $0.014$& $0.009$& $0.017$& $0.007$& $0.015$ \\\hline
    \texttt{diskbb} & $kT_{in} \,(\mathrm{keV})$ & $1.44\pm0.03$& $1.28\pm0.01$& $1.46\pm0.02$& $1.29^{+0.04}_{-0.03}$& $1.52\pm0.04$& $1.27\pm0.02$\\
    & $N_{dbb}$ & $67^{+6}_{-5}$& $74\pm3$& $61^{+4}_{-3}$& $73^{+11}_{-9}$& $54^{+8}_{-6}$& $74^{+5}_{-4}$ \\ \hline
    \texttt{diskline} & $B_{10}$ & $-2.4^{+0.3}_{-0.2}$ & $-1.8^{+0.2}_{-0.1}$& $-2.1\pm0.2$& $-1.7^{+0.8}_{-0.4}$& $-1.4$& $-1.8\pm0.2$ \\
    & $N_{dl}\,(10^{-3})$ & $4.5\pm0.1$& $2.2\pm0.3$& $4\pm1$& $2.2^{+1}_{-0.8}$& $1^{+17}_{-1}$& $2.0\pm0.4$ \\ \hline
    \texttt{cutoffpl} & $\Gamma$ & $2.14\pm0.09$& $2.06\pm0.04$& $2.13\pm0.07$& $2.3\pm0.1$& $2.1^{+0.2}_{-0.1}$& $2.00\pm0.06$\\
    & $E_{cut}\,(\mathrm{keV})$ & $60^{+20}_{-10}$& $49^{+5}_{-4}$& $56^{+11}_{-8}$& $110^{+220}_{-50}$& $60^{+30}_{-10}$& $44^{+5}_{-4}$ \\
    & $N_{cpl}$ &  $3.5^{+0.8}_{-0.7}$& $1.6\pm0.1$& $3.1\pm0.5$& $2.7^{+0.8}_{-0.7}$& $2.4^{+1.1}_{-0.7}$& $1.3^{+0.2}_{-0.1}$ \\ \hline
    & $\chi^2/\nu$ &  $1803.73/1699$& $2657.4/2446$& $2034.98/1863$& $1312.25/1249$& $1039.48/1112$& $2243.91/2073$ \\
    & & $=1.0616$& $=1.0864$& $=1.0923$& $=1.0506$& $=0.9424$& $=1.0825$ \\
\end{tabular}
\end{table*}
        
\begin{table*}[h]
\raggedright
\setlength{\tabcolsep}{2pt}
\def\arraystretch{1.25}
\begin{tabular}{ll||llll|l|l}
    \textbf{Model 1} & & \textbf{80302302004}& & & &\textbf{80302302006} & \textbf{80302302008} \\ 
    & & bright state 1 & dim state 1 & bright state 2 & dim state 2 & entire obs. & entire obs. \\  \hline \hline
    \texttt{ztbabs} & $N_H \,(10^{23}~\mathrm{cm}^{-2})$ & $13.1\pm0.3$& $11.3\pm0.3$& $12.7\pm0.8$ & $11.3\pm0.5$& $9.4^{+0.3}_{-0.2}$& $10.8\pm0.2$ \\ 
    & $z$ & $0.008$& $0.016$& $0.013$ & $0.018$& $0.011$& $0.010$ \\\hline
    \texttt{diskbb} & $kT_{in} \,(\mathrm{keV})$ & $1.40\pm0.02$& $1.31\pm0.03$& $1.44\pm0.05$& $1.28^{+0.03}_{-0.04}$& $1.37\pm0.03$& $1.33\pm0.02$\\
    & $N_{dbb}$ & $74\pm4$& $70^{+4}_{-3}$& $56^{+8}_{-6}$ & $65^{+7}_{-6}$& $20\pm1$& $36\pm2$\\ \hline
    \texttt{diskline} & $B_{10}$& $-2.0^{+0.3}_{-0.2}$& $-1.6^{+0.3}_{-0.2}$& $-2.1^{+0.7}_{-0.3}$ & $-1.8^{+0.4}_{-0.2}$& $-1.80^{+0.10}_{-0.09}$& $-1.7\pm0.1$\\
    & $N_{dl}\,(10^{-3})$ & $2.8^{+0.9}_{-0.8}$& $1.7^{+0.4}_{-0.3}$& $4\pm2$& $2.2^{+0.7}_{-0.6}$& $2.4\pm0.2$& $1.6\pm0.2$ \\ \hline
    \texttt{cutoffpl} & $\Gamma$ & $1.95\pm0.08$& $1.83\pm0.07$& $2.0\pm0.2$& $2.0^{+0.10}_{-0.09}$& $1.66\pm0.05$& $1.75\pm0.05$ \\
    & $E_{cut}\,(\mathrm{keV})$ & $42^{+7}_{-5}$& $34^{+4}_{-3}$& $50^{+30}_{-10}$& $46^{+8}_{-11}$& $32\pm2$& $36\pm3$ \\
    & $N_{cpl}$ & $1.8\pm0.3$& $0.9\pm0.1$& $1.9^{+0.8}_{-0.6}$ & $1.2^{+0.3}_{-0.2}$& $0.46\pm0.05$& $0.54\pm0.06$ \\ \hline
    & $\chi^2/\nu$ & $1916.61/1840$& $2103.89/1941$& $1225.44/1201$& $1702.75/1572$& $2823.57/2362$& $2445.8/2346$ \\
    & & $=1.0416$& $=1.0839$& $=1.0204$& $1.0832$& $=1.1954$& $=1.0425$ \\
\end{tabular}
\end{table*}
        
\begin{table*}[h]
\raggedright 
\setlength{\tabcolsep}{2pt}
\def\arraystretch{1.2}
\begin{tabular}{ll||llllll|l}
    \textbf{Model 1} & & \textbf{80302302010}& & & & & & \textbf{90401317002}   \\ 
    & & dim state 1& bright state 1 & bright state 2 & dim state 2 & dim state 3 & bright state 3 & entire obs. \\  \hline \hline
    \texttt{ztbabs} & $N_H \,(10^{23}~\mathrm{cm}^{-2})$ &$10.3^{+0.5}_{-0.4}$ &$11.0^{+0.6}_{-0.5}$ &$10.7^{+0.5}_{-0.4}$ &$10^{+2}_{-1}$ &$10.1^{+0.6}_{-0.5}$ &$11.2^{+0.5}_{-0.4}$ &$8.2\pm0.3$  \\
    & $z$ &$0.013$ &$0.020$ &$0.019$ &$0.004$ &$0.026$ &$0.032$ &$0.019$ \\ \hline
    \texttt{diskbb} & $kT_{in} \,(\mathrm{keV})$ & $1.21\pm0.02$& $1.28\pm0.03$& $1.30^{+0.02}_{-0.03}$& $1.22^{+0.06}_{-0.09}$& $1.24\pm0.03$& $1.27^{+0.02}_{-0.03}$& $1.33^{+0.01}_{-0.02}$\\
    & $N_{dbb}$& $80^{+8}_{-7}$& $66^{+7}_{-5}$& $57^{+5}_{-3}$& $70^{+30}_{-10}$& $71^{+8}_{-7}$& $62^{+6}_{-5}$& $17.0^{+1.0}_{-0.8}$ \\ \hline
    \texttt{diskline} & $B_{10}$& $-1.7^{+1.4}_{0.57}$& $-1.1\pm1$& $-2.1$& $-1.7$& $-1.5$& $-1.8^{+1.0}_{-0.5}$& $-1.1^{+1.0}_{-0.8}$ \\
    & $N_{dl}\,(10^{-3})$& $0.4^{+0.3}_{-0.2}$& $0.5\pm0.3$& $0.3^{+0.6}_{-0.3}$& $0.5^{+1.5}_{-0.5}$& $0.4\pm0.3$& $0.7\pm0.4$& $0.14^{+0.09}_{-0.06})$ \\ \hline
    \texttt{cutoffpl} & $\Gamma$& $1.8\pm0.1$& $1.8\pm0.2$& $2.0\pm0.1$& $2.0^{+0.2}_{-0.3}$& $1.6\pm0.2$& $2.1\pm0.1$& $1.90^{+0.0.06}_{-0.07}$ \\
    & $E_{cut} \,(\mathrm{keV})$& $60^{+30}_{-10}$& $40^{+20}_{-10}$& $70^{+40}_{-20}$& $130$& $36^{+15}_{-9}$& $100^{+100}_{-40}$& $100^{+90}_{-40}$ \\
    & $N_{cpl}$& $0.31^{+0.09}_{-0.07}$& $0.4^{+0.2}_{-0.1}$& $0.6^{+0.2}_{-0.1}$& $0.4^{+0.3}_{-0.2}$& $0.20^{+0.09}_{-0.07}$& $0.8\pm0.2$& $0.23\pm0.03$ \\ \hline
    & $\chi^2/\nu$& $1478.19/1425$& $1220.84/1272$& $1569.68/1551$& $612.87/595$& $1224.99/1169$& $1531.76/1455$& $2178.77/2066$ \\
    & & $=1.0373$& $=0.9598$& $=1.0121$& $=1.03$& $=1.0479$& $=1.0528$& $=1.0546$\\
\end{tabular}
\vspace{3 mm}
\caption{Tables of best fit parameters using Model 1, \texttt{dscor*ztbabs*constant*(diskbb+diskline+cutoffpl)}
\label{Tabspecfitcpl}}
\end{table*}

\begin{table*}
\raggedright 
\setlength{\tabcolsep}{2pt}
\def\arraystretch{1.2}
\begin{tabular}{ll||ll|llll}
    \textbf{Model 2} & & \textbf{80301301002}& & \textbf{80302302002}& & & \\ 
    & & bright state & dim state & bright state 1 & dim state 1 & bright state 2 & dim state 2 \\  \hline \hline
    \texttt{ztbabs} & $N_H \,(10^{23}~\mathrm{cm}^{-2})$ & $12.4^{+0.4}_{-0.3}$& $10.8\pm0.2$& $11.7\pm0.3$& $11.2\pm0.4$& $11.6^{+0.4}_{-0.6}$& $10.3\pm0.2$ \\
    & $z$ & $0.012$& $0.014$& $0.009$& $0.017$& $0.007$& $0.015$ \\\hline
    \texttt{diskline} & $B_{10}$ &$-1.6^{+1.6}_{-0.5}$ &$-1.3^{+0.4}_{-0.2}$ &$-1.5\pm0.3$ &$-1.4^{+1.0}_{-0.4}$ &$-0.9\pm0.9$ &$-1.2^{+0.6}_{-0.4}$  \\
    & $N_{dl}\,(10^{-3})$ &$2\pm1$ &$1.5\pm0.3$ &$1.9^{+1.0}_{-0.8}$ &$1.9^{+0.9}_{-0.7}$ &$0.7^{+1.6}_{-0.87}$ &$1.2\pm0.3$   \\ \hline
    \texttt{compPS} & $k T_e\,(\mathrm{keV})$ &$72\pm2$ &$80\pm1$ &$74\pm2$ &$89\pm4$ &$83^{+6}_{-5}$ &$79\pm2$   \\
    & $kT_{in} \,(\mathrm{keV})$ &$-1.45\pm0.04$ &$-1.31\pm0.01$ &$-1.45\pm0.03$ &$-1.32\pm0.03$ &$-1.49^{+0.05}_{-0.06}$ &$-1.31\pm0.02$  \\
    & $\tau_y$&$0.59\pm0.04$ &$0.507\pm0.008$ &$0.57\pm0.03$ &$0.45\pm0.02$ &$0.51\pm0.05$ &$0.51\pm0.01$ \\
    & $R_r$ &$0.3^{+0.2}_{-0.1}$ &$0.29\pm0.07$ &$0.57\pm0.03$ &$0\pm0.1$ &$0.1^{+0.3}_{-0.1}$ &$0.4\pm0.1$ \\
    & $N_{cps}$ &$130\pm20$ &$117\pm6$ &$120\pm10$ &$130^{+20}_{-10}$ &$110^{+2}_{-20}$ &$110^{+9}_{-7}$  \\ \hline
    & $\chi^2/\nu$ &$1774.47/1699$ &$2616.27/2446$ &$2004.43/1863$ &$1315.51/1249$ &$1039.21/1103$ &$2211.8/2211$  \\
    & &$=1.0444$ &$=1.0696$ &$=1.0759$ &$=1.0533$ &$=0.94216$ &$=1.0670$  \\
\end{tabular}
\end{table*}       
        
\begin{table*}
\raggedright 
\setlength{\tabcolsep}{2pt}
\def\arraystretch{1.2}
\begin{tabular}{ll||llll|l|l}
    \textbf{Model 2} & & \textbf{80302302004}& & & & \textbf{80302302006}& \textbf{80302302008} \\ 
    & & dim state 1& bright state 1& dim state 2& bright state 2& entire obs. & entire obs. \\  \hline \hline
    \texttt{ztbabs} & $N_H \,(10^{23}~\mathrm{cm}^{-2})$ & $11.6\pm0.3$& $10.1\pm0.2$& $11.0\pm0.6$& $9.8\pm0.4$& $7.3\pm0.2$& $9.3\pm0.2$ \\
    & $z$ & $0.008$& $0.016$& $0.013$ & $0.018$& $0.011$& $0.010$ \\\hline
    \texttt{diskline} & $B_{10}$ &$-1.3^{+1.3}_{-0.5}$ &$-1.4$ &$-1.5^{+1.5}_{-0.6}$ &$-1.4^{+0.8}_{-0.4}$ &$-1.0^{+0.5}_{-0.3}$ &$-1.0^{+0.9}_{-0.4}$ \\
    & $N_{dl}\,(10^{-3})$ &$1.1^{+0.8}_{-0.7}$ &$1.1\pm0.2$ &$2^{+2}_{-1}$ &$1.4^{+0.6}_{-0.5}$ &$1.0\pm0.2$ &$0.8\pm0.2$ \\ \hline
    \texttt{compPS} & $k T_e\,(\mathrm{keV})$ &$80\pm2$ &$83\pm2$ &$80\pm5$ &$81\pm3$ &$58\pm2$ &$80\pm1$ \\
    & $kT_{in} \,(\mathrm{keV})$ &$-1.43\pm0.02$ &$-1.34\pm0.02$ &$-1.46^{+7}_{-6}$ &$-1.32\pm0.03$ &$-1.49^{+0.04}_{-0.03}$ &$-1.38\pm0.02$ \\
    & $\tau_y$&$0.48\pm0.02$ &$0.46\pm0.01$ &$0.52\pm0.06$ &$0.51\pm0.02$ &$0.91^{+0.06}_{-0.05}$ &$0.58\pm0.02$ \\
    & $R_r$ &$0.4\pm0.1$ &$0.5\pm0.1$ &$0.3\pm0.3$ &$0.4\pm0.2$ &$1.3\pm0.1\pm$ &$0.7\pm0.1$ \\
    & $N_{cps}$ &$110^{+11}_{-9}$ &$97^{+7}_{-6}$ &$90\pm20$ &$100\pm10$ &$27\pm3$ &$51^{+4}_{-3}$ \\ \hline
    & $\chi^2/\nu$ &$1890.66/1840$ &$2092.49/1942$ &$1219.77/1201$ &$1687.81/1572$ &$2653.1/2362$ &$2393.55/2346$ \\
    & &$=1.0275$ &$=1.0775$ &$=1.0156$ &$=1.07367$ &$=1.1232$ &$=1.0203$ \\
\end{tabular}
\end{table*}

\begin{table*}
\raggedright 
\setlength{\tabcolsep}{2pt}
\def\arraystretch{1.2}
\begin{tabular}{ll||llllll|l}
    \textbf{Model 2} & & \textbf{80302302010}& & & & & & \textbf{90401317002}  \\ 
    & & dim state 1& bright state 1 & bright state 2 & dim state 2 & dim state 3 & bright state 3 & entire obs. \\  \hline \hline
    \texttt{ztbabs} & $N_H \,(10^{23}~\mathrm{cm}^{-2})$ & $9.7\pm0.2$& $10.4\pm0.5$& $9.7^{+0.2}_{-0.3}$& $8.4^{+1.0}_{-0.9}$& $98\pm0.5$& $9.8^{+0.3}_{-0.2}$& $7.0\pm0.2$ \\
    & $z$ &$0.013$ &$0.020$ &$0.019$ &$0.004$ &$0.026$ &$0.032$ &$0.019$ \\ \hline
    \texttt{diskline} & $B_{10}$ &$-1.5$ &$-1.5$ &$-1.4$ &$-1.5$ &$-1.5$ &$-1.2^{+1.2}_{-0.6}$ &$-0.6^{+0.6}_{-2.0}$ \\
    & $N_{dl}\,(10^{-3})$ &$0.3\pm0.2$ &$0.5\pm0.3$ &$0.1^{+0.3}_{-0.1}$ &$0.3^{+0.7}_{-0.3}$ &$0.4\pm0.3$ &$0.4\pm0.3$ &$0.11\pm0.06$  \\ \hline
    \texttt{compPS} & $k T_e\,(\mathrm{keV})$ &$153^{+3}_{-6}$ &$118\pm7$ &$131^{+3}_{-2}$ &$160^{+10}_{-20}$ &$147^{+4}_{-9}$ &$130\pm4$ &$150\pm3$  \\
    & $kT_{in} \,(\mathrm{keV})$ &$-1.25^{+0.02}_{-0.01}$ &$-1.30\pm0.03$ &$-1.35^{+0.02}_{-0.01}$ &$-1.28^{+0.07}_{-0.04}$ &$-1.28^{+0.03}_{-0.02}$ &$-1.34\pm0.02$ &$-1.39\pm0.01$ \\
    & $\tau_y$&$0.182^{+0.006}_{-0.005}$ &$0.27\pm0.01$ &$0.264^{+0.008}_{-0.007}$ &$0.18$ &$0.185\pm0.008$ &$0.272\pm0.009$ &$0.339\pm0.007$ \\
    & $R_r$ &$0^{+0.08}_{-0}$ &$0.2\pm0.2$ &$0^{+0.05}_{-0}$ &$0$ &$0.07^{+0.28}_{-0.07}$ &$0^{+0.8}_{-0}$ &$0^{+0.01}_{-0}$ \\
    & $N_{cps}$ &$88^{+1}_{-5}$ &$82^{+10}_{-9}$ & $70^{+5}_{-2}$& $70\pm20$ &$79^{+6}_{-9}$ &$72\pm5$ &$22\pm1$ \\ \hline
    & $\chi^2/\nu$ &$1482.36/1425$ &$1219.41/1273$ &$1574.89/151$ &$614.49/596$ & $1228.62/1169$ &$1542.29/1455$ &$2207.33/2066$  \\
    & &$=1.0403$ &$=0.9579$ &$=1.0154$ &$=1.031$ & $=1.051$ &$=1.060$ &$=1.0684$  \\
\end{tabular}
\caption{Tables of best fit parameters using Model 2, \texttt{dscor*ztbabs*constant*(diskline+compPS)}
\label{Tabspecfitcps}}
\end{table*}

\end{document}